\documentclass[11pt]{article}

\usepackage{amsmath}
\usepackage{amssymb}
\usepackage{slashed}
\usepackage{graphicx}
\usepackage{graphics}
\usepackage{epsfig}
\usepackage{color}
\usepackage{xcolor}
\usepackage{soul}
\usepackage{hyperref}
\usepackage{upgreek}
\usepackage{authblk}
\usepackage{blindtext}
\usepackage[section]{placeins}
\usepackage{mathrsfs}
\usepackage{dsfont}
\usepackage[normalem]{ulem}

\newcommand{\be}{\begin{equation}}
\newcommand{\ee}{\end{equation}}
\newcommand{\bea}{\begin{eqnarray}}
\newcommand{\eea}{\end{eqnarrray}}

\newcommand{\dm}{n}

\newcommand{\bep}{\boldsymbol \upepsilon} 
\newcommand{\ca}{\mathcal A} 
\newcommand{\cA}{\mathcal A} 
\newcommand{\bM}{\mathbb M} 
\newcommand{\re}{\mathbb R} 
\newcommand{\pd}{\partial} 
\newcommand{\mbh}{\mathbb H}

\newcommand{\tc}[1]{t{(#1)}}

\newcommand{\av}[1]{\langle {#1} \rangle}
\newcommand{\vol}{\mathrm{vol}}
\newcommand{\cH}{\mathcal H}
\newcommand{\pprec}{\prec\prec} 
\newcommand{\rmin}{{\mathbf r}_m}
\newcommand{\cmin}{\mathbf{e}}
\newcommand{\ppd}{\mathbf d} 
\newcommand{\Ns}{N_{\Sigma}}

\newcommand{\cO}{\mathcal O} 
\newcommand{\Eps}{\epsilon_{\Sigma}}
\newcommand{\ro}{r_0} 
\newcommand{\Nh}{N_{(h)}}
\newcommand{\dmp}{D(M_p)}

\newcommand{\bB}{\mathbb B} 
\newcommand{\fut}[1]{\mathrm{fut}(#1)}
\newcommand{\past}[1]{\mathrm{past}(#1)}
\newcommand{\dd}{\widetilde {\mathbf {d}}} 
\newcommand{\dV}{\mathbf V}
\newcommand{\bN}{\mathbf N}
\newcommand{\DD}{\mathbf d}

\newcommand{\dimm}{n}
\newcommand{\Vas}{{V_{\rm AS}}}
\newcommand{\dflat}{d_{\re^2}}
\newcommand{\dg}{d_{\gamma}}
\newcommand{\dpl}{d_{W_k}}
\newcommand{\dist}{d}
\newcommand{\gkwk}{\gamma^{(W_k)}}

\newcommand{\mco}{\ell}
\newcommand{\lco}{L}
\newcommand{\dsig}{d_h}

\newcommand{\prd}{\widetilde {d}}

\newcommand{\dcont}{\dsig}
\newcommand{\cN}{\mathcal N}
\newcommand{\NA}{N_\ca}
\newcommand{\WW}{\chi}
\newcommand{\ww}{c} 
\newcommand{\cW}{\mathcal W} 
\newcommand{\nin}{\notin}
\newcommand{\wD}{\widetilde {\Delta}}
\usepackage{fullpage}

\author[1]{Astrid Eichhorn \thanks{a.eichhorn@thphys.uni-heidelberg.de}}

\author[2]{Sumati Surya}

\author[1]{Fleur Versteegen \thanks{f.versteegen@thphys.uni-heidelberg.de}} 

\affil[1]{\textit{{Institut f\"ur Theoretische
  Physik, Universit\"at Heidelberg, Philosophenweg 16, 69120
  Heidelberg, Germany}}}
\affil[2]{\textit{Raman Research Institute, C.V. Raman Avenue, Sadashivnagar, Bangalore
560 080, India}}
\title{Induced Spatial Geometry from Causal Structure}

\date{}
\begin{document}

\baselineskip=0.5cm

\maketitle
\begin{abstract}  
 Motivated by the Hawking-King-McCarthy-Malament (HKMM) theorem and the associated reconstruction of spacetime geometry from its causal structure $(M,\prec)$ and local volume element $\bep$, we define a one-parameter family of spatial distance functions on a Cauchy hypersurface $\Sigma$ 
using only $(M,\prec)$ and $\bep$. The parameter   corresponds to  a ``mesoscale'' cut-off which,  when appropriately chosen,  provides a distance function which approximates the induced spatial distance function to leading order.   This admits a straightforward generalisation to the discrete analogue of a Cauchy hypersurface in a  causal set. For causal sets which are  approximated by continuum spacetimes, this distance function approaches the continuum induced distance when the mesoscale  is much smaller than the scale of the extrinsic curvature of the hypersurface,  but much larger than the discreteness scale. We verify these expectations by  performing  extensive numerical simulations of causal sets which are approximated by simple spacetime regions in 2 and 3 spacetime dimensions. 
\end{abstract} 

\section{Introduction}
Underlying every causal Lorentzian spacetime $(M,g)$ is a partially ordered set (or poset) $(M,\prec)$ which is  its causal structure.  The centrality of $(M,\prec)$ in a causal  Lorentzian spacetime has been  recognised for more than half a century \cite{finkelstein, kp,gkp,zeeman} and underlined by the  Hawking-King-McCarthy and Malament (HKMM) theorem \cite{hkm, mal},  which states that under very weak causality conditions $(M,\prec)$ determines $(M,g)$  upto a conformal isometry. Formally, this implies that  $(M,g)$ is equivalent to $(M,\prec)$ plus a local volume element $\bep$, the volume form. Given the standard conception of spacetime as a (mild) generalisation of Riemannian geometry, this suggests an  alternative and radical order theoretic conception with  $(M,g)$ replaced by   $(M,\prec,\bep)$. One might regard unimodular gravity, where $\bep$ is fixed, as a modest first step in such a direction.  The nature of gravity as a theory of dynamical causal structure, encoded in the remaining nine components of the metric, is highlighted by the  dynamical equivalence of General Relativity with unimodular gravity \cite{unimodular}. 

An explicit   reconstruction of  familiar features of spacetime from  $(M,\prec,\bep)$ is challenging  in practice, but there are some simple examples for  which this has been done.    Though  $(M,\prec)$ comes with no prescribed differentiable structure,   the spacetime topology (and thence dimension) can be reconstructed from the causal relations in most cases. In strongly causal spacetimes this is done in a  relatively straightforward manner since the topology generated by the causal diamonds or Alexandrov sets can be shown to be equivalent to the manifold topology \cite{penrose}. More generally, the manifold topology can be recovered using the  convergence of totally ordered subsets of zero spacetime volume \cite{ps}.  These  zero volume  totally ordered subsets in turn have a ready geometric interpretation as non-focussing segments of null geodesics. Attempts at  proceeding  beyond this have proven to be non-trivial and there is no known {\it general}  prescription in the continuum for  timelike and spacelike geodesics,  let alone curvature invariants\footnote{An outline of how to do the full reconstruction for  $\bf M^4$ is given in \cite{Sorkin:2003bx}.}. 

Such a geometric reconstruction is  of particular relevance to causal set theory (CST) which is an approach to quantum gravity which adheres most closely to  the order theoretic paradigm suggested by the HKMM theorem\cite{Bombelli:1987aa, Henson:2010aq, Dowker:2005tz, Dowker:aza, Surya:2011yh}.  In CST, the additional assumption of spacetime discreteness allows a finite  cardinality to be associated with every causal interval in the causal set, and provides the analog of a local volume element $\bep$.  The resulting locally finite poset or {\sl causal set} $C \subset (M,\prec)$ therefore possesses all the information required to recover the spacetime geometry in the continuum approximation, i.e.,  $C \approx (M,\prec,\bep) $.  Indeed, some aspects of geometric reconstruction become simpler with the assumption of discreteness.  For example, a time-like geodesic segment between a (sufficiently nearby) pair of events  corresponds to the largest totally ordered set between them. This in  turn gives the proper time between the events  in units of the discreteness scale \cite{bg}. Spatial geodesics are more difficult to construct, but again, discreteness provides a way of 
 achieving this \cite{RW}.

Despite being discrete, manifold-like causal sets also satisfy local Lorentz invariance in that there are no preferred directions 
\cite{LLI}.  
Unlike a regular lattice, this implies that  a causal set underlying any continuum spacetime is not a fixed or  bounded valency graph. For example,  a causal set approximated by  flat spacetime has an infinite number of nearest neighbours, lying within past and future invariant hyperboloids hugging the light cone.     This in turn gives rise to an inherent non-locality characteristic of manifold-like causal sets \cite{Sorkin:2007qi}. 
While this results in some very interesting new phenomenology \cite{Belenchia:2014fda, Belenchia:2015aia, Saravani:2015rva, Belenchia:2015ake},  nonlocality makes the spatial geometry  much harder to construct. In a globally hyperbolic spacetime, for example, it is natural and advantageous to use a foliation by spatial (Cauchy)  hypersurfaces, thus splitting spacetime into ``space+time''.  This apparent sacrifice of covariance  is  useful since it allows us to obtain physically interesting predictions in classical gravity, for example, those for gravitational wave forms. Reconstruction of spatial geometry in a causal set is part of the broader question in CST of  how exactly non-locality and covariance together conspire to give rise to a  local dynamics resembling General Relativity in the continuum approximation.

 In this work we address the more modest question of whether the induced spatial geometry on a  Cauchy hypersurface can be recovered in an  appropriate sense from the causal order and the volume element, both in the continuum and in the causal set.  We take our cue from  \cite{homologyone,homologytwo}  where  
the homology  of a compact Cauchy hypersurface $\Sigma$ was constructed from the ambient causal structure: intersections of  causal intervals with $\Sigma$ provide a finite open covering of $\Sigma$,  from which a ``nerve'' simplicial complex can be formed. The  homology of this simplex was shown in  \cite{homologyone,homologytwo}  to be equivalent to that of $\Sigma$ when the covering is   ``fine  enough''  compared to the curvature scale of $\Sigma$.    

In Section \ref{continuum} we make a similar use of the ambient causal structure to 
obtain a one-parameter family of  distance functions $\dist_\mco$  on a compact Cauchy hypersurface $(\Sigma,h)$ in a globally hyperbolic spacetime  $(M,g)$. The construction uses  piecewise-flat approximations of Lorentzian geometry and is analogous to the standard Riemannian construction of approximate geometry  from piecewise linearisations (see Fig.~\ref{linearapprox}).  $\dist_\mco$ is obtained purely from $(M,\prec, \bep)$ and the embedding $(\Sigma, h)\subset (M,g)$, where $\mco > 0$ is a ``mesoscale'' cut off.  When  
{$\mco<<\ell_K$,}
the extrinsic and intrinsic   curvature scale of $(\Sigma, h)$, we find that for any $\epsilon>0$, there exists a small enough $\mco$ such that the difference between $\dist_\mco$ and the induced distance function obtained from $h$ is less than $\epsilon$.  In Section \ref{discrete} we show that  this construction is  readily carried over to causal sets and provides a one-parameter family of  distance functions $\ppd_\mco$ on the discrete analogue of a Cauchy hypersurface, an inextendible antichain $\cA$.  For a causal set that is approximated by a continuum spacetime with volume cut-off $\rho^{-1}$,  $\ppd_\mco$  limits to $\dist_\mco$ when $\mco$ is chosen appropriately.   As was shown in \cite{Eichhorn:2017djq} causal sets exhibit a  version of  discrete ``asymptotic silence'' where events which are very close in the continuum are in fact  spatially distant in the causal set.  We refer to the scale at which this effect sets in 
 as
$\ell_{DAS} $, which itself is larger than $\ell_\rho=1/\rho^{1/n}$, the linear discreteness scale.
We argue that $\ppd_\mco$ will limit to $\dist_\mco$ and thence the induced continuum spatial distance $d_h$, when the three scales involved  are ``well separated'' ,  i.e., $\ell_{DAS} << \mco<< \ell_K$,  for distances between elements in $\cA$ that are larger than $\ell_{DAS}$, the scale of ``asymptotic silence''.  Using extensive numerical simulations for causal sets that are approximated by some simple 2 and 3 dimensional geometries in Section \ref{sec:numsim},  we verify that  $\ppd_\mco$ converges to $\dist_\mco$ at distances larger than $\ell_{DAS}$   even for relatively small causal sets,  as long as there is a sufficient
separation of scales.  Our numerical simulations yield quantitative estimates of the corresponding minimum and maximum choices of $\mco$ in relation to $\ell_{DAS}$ and $\ell_K$.

This distance function provides a new observable for CST which can be used for example in
studying the Hartle-Hawking wavefunction in the sum over histories formulation. In \cite{HaHa} the  cardinality or spatial volume of the final spatial geometry was used as an observable when studying the Hartle-Hawking wave function. Using the distance function instead should give a stronger constraint on the spatial geometry, thus allowing for a clearer continuum interpretation. More generally, the transition amplitudes between different spatial geometries in the gravitational path integral could be calculated. This  opens the door to direct comparisons with  other path integral approaches to quantum gravity,  for example Spin Foams and Causal Dynamical Triangulations, in  which 
 such  transition amplitudes are calculated. 

The discretisation suggested by CST can itself be used as a regularisation of spacetime both for quantum field theory and quantum gravity. This  is analogous to the (causal) dynamical triangulations approach to computing the path-integral for quantum gravity, which makes  it amenable to Monte-Carlo simulations \cite{Ambjorn:2001cv,Laiho:2016nlp}. From this point of view, showing the existence of a universal phase transition (i.e., tied to a higher-order phase transition in the space of bare couplings) is tantamount to establishing the existence of an asymptotically safe \emph{Lorentzian} path integral (see the discussion in \cite{Eichhorn:2017bwe}). Therefore, understanding how the causal set encodes geometry may be useful for a broader range  of viewpoints on quantum gravity. 

By providing a spatial distance function, our work makes it possible to study the spectral dimension arising from a purely spatial diffusion process. As the number of nearest neighbours determined by a Lorentzian metric diverges as a function of the spacetime volume, a diffusion process along the links of a causal set necessarily results in an increasing spectral dimension for short diffusion times \cite{AstridSebastian}.  A spatial diffusion process will make it possible to compare with the dimensional reduction observed in other quantum gravity approaches, e.g. \cite{Ambjorn:2005db,Lauscher:2005qz} and indicated by other measures of dimensionality in causal sets \cite{Carlip:2015mra,Abajian:2017qub}.

\section{Induced Spatial Distance in the Continuum from $(M,\prec, \bep)$ } 
\label{continuum} 

\begin{figure}[!t]
\centering{\includegraphics[width=0.5\linewidth,clip=true,trim=6cm 7cm 6cm 5cm]{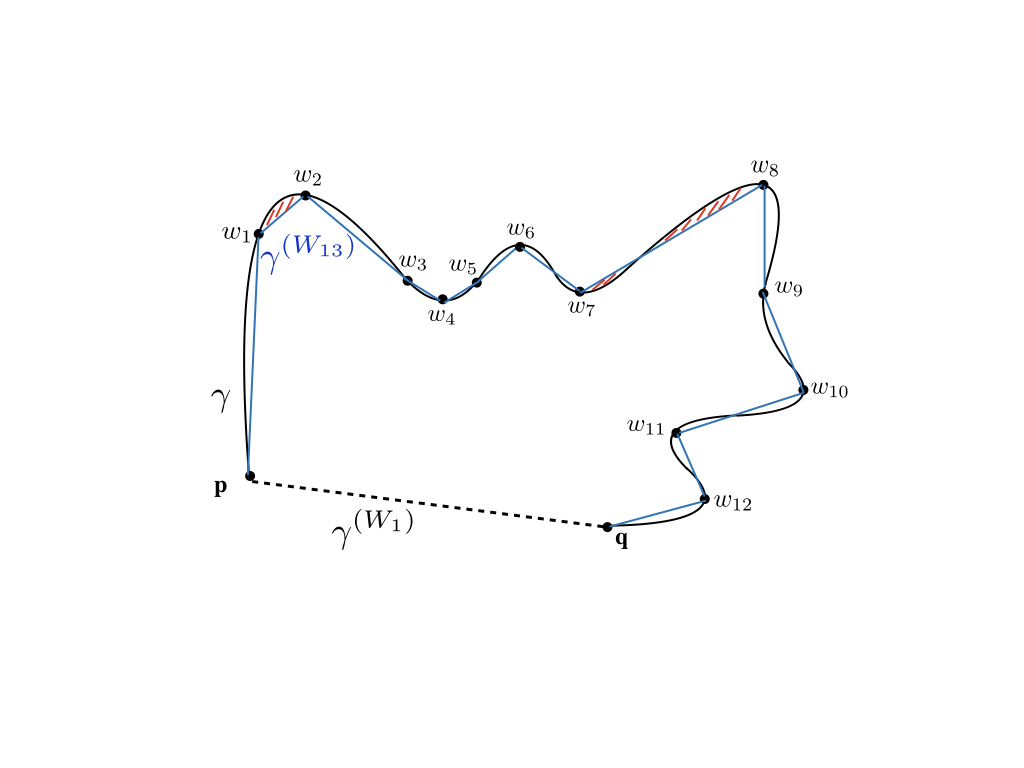}
\caption{\label{linearapprox}  Here we show a curve $\gamma$ in $\re^2$, with two different discretisations, $\gamma^{(W_1)}$(dashed curve) and $\gamma^{(W_{13})}$ (blue curve).  The former is a very crude approximation and the latter is a much better approximation. The red shaded regions show two different  suspended areas $A_{1,2}$ and $A_{7,8}$, which are much smaller than the large suspended area between $\gamma$ and $\gamma^{(W_1)}$.}}
\end{figure}

To set the stage,  consider the simple example in Riemannian geometry of a curve embedded in the 2d plane, $\gamma:[0,1]\rightarrow \re^2$, with $\gamma(0)=p, \gamma(1)=q$, as   shown in  Fig.~\ref{linearapprox}. Our aim is to construct a distance function $\dist(p,q)$  along  $\gamma$ using {\it only} the ambient flat metric on $\re^2$,  via  a piecewise linearisation of $\gamma$ such that in an appropriate 
limit (explained below)
it  approximates the  intrinsic length $\dg(p,q)$ along $\gamma$.    We proceed as follows. 

Consider a {\it finite} set   $W_k\equiv (w_0, w_1\ldots, w_k )$,    $w_i \in \gamma,  w_0=p$ and $w_k=q$, and let  $\gamma_{i,i+1}$ be  the straight line segment from $w_i$ to $w_{i+1}$ in $\re^2$, as shown in Fig.~\ref{linearapprox}.   We define the piecewise linearisation $\gkwk$ of $\gamma$ associated with $W_k$ by 
\begin{equation}
 \gkwk\equiv \bigsqcup_{i=0}^{k-1} \gamma_{i,i+1}, 
\end{equation} 
which is a  continuous, but only piecewise differentiable  curve between the consecutive points of $W_k$.  We can then define the piecewise linear distance 
\begin{equation} 
\dpl(p,q) \equiv \sum_{i=0}^{k-1} \dflat(w_i,w_{i+1}), 
\label{pldist} 
\end{equation} 
where $\dflat(w_i,w_{i+1})$ is the ambient (flat)  distance in  $\re^2$.  The question is how well $\dpl(p,q)$ approximates the intrinsic distance $\dg(p,q)$.  For $k=1$  this approximation can be very poor for a $\gamma$ with non-zero curvature as shown in Fig.~\ref{linearapprox}. In order to { improve}  the approximation, not only must one increase $k$,  but also make sure that  regions of $\gamma$ with  larger curvature are better represented in the discretisation. To formalise this we consider a finite open convex collection   
$\cO=\{O_j\}, \, \, j <\infty$,  where  $O_j$ are convex\footnote{Convexity means that there is a unique geodesic (which in this case is a straight line)  between any pair $r,s \in O_j$ which is contained in $O_j$.}  open sets in $\re^2$,  such that $\cO$   ``covers '' $\gamma$ as a subset of $\re^2$:  $\gamma \subset \bigcup_jO_j$.   One can then  bound the accuracy of the linearisation of $\gamma$ to arbitrary precision, since for every $\epsilon >0$ there exists an $\cO$ and a set  $W_k$, $k< \infty$, such that  $w_i,w_{i+1} \in O_i\in \cO$, and  $|\dflat(w_i,w_{i+1}) - d_\gamma(w_i,w_{i+1})| < \epsilon$.  As expected, such an $\cO$ will  be ``denser''  along a section  of $\gamma$ with higher extrinsic curvature, and sparser in sections with smaller extrinsic curvature, as shown in Fig.~\ref{linearapprox}.

 In order for a path distance $\dpl$ to be a  distance function on $\gamma$, we need to additionally define $\dpl$ for every pair,  $p',q' \in \gamma $, and moreover to ensure that it satisfies the triangle inequality. 
While the inequality is saturated {\it along} a given $\gamma_k$: $\dpl(w_{i},w_{i+2})\equiv\dflat(w_i,w_{i+1})+\dflat(w_{i+1},w_{i+2}) =\dpl(w_i,w_{i+1})+\dpl(w_{i+1},w_{i+2})$, we cannot use a fixed $W_k$ to define the distance function for every pair of points on $\gamma$, since some points are not included in $W_k$.  Instead,  one has to  consider all  discretisations of all possible segments of $\gamma$. 

The process we describe now is analogous to the standard construction of a distance function in a Riemannian metric space $(\Sigma, h)$ \cite{petersen}, where one starts with the ``path-length''  $l(\alpha)\equiv \int_{\alpha} \sqrt{h_{ij}dx^idx^j}$ for a given path $\alpha$ from $p$ to $q$ in $\Sigma$ and  then minimises $l(\alpha)$ over all possible $\alpha$ to obtain
the distance $d(p,q)$. This ensures that the distance function $d$ satisfies the triangle inequality.

In order to get a true distance function on $\gamma$, we therefore minimise $\dpl$  over the set $\Gamma$ of  {\it all} finite discretisations of $\gamma$.  
However, because we are using the ambient space to approximate the induced distance function,  we see that minimising  over all discretisations  in $\Gamma$  would typically  lead to a grossly inaccurate distance function. For example, for a ``C'' shaped curve in $\re^2$,  the shortest distance between the end points corresponds to $k=1$, which is the worst possible approximation to the induced distance. To improve the accuracy,  one has to limit the discretisations so that consecutive $w_i \in W_k$ lie within a ``small enough'' open interval in $\gamma$ {with respect to the local curvature scale of $\gamma$}.  One way to do this in $2$d,  without explicitly referring to the curvature of $\gamma$,   is by requiring that the ``suspended'' area $A_{i,i+1}$ of the closed region bounded by the segment $\gamma(w_i, w_{I+1})\subset \gamma$ between $w_i,w_{i+1}$ along $\gamma$ and the straight line joining $w_i, w_{i+1}$  in $\re^2$,  is {sufficiently} small.  We can then restrict $\Gamma$ to  those discrete paths for which  $A_{i,i+1}<\mco$, a ``mesoscale'' cutoff. Varying over such paths will give us the desired triangle inequality, but also an approximation with an error bounded from above by a monotonically increasing function of $\mco$. In this simple case, it is clear that the shortest distances will tend to ``stretch'' this bound, in the sense that the paths with the smallest number of $k$ will minimise this  distance, thus limiting the accuracy.  However, in the more general Lorentzian cases we will look at,  the shortest distance path could correspond to increasing $k$ and hence the accuracy, arbitrarily. 

\begin{figure}[!t]
\centering{\includegraphics[width=0.6\linewidth,clip=true,trim=6cm 10cm 6cm 7cm]{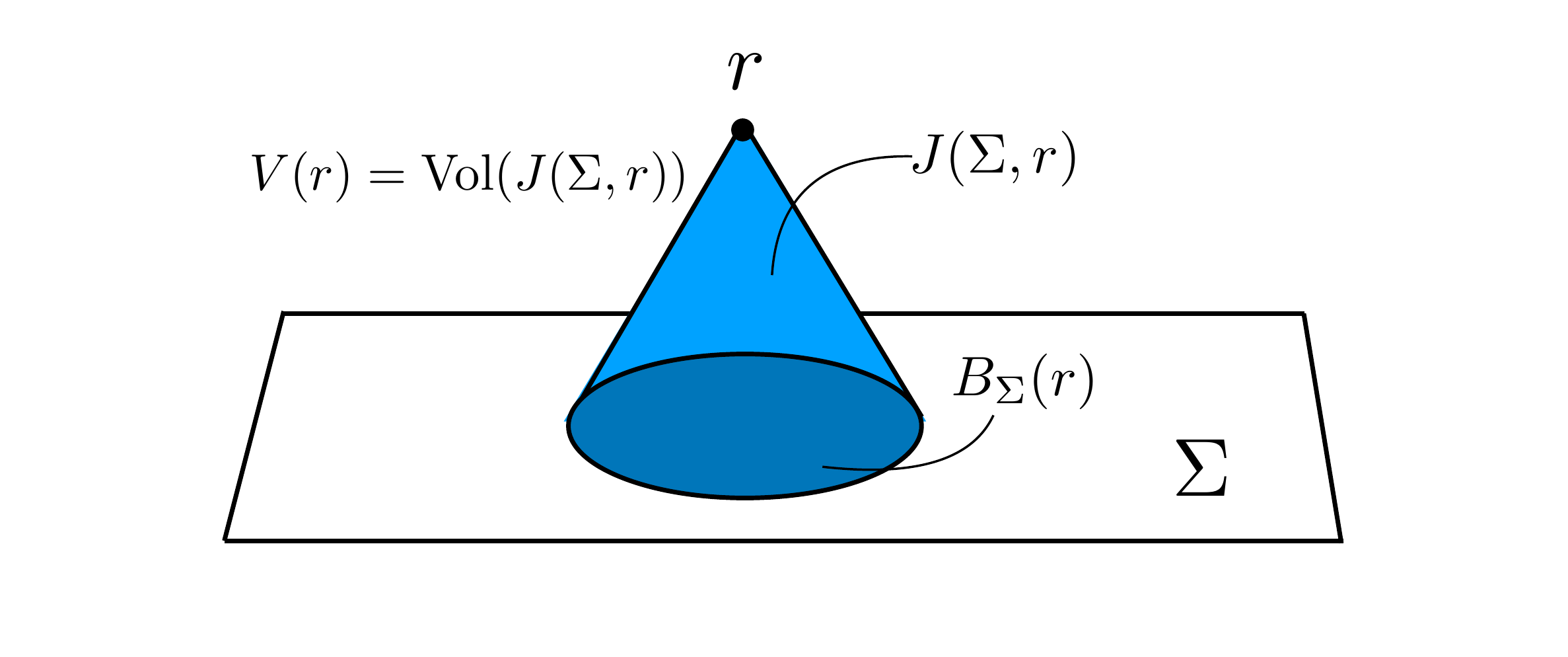}
\caption{\label{beam} An example of the sets $J(\Sigma, r)$ and $B_\Sigma(r)$.}}
\end{figure}

In this construction,  we have used the ``ambient''  geometry of $\re^2$ to define not only an  induced distance on $\gamma$, but also to define a mesoscale cut-off $\mco$ by utilising an ``area'' function in $\re^2$.  In higher dimensions, for example a surface  in $\re^3$,  it is not straightforward  to define $\mco$,  since the higher dimensional analog of a suspended area is harder (and less natural)  to define. However, in the Lorentzian construction that we will now describe,  there is no dimensional obstruction per se since the causal structure gives us a natural ``suspended'' volume. Nevertheless, translating this volume into an ambient  distance function is not straightforward.  For an inertial Cauchy hypersurface $\Sigma$  in  Minkowski spacetime,  the suspended volume is the half cone which has an exact relation with the actual distance function on $\Sigma$, cf.~Fig.~\ref{beam}. For more general $\Sigma$ and for more general spacetimes, this relation is not exact, but can be made ``close enough'' by making the mesoscale $\mco$ ``small enough'' with respect to the local  curvature scale.  This will be the essence of our construction.

We will restrict our attention to an $n$-dimensional globally hyperbolic spacetime $(M,g)$ with compact Cauchy hypersurfaces $\Sigma$ and induced Riemannian metric $h$.  Associated to $(M,g)$ is the triple $(M,\prec,\bep)$ where $\prec$ denotes the causal relation.  We wish to define a distance function $\dist:\Sigma \times \Sigma \rightarrow [0, \infty)$  given only $(M,\prec,\bep)$ and $\Sigma$. Note that both $M$ and $\Sigma$ are taken to be {\it sets} of events, with no further structure overlaid on them. The causal relation  $\prec$ allows us to define the {\sl causal future} and {\sl causal past}  of $p$, $J^+(p) \equiv \{r| p \prec r\} $ and $J^-(p) \equiv \{s| s \prec p\} $, respectively  {as well as the} {\sl causal interval}  $J(p,r)\equiv J^+(p) \cap J^-(s)$.  The causal past  and future of a set of events $S \subset M$ is $J^\pm(S)=\bigcup_{s \in S} J^\pm(s)$, and the causal interval between $S,S'\subset M$ is $J(S,S')=J^+(S) \cap J^-(S')$. Using the volume element $\bep$ we can get the volume $\vol(.)$  of these  sets.  Of particular interest to us are the sets $J(\Sigma, r)$ and the  {\sl beam} $B_\Sigma(r)\equiv J^-(r)\cap \Sigma$  as well as  the spacetime  volume  
\begin{equation} 
V(r)\equiv\vol(J(\Sigma,r)),  
\label{volume} 
\end{equation}  
an example of which is shown in Fig.~\ref{beam}. 
The idea is to  use this suspended spacetime volume $V(r)$ to extract an appropriate predistance function on $\Sigma$. 

\begin{figure}[!t]
\centering{\includegraphics[width=0.5\linewidth,clip=true, trim=11cm 11cm 4cm 6cm]{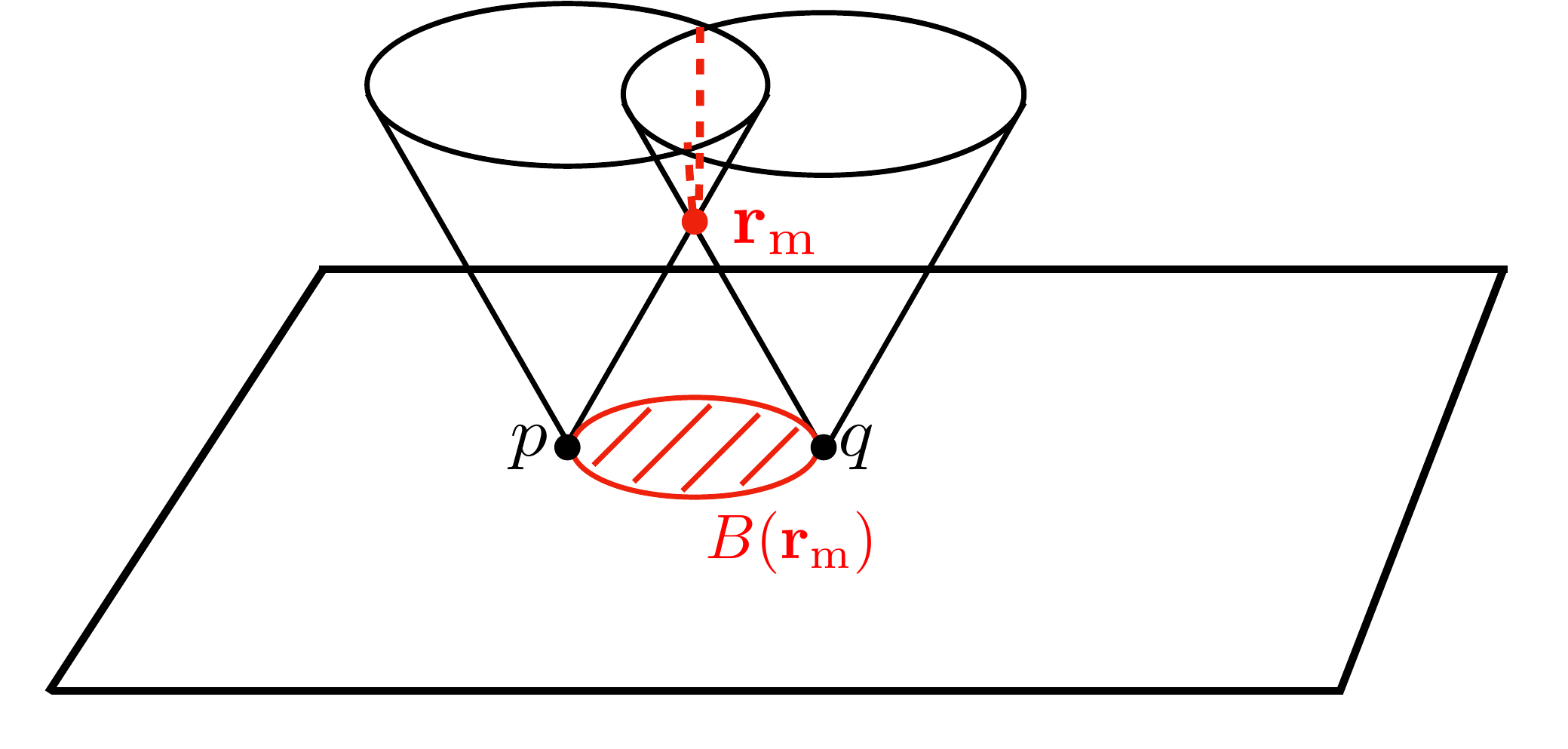}
  \caption{\label{fig:Hpq} The intersection of the future light cones from $p,q\in \Sigma$, in $3$d Minkowski spacetime with an inertial $\Sigma$, i.e. $K=0$. The points on $\cH(p,q)$ (denoted by the dashed red line) are null related to both $p$ and $q$. $\rmin $ is the unique event in $ \cH(p,q)$ for which $V(r)$ takes a minimum value. Using Equation (\ref{zetadef}), $V(\rmin)$ can be related to the diameter of $B_\Sigma(\rmin)$ which is the distance between $p$ and $q$. }}
\end{figure}

Consider first the simple example of $n$ dimensional Minkowski spacetime $(M\equiv \bM^n,\eta)$ with inertial Cauchy hypersurface $(\Sigma, h)$ with $\Sigma \simeq \re^{n-1}$ and $h=\delta$, the flat $n-1$ dimensional Riemannian metric.  For $r \in J^+(\Sigma) $, $V(r)$ takes the simple form 
\begin{equation} 
V_\eta(r)=\zeta_\dimm T^\dimm,
\label{zetadef} 
\end{equation} 
where $T$ is the proper time from $r$ to $\Sigma$ which is the height of the cone in Fig.~\ref{beam}, and 
\be\label{zeta}
\zeta_\dm= \frac{\pi^{(\dm-1)/2}}{ \dm\, \Gamma \left(\frac{\dm+1}{2} \right)}.
\ee
The beam $B_{\Sigma}(r)$ is geometrically a ball $(\bB^{n-1}, \delta)$ whose boundary $\partial B_{\Sigma}(r) \simeq S^{n-1}$ is a round sphere of diameter $2T$. If $p,q \in \partial B_{\Sigma}(r)$ are antipodal points  then $\dist(p,q)=2T$. 
Eq.~\eqref{zetadef}  allows us to write  
 the distance
in terms of $V_\eta(r)$ and the spacetime dimension $\dimm$.  Thus we have related  a linear dimension of $B_{\Sigma}(r)$, its diameter,  to the spacetime volume $V_\eta(r)$: this is the essence of our construction for a distance function.  As a first step, we construct a predistance function using this relation.  Our construction is illustrated in Fig.~\ref{fig:Hpq}. 

\begin{figure}[!t]
\centering 
\includegraphics[width=0.9\linewidth,clip=true, trim=0cm 11cm 0cm 0cm]{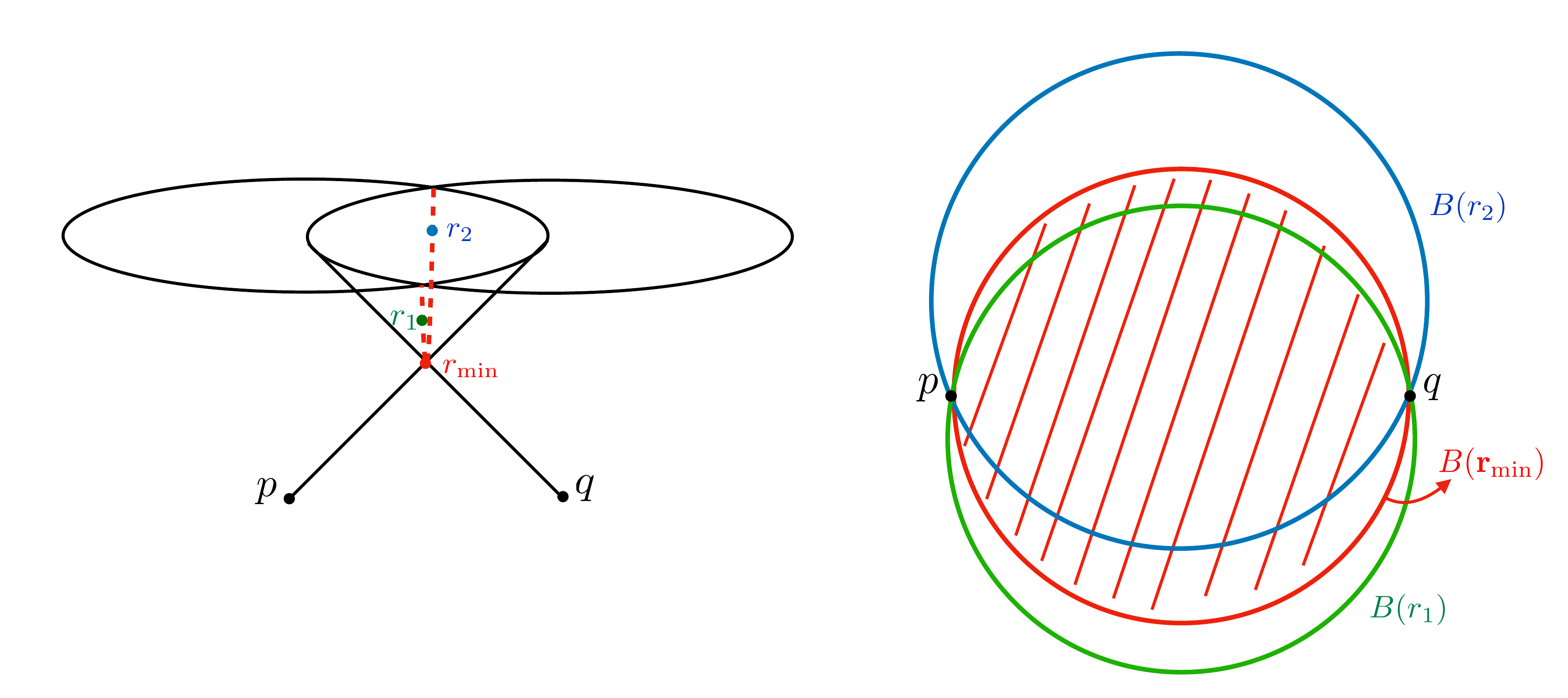}
\caption{\label{fig:diffbeams}We show the three beams related to the points $\rmin$, $r_1$ and $r_2$ which all lie in $\cH$. The locations of the three points in $\cH$ are indicated in the left part of the figure, which contains the relevant part of Fig.~\ref{fig:Hpq}.}
\end{figure}

We work in Cartesian coordinates $(t,x_1, \ldots x_{n-1})$ with  $(\Sigma, \delta)$  at $t=0$. Let $p,q \in \Sigma$, with  $p$ at  the origin $(0,0,0,\ldots, 0)$ and $q$ on the positive $x_1$ axis at $(0, x_q, 0, \ldots 0)$.      
The events in the set $\cH(p,q)\equiv \pd  J^+(p)\cap \pd  J^+(q) $   are  null related to both $p$ and $q$.  In these coordinates the two null hypersurfaces $\cH_p \equiv \pd J^+(p)$   and $\cH_q \equiv \pd J^+(q)$ are 
\begin{eqnarray} 
\cH_p &=& \left\{r{\Big|\,} \sum_{i=1}^{n-1} x^2_i(r)-{\tc{r}}^2=0 \right\}, \nonumber \\ 
\cH_q&=&\left\{r{\Big|\,} (x_1(r)-x_q)^2+\sum_{i=2}^{n-1}(x^2_i(r)-t^2(r)=0 \right\},  
\label{hphq}
\end{eqnarray} 
so that 
\begin{equation} 
\cH(p,q)= \left\{r{\Big|\,} x_1(r)=\frac{x_q}{2},  \sum_{i=2}^{n-1}x^2_i(r)-t^2(r)=-\biggl(\frac{x_q}{2}\biggr)^2 \right\}, 
\label{hpq} 
\end{equation} 
 which is  an $n-2$ dimensional spacelike hyperboloid $\mbh^{n-2}$.  
For any $r=(\tc{r},x_1(r), \ldots, x_{n-1}(r))$  with $\tc{r}>0$,   $V(r)$ increases monotonically with $\tc{r}$. At  $\rmin=(\frac{x_q}{2},\frac{x_q}{2},0, \ldots, 0)$,  $\tc{r}$ takes on its  smallest value for all $r\in \cH(p,q)$.  At this unique minima of $\tc{r}$ on $\cH(p,q)$,   $V(\rmin)$ also takes on its smallest value. We can  thus define the {\sl predistance  function} as 
\begin{equation} 
\prd(p,q)= 2 \biggl(\frac{V(\rmin)}{\zeta_\dimm}\biggr)^{\frac{1}{\dimm}},
\label{predist}  
\end{equation} 
which, being equal to $2\tc{\rmin}=x_q$ {\it is} the  distance function $d_\delta$ on $\Sigma$ (and hence satisfies the triangle inequality). In Fig.~\ref{fig:diffbeams} we show how the past light cones  from different events  in $\cH$ intersect $\Sigma$.  Note that $d_\delta(p,q)$ is  the diameter of the beam $B_{\Sigma}(R)$ only when $r=\rmin$. 

\begin{figure}[!t]
\centering
\includegraphics[width=0.75\linewidth, clip=true, trim=0cm 0cm 0cm 16cm]{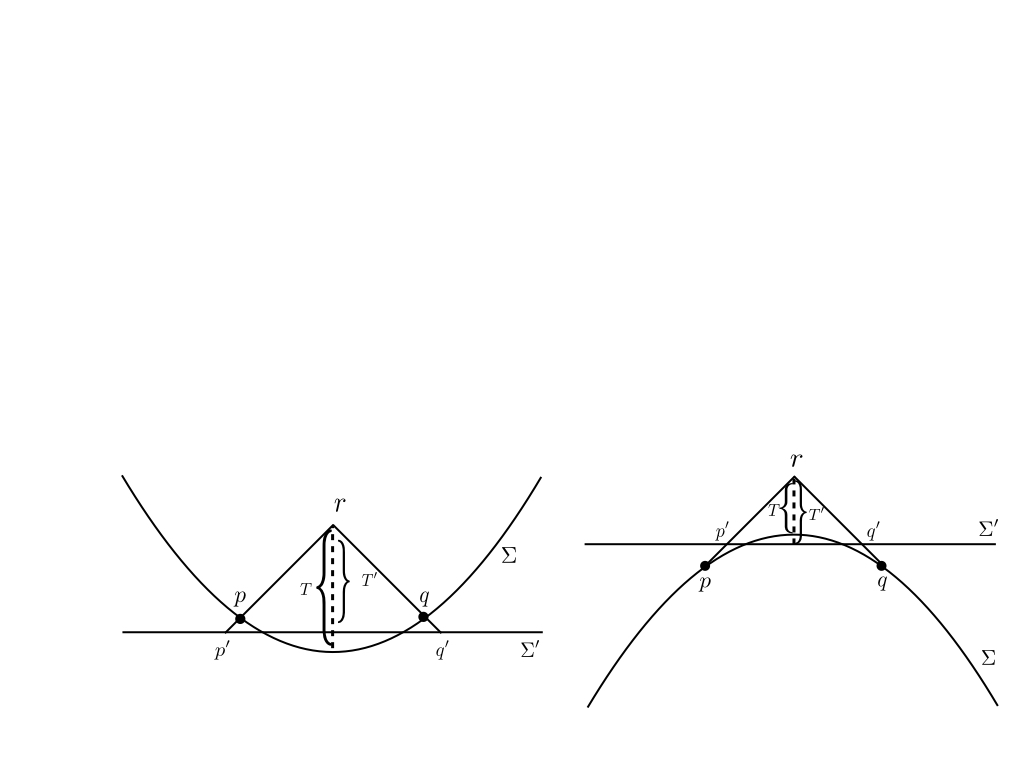}
\caption{\label{fig:ext_curv_ill} $\Sigma'$ associated with $\Sigma$ of negative and positive extrinsic curvature.}
\end{figure}

This definition is at the heart of our construction even in the general case, though we must  proceed with caution.  Again, let us continue with the simple example of Minkowski spacetime. Instead of being an inertial hypersurface, let  $(\Sigma,h)$ be a Cauchy hypersurface in $\bM^\dimm$, with constant extrinsic curvature $K_{ab}$, i.e., a hyperboloid $\mbh^{n-1} \subset \bM^n$.  Because the ambient spacetime is still $\bM^\dimm$  we can again construct $\cH(p,q)$ for every pair $p,q \in \Sigma$, and the same argument shows that there exists an $\rmin \in \cH(p,q)$ which minimises the suspended volume $V(r)$ for $r \in \cH(p,q)$.    The volume $V(\rmin)$ is that of  the region $J(\Sigma,\rmin)$ which is not the  half cone with a flat base but an inverted  ``icecream cone'' which is either scooped out ($K>0$) or topped up ($K<0$) (see Fig.~\ref{fig:ext_curv_ill}).  Let us define the predistance  $\prd(p,q)$ as in Eq.~(\ref{predistance}). This is clearly {\it not} the induced distance $\dsig(p,q)$, since the beam $B_\Sigma(\rmin)$  is no longer a flat open ball.  What does $\prd(p,q)$ measure?   Consider an inertial frame $(\Sigma ', \delta_{ab})$ for which $J(\Sigma',\rmin)$ is a  half cone with a flat base, such that $\vol(J(\Sigma',\rmin))=V(\rmin)$.  The diameter of $\pd B_{\Sigma'}(\rmin)$ is therefore  given by $2T'$,  where  $T'$ the proper time from $\rmin$ to $\Sigma'$,  and moreover,  $V(\rmin)={\zeta_\dimm}{T'}^\dimm$. Thus, $\prd(p,q)$ is in fact the  induced distance on $\Sigma '$ between a pair of antipodal points on $\pd B_{\Sigma'}(\rmin)$, i.e., 
\begin{equation} 
\prd(p, q) = 2T' = 2(T +(\mathrm{sgn} K) \epsilon(p,q)),  
\label{flatK}
\end{equation}
where $T$ is the proper time from $\rmin$ to $\Sigma$ and  $\epsilon(p,q)$ depends on $p,q$ and the extrinsic curvature $K$,  cf.~Fig.~\ref{fig:ext_curv_ill}.
It is clear that at least locally, $\epsilon$ can be made as small as required by choosing a region $\cN \subset \Sigma$  which is small enough with respect to the curvature scale. In other words,  for every $\epsilon'>0$ there exists an $N \subset \Sigma$ such that for every pair  $p,q\in N$, $\epsilon(p,q) <\epsilon'$.   

In our simple example we can also explicitly calculate the intrinsic distance $\dsig(p,q)$ in $(\Sigma, h)$ and compare with Eq.~(\ref{flatK}).  Choosing a  coordinate system with $p=(t_q,-x_q, 0, \ldots,0)$, $ q= (t_q,x_q, 0, \ldots,0)$ where $\tau^2=t_q^2-x_q^2$, we see that     
\begin{equation} 
\dsig(p,q) =  2\tau \int_{0}^{x_q} \frac{1}{\sqrt{x^2+\tau^2}}dx  = 2\tau \ln\biggl(\frac{x_q + \sqrt{\tau^2+x_q^2}}{\tau}\biggr), 
\end{equation} 
which for small enough $\frac{x_q}{\tau}$ (or equivalently  small enough $K \tau$)
\begin{equation} 
  \dsig (p,q) \approx  2(x_ q +  \frac{x_q^2}{2\tau}  + \ldots). 
  \end{equation} 
In these coordinates,  $\rmin=(\tau',0,0, \ldots, 0)$, with $T=\tau'-\tau$. Since $\rmin$ and $p,q$ are null related, moreover, 
\begin{equation} 
\biggl( \tau'- \sqrt{\tau^2 + x_q^2} \biggr)^2=x_q^2, 
\end{equation} 
which to leading order gives $T= x_q +  \frac{x_q^2}{2\tau}  + \ldots.$, so that 
\begin{equation} 
\dsig(p,q) = 2T + x_q \,\mathcal{O} \biggl(\frac{x_q^2}{\tau^2}\biggr). 
\end{equation}  
Hence we can substitute  
\begin{equation} 
\prd(p,q)=\dsig(p,q)+(\mathrm{sgn} K)  2 \epsilon(p,q). 
\label{signK}
\end{equation} 
where $\epsilon(p,q)$ now includes all higher order corrections.

We are ready to now consider a  general globally hyperbolic spacetime $(M,g)$ with compact Cauchy hypersurface $\Sigma$.  
The idea behind the following construction is to restrict to piecewise linear sections of the spacetime, where one can make the approximation that each section is ``approximately flat'' such that we work with   caustic free   past light cones ruled by null geodesics. The natural choice for such a region is a  Riemann Normal Neighbourhood (RNN).

Let   $p\in \Sigma$ and  $M_p$ an RNN in $(\Sigma, h)$ whose domain of dependence\footnote{ The future/past domain of dependence $D^\pm(S)$ of a subset $S \subset M$ is the set of events  $p \in M$, such that \emph{every} past/ future-inextendible causal curve through $p$ intersects $S$. The domain of dependence is then the union $D(S) \equiv D^+(S) \cup D^-(S)$. } $D(M_{p})$ is itself a proper subset of an RNN $N_p \subset (M,g)$ of $p$.  For any $q \in M_p$,  the set $J^+(p)\cap J^+(q)\cap D(M_p)$ is non-empty since $N_p$ is diffeomorphic to an open convex set in $\re^\dimm$. As in flat spacetime, the  events in   $\cH(p,q)\equiv \pd  J^+(p)\cap \pd  J^+(q) \cap D(M_p)$  are  null related to both $p$ and $q$.  $\cH(p,q)$  is moreover diffeomorphic to an open set in a codimension 2 hyperboloid $\cO' \subset \mbh^{\dimm-2}$. 
It is relatively simple to show  
that $\cH(p,q)$ is spacelike (as is the case in Minkowski spacetime) using results from the theory of causal structure \cite{penrose} and the fact that $\dmp$ is an RNN. The latter ensures that there are no conjugate points along null geodesics in $\dmp$. We leave the proof to Appendix \ref{app.difftop}. 

In analogy with the flat spacetime construction, we define the function 
\begin{equation} 
V(p,q) \equiv  \inf_{r \in \cH(p,q)} V(r). 
\label{minvol} 
\end{equation}
Since $V(r)$ is a continuous function in a globally hyperbolic spacetime and since $\overline{D(M_p)}$ lies in an RNN, the closure of $\cH(p,q)$ is compact. The Extreme Value Theorem in analysis  then implies that $V(p,q)$ is realised at some $\rmin \in \overline{\cH(p,q)}$, i.e.,   $V(\rmin)=V(p,q)$.  
As we show in Appendix \ref{app.difftop}, wlog the minimum value must moreover  lie in the interior of this region, i.e., within $\cH(p,q)$. Thus,  there exists an $\rmin \in \cH(p,q)$ such that  $V(\rmin)=V(p,q)$.   The predistance function $\prd(p,q)$ is therefore again given by   Eq.~\eqref{predist}.

Intuitively one would expect that if $p,q$ are brought closer together that 
Eq.~\eqref{predist} will give  a more accurate distance, as curvature effects become smaller.   In the flat spacetime example, we saw that even when $p,q$ lie  in an appropriately small $M_p$, $\prd(p,q)$  differs from the induced distance $\dsig(p,q)$:  the error comes from identifying $V(\rmin)$ with a different $\Sigma'$ which can,  in principle, be calculated for a constant $K$.    
In the more general setting, the source of the errors are more complicated when using the simple flat spacetime inspired form Eq.~\eqref{predist}. They nevertheless can be tracked to leading order as we now show. 

In   \cite{GHterm}  $V(r)$ was calculated to leading order using a combination of Gaussian normal coordinates (GNCs) and  Riemann Normal coordinates (RNCs). In a Gaussian normal    neighbourhood (GNN) $\Ns$ of $\Sigma$ the  metric takes the simple form  
\begin{equation} 
ds^2=-dt^2+h_{ij}(t,\vec x)dx^idx^k,
\end{equation} 
in GNCs,  where $h_{ij}(t,\vec x)$ is the induced metric on $\Sigma$. Any $r \in \Ns$ is  uniquely mapped to a point $\ro\in \Sigma$ via a (unique) timelike geodesic from $\ro$ whose tangent is normal to $\Sigma$. If $\ro$ is chosen to be the origin of the coordinate system, then the coordinates of $r=(\tc{r},0, \ldots, 0)$.  Consider further,  an  RNN $N_{\ro}\subset N_\Sigma$  about $\ro$ in which  the metric at any $s\in N_{\ro}$ takes the form 
\begin{equation} 
{g_{ab}(s) = \eta_{ab} - {\frac{1}{3}} R_{acbd} x^c(s)x^d(s) + O(x^3(s)). } 
\label{metricRNN}  
\end{equation}    
Note that the GNCs and RNCs are different in general, even if their origins coincide. By transforming the former to the latter the volume $V(r)$ in $N_{\ro}$ was calculated in \cite{GHterm} to leading order to be   
\begin{equation} 
V (r) =V_\eta(r) \left (1+\frac{n}{2 (n+1)}K (\ro)T\right) +O (T^{n+2})
\label{GNCvol},  
\end{equation} 
where $T=\tc{r}-\tc{\ro}$ is the proper time from  $\ro$ to $r$ and $K(\ro)$ is the trace of the extrinsic curvature at $\ro$,  with $V_\eta$ given by   Eq.~\eqref{zetadef}.  There are  a couple of features of this formula which are noteworthy. By dimensional arguments, the Riemann curvature terms only arise at $T^{n+2}$, while the extrinsic curvature term can arise at order $T^{n+1}$. That this latter term is non-zero  means that we can ignore the corrections coming from the Riemann curvature. Moreover, at leading order, $K(\ro)$ is a constant in $\Nh$.

\begin{figure}[!t]
\centering
\includegraphics[width=0.75\linewidth]{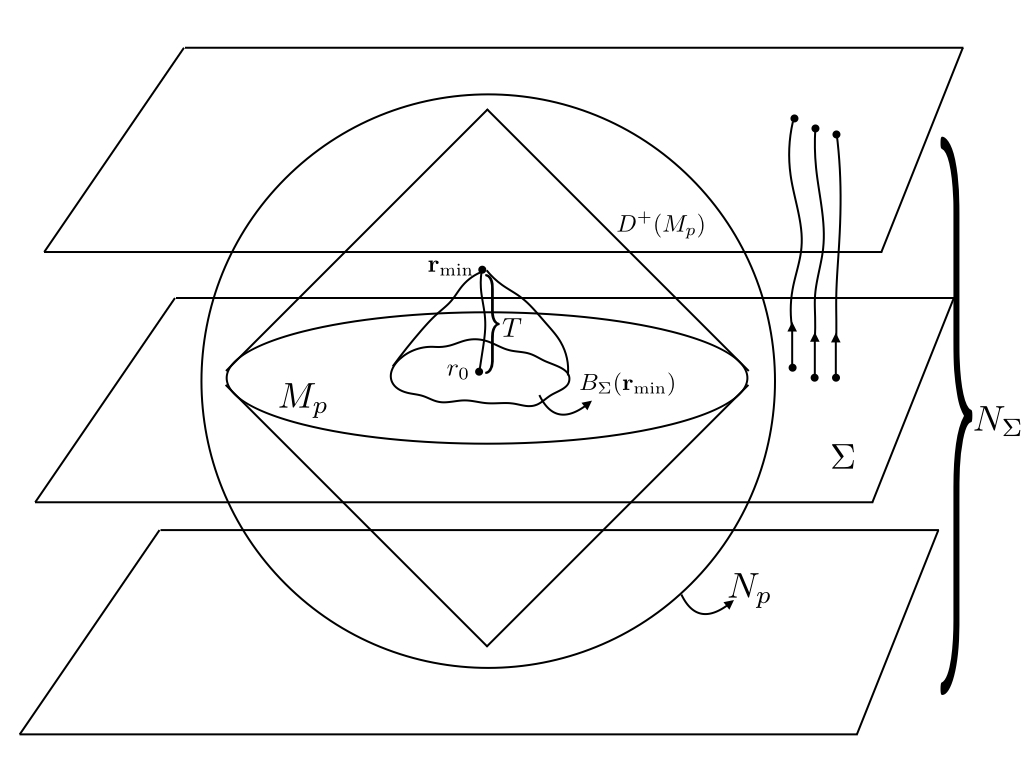}
\caption{\label{fig:GRN}{ We illustrate the GRN and RNN in the case with non-constant curvature.} }
\end{figure}

Let $M_p \subset \Sigma$ be an RNN about $p \in \Sigma$ such that $\dmp \subset N_p\subset \Ns$ where $N_p$ and $\Ns$ are the RNN and GNN defined above. We will call such a neighbourhood of $p$ in $\Sigma$  a Gaussian-Riemann neighbourhood (GRN) (see Fig.~\ref{fig:GRN}). 
For  $q\in M_p$,  we define $\cH(p,q), V(p,q), \rmin$ as before. If we use Eq.~\eqref{predist} to define  $\prd(p,q)$, then from Eq.~\eqref{GNCvol}, 
\begin{equation} 
\prd(p,q) \simeq 2T \left(1+ \frac{1}{2 (n+1)}K (\ro)T\right) +O (T^3). 
\end{equation} 
Moreover, to  leading order the actual induced distance takes the form 
\begin{equation} 
d_h(p,q)= 2T(1+\alpha KT + O(T^2)),
\end{equation}  where $\alpha$ is a fixed dimension dependent parameter.
The difference is then 
\begin{equation} 
\prd(p,q)-d_h(p,q) = 2\left( \frac{1}{2 (n+1)}-\alpha\right) K(\ro)T^2 + O(T^3).
\end{equation}  
As in the case of constant $K$ in flat spacetime, to leading order $\prd$ will overestimate the continuum distance for $K>0$ and underestimate it for $K<0$. Hence, we expect that $\alpha <  \frac{1}{2 (n+1)}$.   
By choosing the GRNs to be sufficiently small, the error can thus be arbitrarily bounded, just as in the case of the curve in $\re^2$.  In other words, for every $\epsilon'>0$ there exists a GRN $N_p$ about $p \in \Sigma$ such that for all $q\in N_p\cap \Sigma$    
\begin{equation} 
\prd(p,q)=\dsig(p,q) \pm (\mathrm{sgn} K) \epsilon(p,q),
\label{error} 
\end{equation}  
with $\epsilon(p,q)<\epsilon'$.  $\epsilon(p,q)$ depends on $K_{ab}$ and $T$ to  leading order, with the sign determined by that of  $K$. The  subleading corrections come from the Riemannian  curvature.

To obtain the distance function, as in the case of the curve in $\re^2$,  we construct  discretised paths  in $\Sigma$.  For any $p,q \in \Sigma$ let   $W_k=(w_0,w_1, \ldots, w_{k-1}, w_{k}) \subset \Sigma $,   where $w_i \in \Sigma$ and $w_0=p, w_k=q$. The path distance is defined as 
\begin{equation} 
\dpl(p,q)= \sum_{i=0}^{i-1} \prd(w_i,w_{i+1}).
\end{equation} 
The distance function is obtained from minimising   over all discrete paths $\gkwk$ from $p$ to $q$ as we did for the curve in $\re^2$, i.e., 
\begin{equation} 
  d(p,q) \equiv \inf_{\gamma } \dpl(p,q), 
\end{equation} 
which satisfies the triangle inequality. Again,   it is clear that without a mesoscale cut-off $\mco$ which restricts the allowed predistances $\prd(w_i,w_{i+1})$ from above, the errors in $d(p,q)$ can be very large.  We thus define the one-parameter family of distance functions $d_\mco(p,q)$ by restricting  to those $\gamma_k$ such that for every $(w_i, w_{i+1}) $ in $\gamma_k$,    $\prd(w_i,w_{i+1})< \mco$. If $\mco$ is small enough {with respect to the local curvature scale of $\Sigma$}, this ensures that $(w_i, w_{i+1})$  lies in a GRN. Thus, for each segment of the path $\dpl(w_i, w_{i+1})$ differs from $\dsig(w_i,w_{i+1})$ by $\epsilon(w_i, w_{i+1})$ upto sign of $K$ as in  Eq.~\eqref{error}.      

Since $\Sigma$ is compact  it admits a \emph{finite} open cover, which implies that  the total error along the path is  bounded. Let us choose a finite  open cover $\cO=\{N_\alpha\}$ such that each $N_\alpha$ is a GRN. Let $\epsilon_\alpha$ denote the upper bound on $\epsilon(p,q)$ for all $p,q \in N_\alpha$. Finiteness of the cover then implies that  
\begin{equation} 
\Eps = \sup_\alpha \epsilon_\alpha,
\end{equation} 
is finite.  Conversely, for every $\Eps>0$ there exists a finite  open covering of GRNs $\cO_{\Eps}=\{ N_\alpha \}$ of $\Sigma$  such that 
$\epsilon_\alpha\leq \Eps$. Thus,  $\cO_{\Eps}$ is an open covering compatible with $\Eps$. If the cut-off $\mco$ is moreover chosen such that $\prd(p,q)<\mco$  for all $p,q \in N_\alpha$,  the error in $d_\mco(p,q)$ remains bounded.

Again, as in the case of the curve in $\re^2$ for  a piecewise discretisation, one does obtain the  continuum path as $k\rightarrow \infty$. However for any finite $k$,   the error $\epsilon$  can be positive or negative (depending on the sign of $K$). Thus, a  coarser $\gkwk$ could be  preferred over a more refined one thus increasing the error. We saw this already in the case of the curve in $\re^2$ and that this error can be limited by the mesoscale cut-off $\mco$. However, again because of the sign,  it is also possible that a finer $\gkwk$ is preferred. As an example consider a constant $K$  hypersurface in Minkowski spacetime which  is everywhere nearly null. The continuum distance along the hypersurface is always smaller than any finite discretisation with the coarsest ($k=1$) discretisation giving the largest estimate. If $k$ is unbounded from above, the worry might be that the errors on individual segments accumulate in the limit. On the other hand, as $w_i$ and $w_{i+1}$ approach each other,   $\epsilon(w_i, w_{i+1})$ decreases monotonically, and hence the accumulation of errors also converges to zero.

The astute reader will notice that despite our initial promise, our arguments have used more information about the manifold than simply its causal structure and the volume element. However,  the  {\it definition} of  the distance function depends only on the suspended volume. We have used the differentiable structure simply to bound the errors with respect to the induced geometry but again, the mesoscale cut-off  is  fixed only by the suspended volume element. As we shall see, all this is readily implemented in the causal set.

\section{Induced Spatial Distance in the Causal Set} 
\label{discrete}

In CST the set of continuum spacetime geometries is replaced by a {\sl sample space} of causal sets\footnote{The sample space can could either be the set of locally finite $N$ element posets, or the set of countable posets which are past finite,  or one of causal sets of a fixed order theoretic dimension.},  including those with no  continuum interpretation.  Guided by the HKMM theorem, a causal set is said to have a continuum interpretation whenever  the order in $C$ corresponds to the causal order of a spacetime $(M,\prec) \subset (M,g)$,  and additionally, the cardinality of any  causal interval  in $C$ corresponds to  its spacetime volume. Such a correspondence is not respected by a regular lattice in a coordinate invariant way, but is, in an average sense,  by a ``random lattice'' generated by a Poisson process.  
More precisely, a causal set $C$ is said to be  {approximated} by the spacetime $(M,g)$ at {sprinkling} density $\rho$ if there exists an embedding $\phi: C \rightarrow M$ such that (i) $e_i \prec e_j  \Leftrightarrow \phi(e_i) \prec \phi(e_j) $, i.e., the order relation in $C$ is the causal order in $(M,g)$ ,   (ii) the set $\{\phi(e_i)\}$ can be obtained from $(M,g)$ via a high probability Poisson process where for a given spacetime region of volume $V$, the probability of finding $n$ elements  of $\phi(C)$ is given by 
\begin{equation} 
P_V(n)= \frac{(\rho V)^n}{n!} \exp^{-\rho V}. 
\label{poisson}
\end{equation}  
For this distribution $\av{n}=\rho V$, which is the required $n \sim V$ correspondence.   Conversely, the ensemble of causal sets underlying a  spacetime $(M,g)$ at  density $\rho$ can be obtained by a  {\sl Poisson sprinkling} into $(M,g)$, where the order is obtained from the causal ordering in $(M,g)$.  

In CST the causal set is thought to be fundamental with the continuum being emergent. The question is then how the familiar continuum geometry of $(M,g)$ manifests itself order theoretically in $C$.  Such an identification gives us physical (covariant) observables in CST which approximate to  familiar continuum geometric quantities, but which  are defined for \emph{all} causal sets not only those with continuum approximations. 

The continuum construction of $d_\mco$ using purely $(M,\prec)$ and $\bep$ 
is readily extended to CST as we will now show, thus providing a potentially new observable for quantum gravity.

Consider a causal set $C$ which is approximated by an $n$-dimensional globally hyperbolic spacetime $(M,g)$ with compact Cauchy hypersurfaces at sprinkling density $\rho$.   The analogue of a spatial hypersurface is an inextendible antichain $\ca \subset C$ which  is a set of mutually  unrelated elements in $C$ such that no other elements in $C$ are spacelike to all the elements in it, i.e., all other elements in $C$ share a relation  with at least one element in $\ca$.  Devoid of relations, $\ca$ contains no intrinsic geometric or topological information.  Taking our cue from  \cite{homologyone,homologytwo} we will use  the ambient causal set $C \supset \ca$ to construct a distance function on $\ca$, just as we did in the continuum. 

Before we proceed to do so, we must first ask in what sense $\ca$ corresponds to a continuum spatial geometry $(\Sigma, h)$.  Given the embedding $\phi: C \rightarrow (M,g)$,  $\phi(\ca) \subset (M,g)$ is simply a set of spacelike events in $(M,g)$ contained not in one, but an {uncountably} infinite family of spatial geometries. How can one pick the ``most''  representative of these?  At best, $\ca$ should be able to provide a piecewise linear representation
of these geometries, where geometric structure on  scales smaller than the induced  spatial discreteness  scale is deemed irrelevant. It is precisely {\it this}  piecewise linear geometric structure that we hope to  capture using the distance function.  For a sprinkling of Planckian density, this means that Planckian  (extrinsic) curvature that might exist in the continuum manifold is not represented in the causal set, as it is considered unphysical in CST.

For every $c\in \fut{\ca}$ define 
\begin{equation} 
\bN(c)=|\fut{\ca}\cap \past{c} |,
\end{equation}
where $\fut{.}$  and $\past{.}$ denote the future and past, respectively of a subset of $C$, and $|.|$ denotes  the cardinality of the set.   
Unlike in the continuum,  null related events in $C$ almost surely do not exist, in the sense of probability theory. Thus, we cannot use an analogue of $\cH(p,q)$ but must jump instead to  Eq.~(\ref{minvol}). {Using the continuum-discrete identification $\bN(c)=\rho \dV(c)$}, define 
\begin{equation} 
\dV(a,b) \equiv  \inf_{c\in F(a,b)} \dV(c).  
\label{minvol} 
\end{equation}
where $F(a,b) \equiv \fut{a}\cap \fut{b}$ is  the {\sl common future}  of $a,b$.  
Unlike in the continuum case, $F(a,b)$ is {{at most} only} countably infinite. Moreover, since $\ca$ is finite (because $\Sigma$ is compact), for every $\dV' >0$ there are a finite number of elements in $c \in F(a,b)$ with $\dV(e)<\dV'$.  Thus, there exists an  $\cmin$ for which $\dV(\cmin)=\dV(a,b)$.   
The expression for the predistance function is then identical to the continuum case: 
\begin{equation} 
\dd(a,b) \equiv   2 \biggl(\frac{\dV(\cmin)}{\zeta_n}\biggr)^{\frac{1}{n}}.  
\label{predistance} 
\end{equation} 
This is the predistance function that was used in \cite{Eichhorn:2017djq} to obtain a dimension estimator and to demonstrate a behaviour akin to asymptotic silence that we will refer to as ``discrete asymptotic silence'' (DAS), in the following. 

In comparing with the continuum, we already see an important difference. 
Because of the absence of null related events and due to the probabilistic nature of the continuum approximation, $\dV(\cmin)$  is almost surely
``too large'', so that  the beam $B_{\ca}(c) \equiv \past{c}\cap \cA$   encompasses a 
larger  continuum spatial volume and hence overestimates the spatial distance.   Indeed, for continuum distances around the discreteness scale, this is the source of the DAS studied in \cite{Eichhorn:2017djq}.
It is only for larger distances that one hopes to recover the continuum distance. Note that the predistance cannot lead to an underestimation of the continuum distance for the case of Minkowski spacetime and vanishing extrinsic curvature, because no element $e\, \in \, J^-(\rmin)$ is in $F(a,b)$.

That the above allows a  violation of the triangle inequality is clear from the analogy with the continuum cases with non-vanishing extrinsic curvature. Hence one needs to again follow the minimising prescription as in the continuum. Starting with the predistance function $\dd(a,b)$ we minimise over ``paths'' on $\ca$ from $a$ to $b$.  As in the continuum,  define  the $k$-element ordered set  $\WW_k \equiv (\ww_0, \ww_1, \ldots, \ww_k)$ where $\ww_i\in \ca$ and  $\ww_0=a, \ww_k=b$, and each of the $\ww_i$ are distinct. The path distance along  $\WW_k$ is 
\begin{equation} 
\dd_{\WW_k} (a,b) \equiv \sum_{i=0}^{k-1} \dd(\ww_i, \ww_{i+1}). 
\end{equation} 
The distance function is then 
\begin{equation} 
\DD(a,b) \equiv \inf_{\WW \in \cW} \dd_\WW(a,b),  
\end{equation} 
where $\cW$ is the set of all finite ordered subsets of $\ca$.  
This prescription is completely intrinsic to the causal set $C$. However, if we are interested in reconstructing the geometry of the continuum, then a  comparison with the continuum suggests that this minimisation process could lead to very large errors without a mesoscale cut off $\mco$. The mesoscale cut-off corresponds to the  extent of ``thickening'' of the antichain in the language of \cite{homologyone,homologytwo}. There too, one needs to define a mesoscale regime that lies well above the discreteness scale and well below the local curvature scale. In this regime there is a one parameter family of simplicial complexes labelled by the discrete thickening  volume $\mathbf{v}=\zeta_n (\mco/2)^n$ whose homology is stable.  For small $\mathbf{v}$, the homology fluctuates strongly as a function of $\mathbf{v}$ before rapidly settling into a ``stable'' parameter range.   While the value of  the 
cut-off is adhoc from the discrete point of view, it is only by varying it that we can determine if our distance function converges.
We therefore  restrict $\cW$ to those $\WW_k$ for which $\dd(\ww_i, \ww_{i+1})< \mco $ and denote the distance function associated to each $\mco$ by $\DD_\mco(a,b)$.

 Apart from the continuum errors, there are errors  in $\DD_\mco(a,b)$ which arise from the probabilistic nature of the discrete-continuum correspondence. The key difference between the discrete and the continuum cases  is the existence of DAS, which is significant for scales smaller than $\ell_{DAS}$ (which itself is larger than $ \ell_\rho$, the discreteness scale). Specifically, in a given discrete realisation of a causal set  $C$ approximated by $\bM^n$,  there could exist a pair  $a,b $ in a maximal antichain $ \cA$ for which the associated $\cmin$ is very far into the future.  In other words,  the continuum volume, $V(\cmin)>> \rho \dV(\cmin)$.  This means that for  $\rmin \in \cH(a,b)$,  the continuum interval $J(\rmin, \cmin)$ is empty \footnote{Note that this does not need $\Sigma$ or $\cA$ for  its definition in $\bM^n$.}. However since large voids are highly suppressed in  the Poisson sprinkling it is very unlikely that this is the case.   On the other hand, at  small scales, it is essentially this effect which leads to the onset of DAS. 
 At small $d_h(p,q)$ this leads to a significant overestimation of $d_h(p,q)$ by  $\DD_\mco(p,q)$. Accordingly, $\mco$ should not be made too small.
 
  For every continuum spacetime $(M,g)$ and a $(\Sigma, h)$, there exists a dense enough causal set for which the curvature scale $\ell_K$ is such that   $\ell_{DAS} << \ell_K$ and hence  a stable regime for $\mco$ exists  in between the two. Of course, from the CST perspective  sub-Planckian structures in continuum spacetimes are irrelevant to  the physical causal set.  However, it will still be the case that  $\ell_{DAS}$   is much smaller than an appropriately  coarse grained curvature scale.

In the next section we demonstrate the robustness of our construction with extensive numerical simulations. 

\section{Numerical Simulations}
\label{sec:numsim}

In this section we present results from  numerical simulations for  a set of simple manifold-like causal sets. 

We begin by simulating a manifold-like causal set $C$  {via a Poisson} sprinkling into a prescribed finite spacetime region in Minkowski spacetime $\bM^n$,  for $n=2,3$.  Our code constructs the distance function on {a}  chosen antichain $\ca$ in $C$, as described in Section \ref{discrete}.  In order to test our construction, it is important to find an appropriate  ``closest''  spatial geometry with which to compare.  Doing this is tricky, because as we noted in the previous section, there is an uncountably infinite family of spatial geometries that contain $\cA$; from the CST point of view, $\ca$ simply provides an appropriate  piecewise linearisation of this family.  
For this reason, we choose $\ca$ to be the  antichain  to the past of all other elements in the causal sets, with the assumption that for sufficiently high sprinkling density it will ``hug'' the initial boundary. The approximating continuum spatial geometry is taken to be that of the past boundary of the region we are sprinkling into. By choosing different  boundary geometries with and without extrinsic curvature,  one can thus make  appropriate comparisons\footnote{It is important to stress this practicality: while the prescription applies to {\it any} inextendible antichain, the comparison with a continuum geometry is much harder to do numerically {in the general case}, since the piecewise linear geometry approximating $\ca$ will first need to be reconstructed. }.

It is clear that in order to get a 
 meaningful
comparison with the continuum, not only should the  cardinality $N$ of $C$ be sufficiently large, but also the cardinality $\NA$  of $\cA$.  
For a large $\NA$ 
the procedure of minimisation over paths becomes computationally daunting, since the number of paths of size $k$ for each pair $a,b \in C$  is given by $(\NA-2)(\NA-3) \ldots (\NA-k)$ and  hence grows factorially.  To make the process more efficient, we must reduce the redundancies and devise an efficient algorithm that  quickly converges to the shortest path. 

Again, the continuum example of an inertial $\Sigma$ in $\bM^\dimm$ is helpful. For any $p,q$ in $\Sigma$, consider the  beam $B_{\Sigma}(\rmin)$.  Even though the predistance function is already a distance function in this case, we re-examine the distance minimisation procedure. If we pick a discrete path $\gkwk$ such that there is a  $w_i \in W_k$  which does not lie in $B_{\Sigma}(\rmin)$, then $\gamma_k$ will clearly be longer than the actual distance. To converge to the actual distance faster, we can thus simply ignore these paths, and consider only those for which $W_k \subset B_{\Sigma}(\rmin)$. 

\begin{figure}[!t]
\centering
\includegraphics[width=0.75\linewidth, clip=true, trim=0cm 15cm 9cm 0cm]{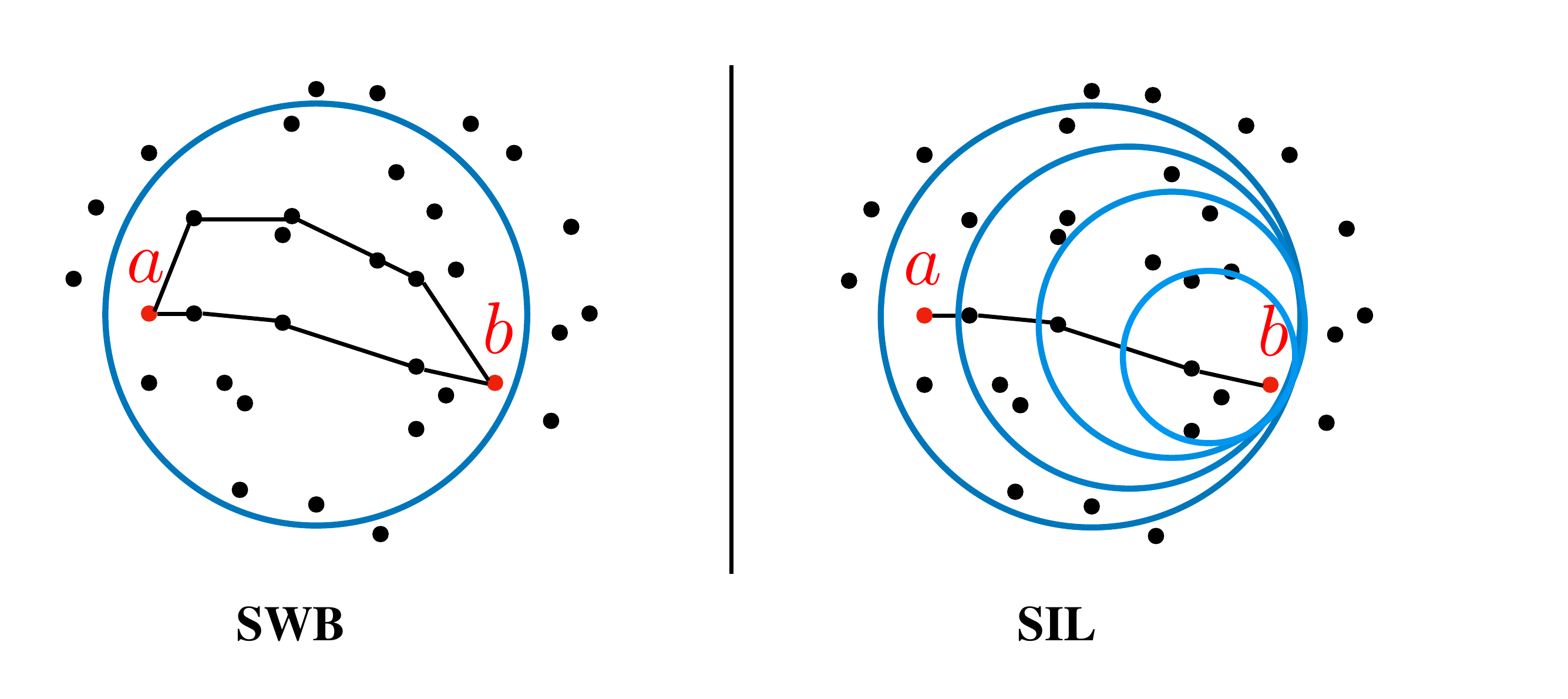}
\caption{\label{fig:swbsil}  We illustrate the SWB algorithm (left) and the SIL algorithm (SIL) for the calculation of $\DD(a,b)$. SWB contains a minimisation over all paths in $B_\ca(a,b)$. SIL ``steps into the light" repeatedly.}
\end{figure}

In the causal set, as we showed above, associated with every $a,b \in \cA$ there is a volume minimising element  $\cmin \in F(a,b)$ and its  beam 
$B_\cA(\cmin) \subset \cA$.  Following the above continuum intuition, we minimise the distance from $a$ to $b$ over paths that lie only inside $B_\cA(\cmin)$. We shall refer to this as  ``staying within the beam''(SWB).  We may further whittle down the set of allowed paths by repeating this procedure at every step by ``stepping into the light" (SIL).  We illustrate  the two procedures in Fig.~\ref{fig:swbsil}.

In the SWB algorithm, since for every $\ww_i \in B_\cA(\cmin)$ the beam for $(a,\ww_i)$ is roughly contained in  $B_\cA(\cmin)$, the process of minimisation also calculates the  distance $d(a,\ww_i)$ in the process of calculating $d(a,b)$. In both SWB and SIL, we use the Dijkstra algorithm \cite{dij}  to find the shortest path between two nodes  in a graph.  For both algorithms we  cross check that there is no significant difference  with these restrictions compared to the  full minimisation procedure. 
The maximum size of the individual steps is {moreover} limited by imposing a mesoscale cutoff $\mco$. We will determine an appropriate range of values for $\mco$ by demanding a convergence to the continuum results at large enough $d(a,b)$.

The SWB algorithm is easily implemented since all we have to do for every pair $a,b$, is to find the beam $B_\ca(a,b)=B_\ca(\cmin)$ and to do the minimisation of paths within the set. In the SIL algorithm, we proceed as follows. Lets pick the first element  $\ww_1 \in B_\ca(a,b)$. The next element $\ww_2$ is then picked from $B_\ca(\ww_1, b)\cap B_\ca(a,b)$. Continuum  intuition suggests that $B_\ca(\ww_1, b)$ is contained in $B_\ca(a,b)$ but because of the random nature of a causal set, this need not be the case. 
Discrete paths  that stray out of the beam could be  shorter than those that are contained within the beam because of ``under densities" that occur due to the statistical nature of the number-volume correspondence. 
The same principle is applied at every consecutive step of the procedure, and this gives us an SIL path $\WW_k$ as shown in Fig.~\ref{fig:swbsil}. The shortest path from $a$ to $b$ is then found by minimising over all SIL paths $\WW_k$ that can be constructed from the elements that lie within $B_{\cA}(\cmin_k)$ and within the successively shrinking beams of each step, provided that each "step-size" from an element $\ww_i$ to $\ww_j$ is smaller than the cut-off $\mco$.

Note that by implementing the cut-off, it becomes possible to assign a notion of ``nearest neighbours" in the sense that two elements $\ww_i$ and $\ww_j$ can be thought of as nearest neighbours when the predistance is smaller than $\mco$. Likewise, two elements are next-to-nearest-neighbours when their predistance is smaller than $2 \mco$ etc. The notion of neighbours can be used as an algorithmic short-cut when obtaining numerical data.

 To confirm that our algorithm gives us the desired results we perform two checks. We first compare the SIL and SWB results for randomly selected pairs of elements in $\cA$. And secondly we compare {both} with paths that are not constrained in any way, or those that can ``step out of the light'' (SOL) at any step.  Our simulations show that these  algorithms are in quantitative agreement. This confirms that the  intuition from the continuum carries over. We therefore use only the computationally most efficient {SWB/SIL} algorithm.

The  simulations we now present  are restricted to regions in $2$ and $3$ dimensional  Minkowski spacetime with either extrinsically ``flat'', i.e., $K=0$  initial boundaries, or ones with $K \neq 0$. Furthermore, the sprinkling density is set to one, i.e., $\rho = 1$ such that we implicitly work in Planckian units. Specific information on the numerics such as the size of the causal set, the size of the anti-chain etc., is relegated to Appendix~\ref{app:num}.

We define the error, 
\be
\Delta(a,b) = \frac{\DD(a,b)-d_h(a,b)}{d_h(a,b)}, \label{eq:Deltadef}
\ee
where $d_h(a,b)$ is the induced continuum distance on the initial boundary.  At small $d_h(a,b)$ we expect $\Delta(a,b)\gg 0$, because of DAS, but that there exists a  ``large enough''  $d_h(a,b)$ for which $\Delta(a,b) \rightarrow 0$. It is also useful to  compare the distance with the predistance to demonstrate the importance of the path minimising procedure. The error  for the predistance is defined analogously as 
\be
\widetilde{\Delta}(a,b) = \frac{\dd(a,b) - d_h(a,b)}{d_h(a,b)}.
\ee

\subsection{Two dimensions}

\begin{figure}[!t]
\includegraphics[width = 0.45\linewidth, keepaspectratio]{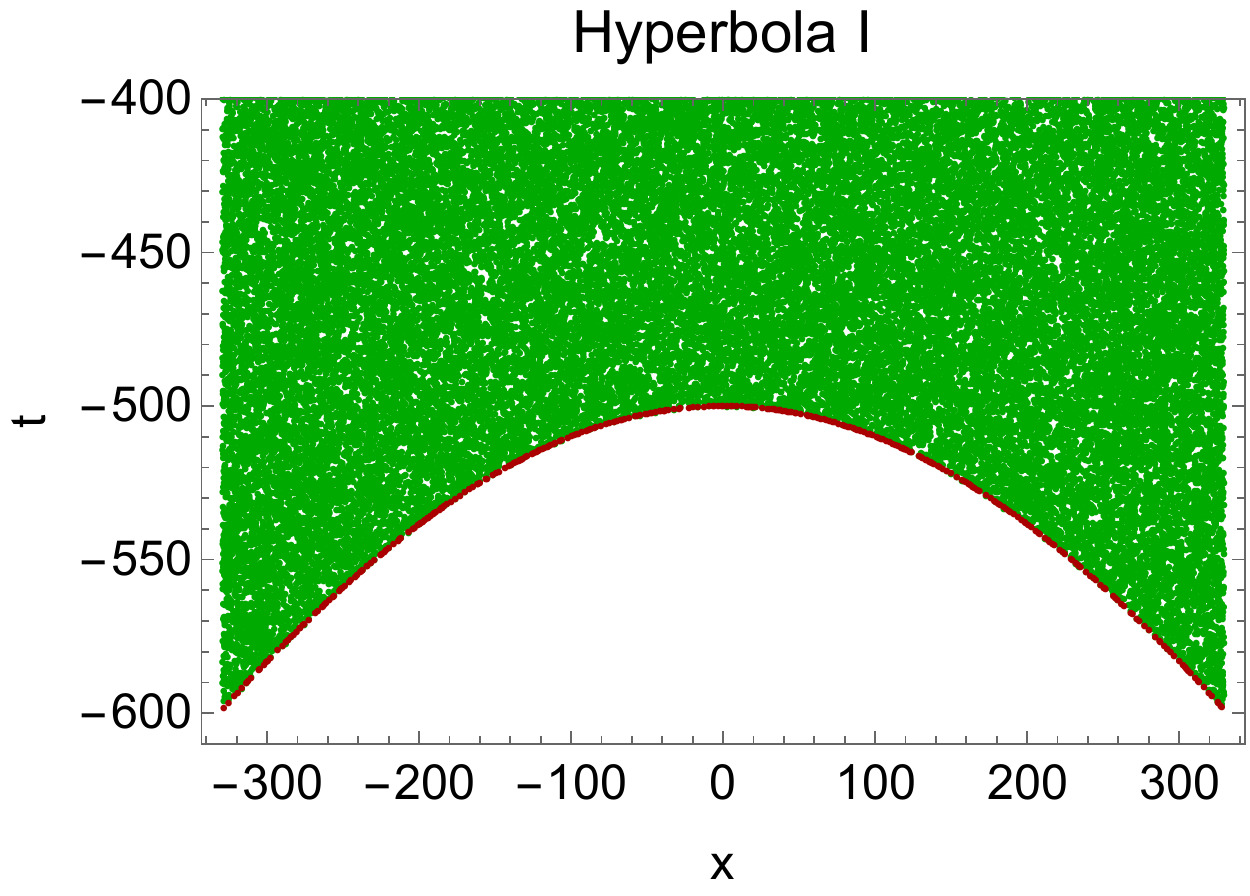}\quad\quad \includegraphics[width = 0.45\linewidth, keepaspectratio]{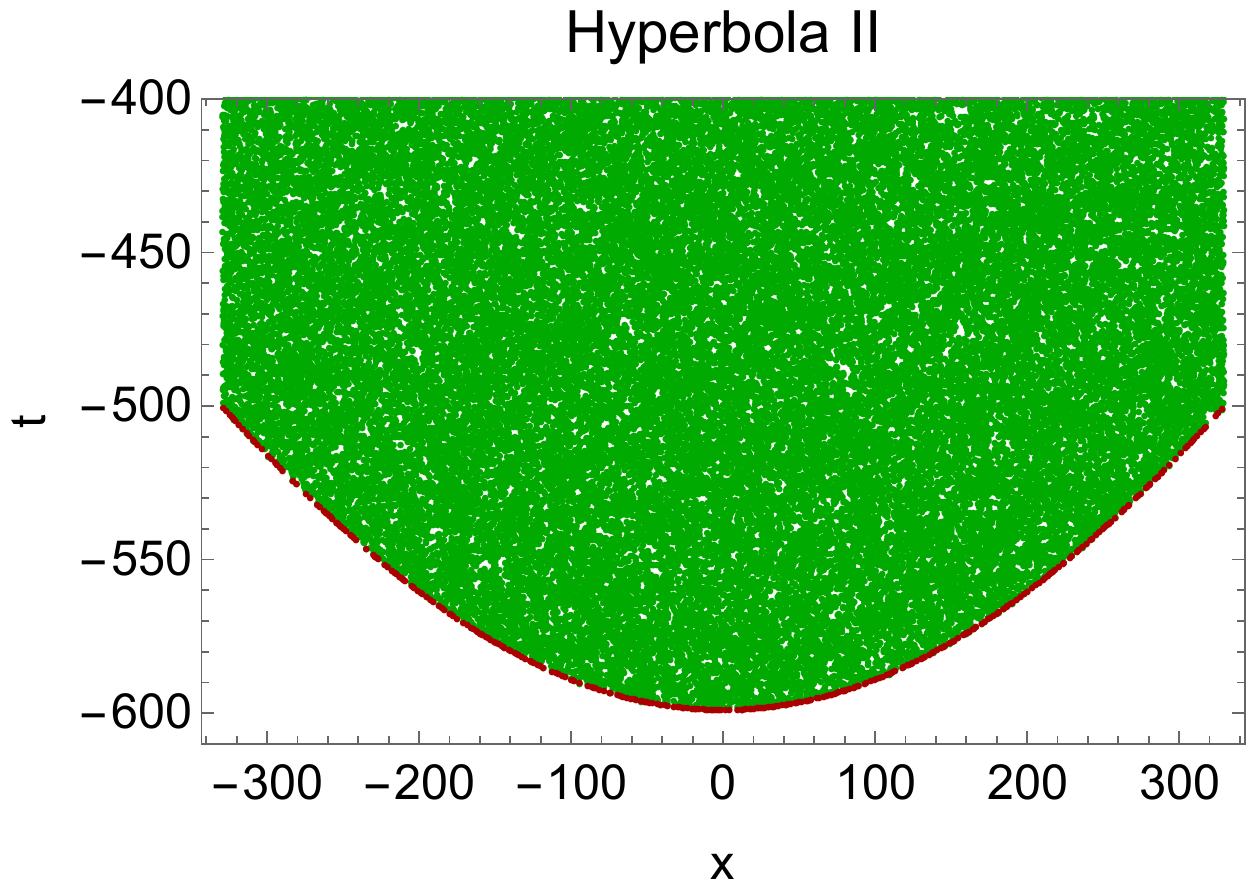}
\caption{\label{fig:HypSet} Sprinkling into regions of 2d Minkowski spacetime,  with initial boundaries of positive and negative constant curvature, for a specific radius of curvature   $\ell_K = 500$. The antichain (red dots) hugs the spacelike boundary. Here only a part of the full range of the $t$-axis has been depicted, whereas for the simulations the $t$-axis runs from $-600$ to $+600$.}
\end{figure}

The simplest example of a manifold-like causal set is one that is approximated by a region in $\bM^2$.  We will consider a rectangular region in $\bM^2$, from which a subregion is removed, to create  an initial boundary of different extrinsic curvatures.  As described  earlier, we choose the antichain to be the minimal one or initial one, with the assumption that for sufficiently high sprinkling density it will ``hug'' the initial boundary.  This allows us to make detailed comparisons with the continuum induced distance on the boundary. We show an example of this in Fig.~\ref{fig:HypSet}.

Note that our choice of region is not ideal since it is not causally convex and allows for non-trivial edge effects. In particular, for pairs of points on the minimal antichain that are close to the edge of the region, $V(\rmin)$ will necessarily be smaller than if one were looking at a strictly larger region. These edge effects are not that important in 2d, but will  become progressively more important in higher dimensions, where there are many more such ``edge''  pairs.

We begin with the full rectangular region, with flat $K=0$  initial boundary and continue onto the more challenging cases where the bottom of the rectangle has been scooped out, yielding initial boundaries  with varying types of extrinsic curvature. The latter include hyperbolae of constant extrinsic curvature as well as those of varying extrinsic curvature.  The continuum distance $d_h$ must be calculated for each of these cases in order to find $\Delta$ and $\wD$.  Specifically, we examine different types of extrinsically curved boundaries and find the range of $\mco$ for which the error is minimised.

\begin{figure}[!t]
\includegraphics[width = 0.45\linewidth, keepaspectratio]{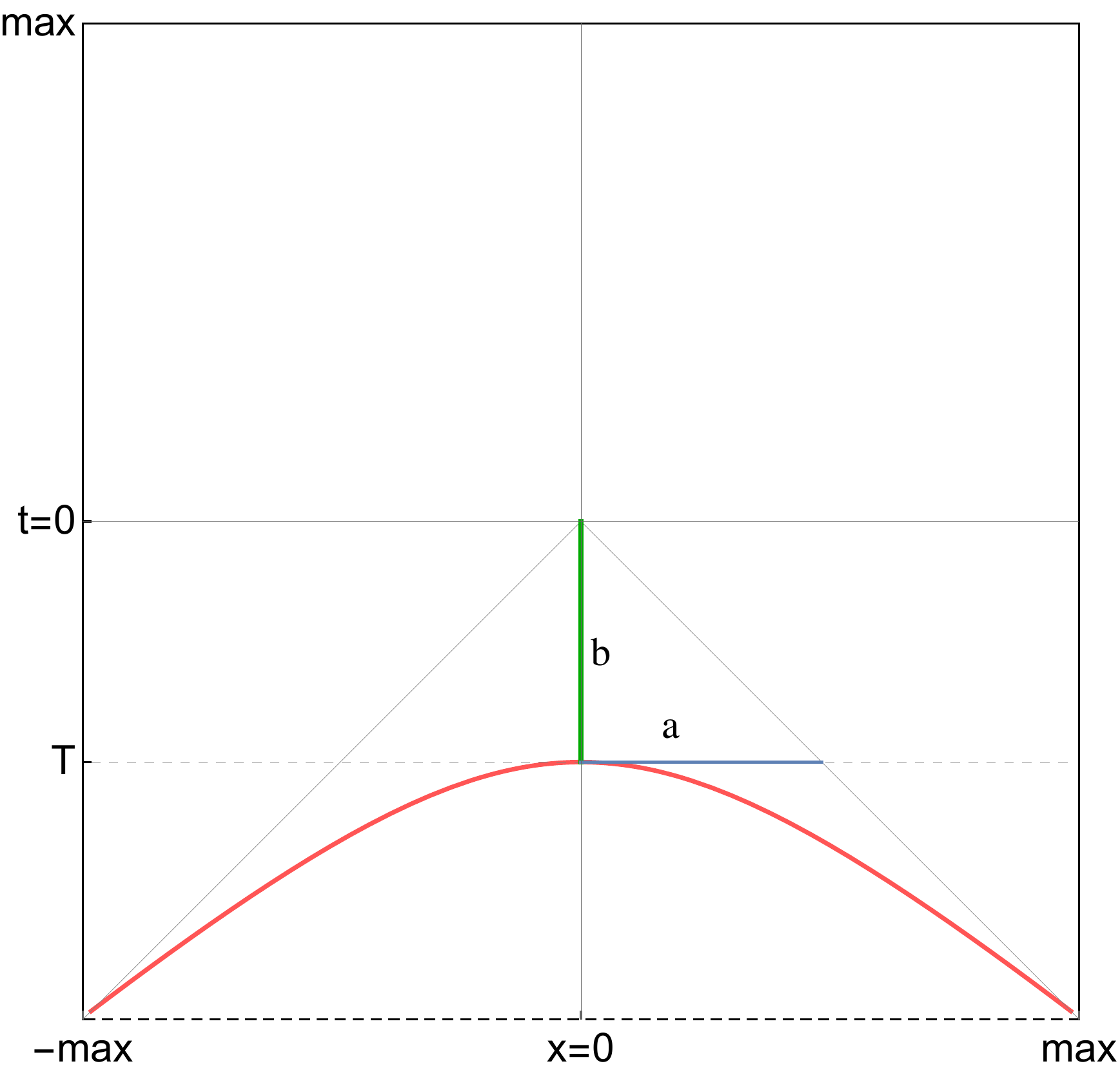}\quad\quad \includegraphics[width = 0.45\linewidth, keepaspectratio]{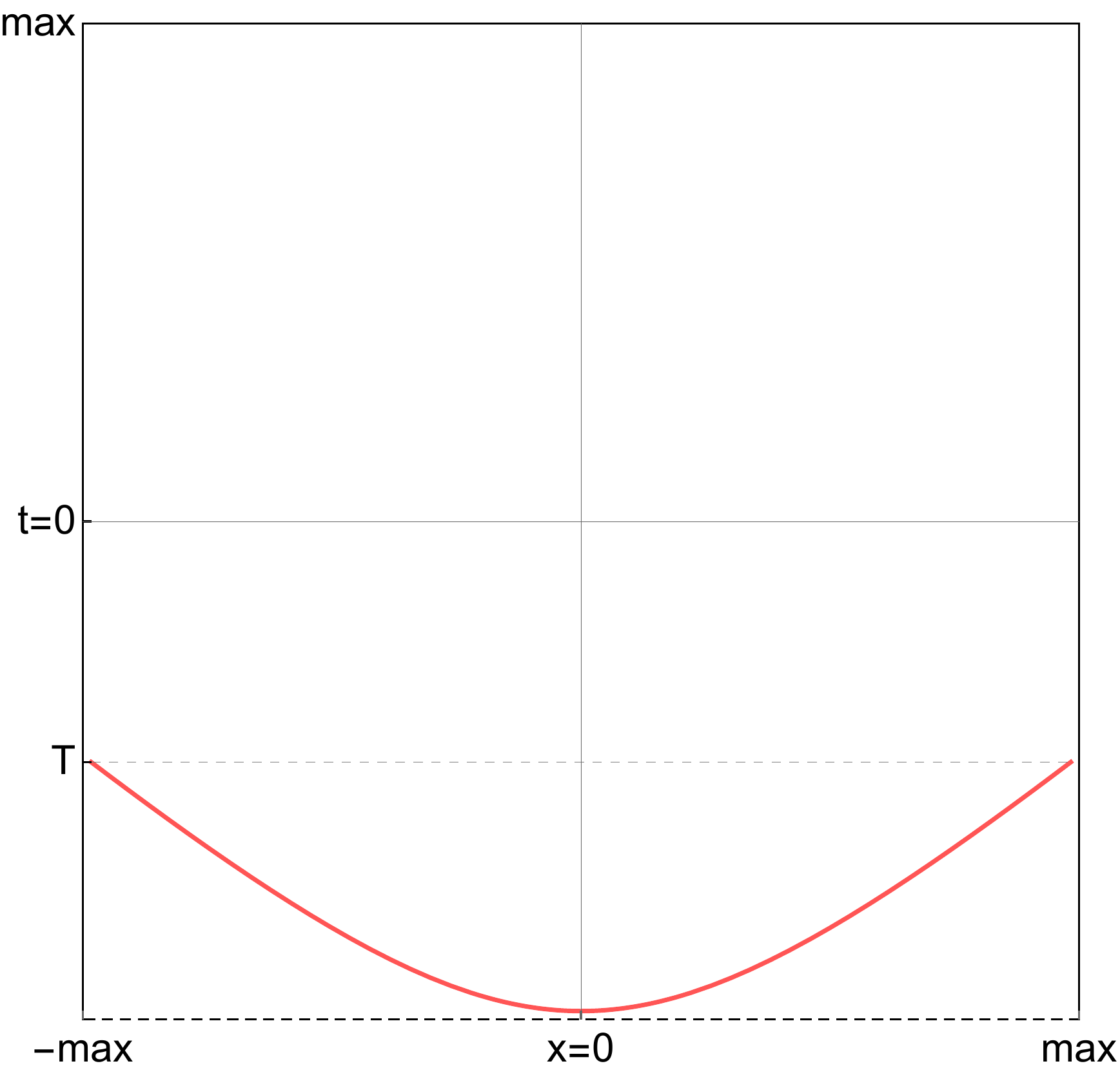}
\caption{\label{fig:paraHypI} The $H_+$ hypersurface is shown on the left and the $H_-$ on the right, with an illustration of the parameters $a$ and $b$, and their relationship to $T$ and $max$.}  
\end{figure}

Let us focus first on hyperbolae, with positive or negative curvature, which we denote by $H_\pm$, given by the equation
\begin{equation}
\label{eq:hyp}
\frac{(x-x_0)^2}{a^2} - \frac{(t-t_0)^2}{b^2} = -1,
\end{equation}
where $a$ and $b$ are the distance from the origin to the vertex and the distance from the vertex to the conjugate axis, respectively, as shown in Fig.~\ref{fig:paraHypI}  and $x,\, t$ are the coordinates for any point on the hyperbola.  The extrinsic curvature $K_{a b} = \nabla_{a}\, N_b$, where $N_b$ is the tangent vector to the hyperbola and $\nabla_a$ the covariant derivative, can be calculated.  In 2 spacetime dimensions this is just a scalar, which we denote by $K$, and yields the radius of curvature $\ell_K = \frac{1}{K}$. In the case of a non-constant curvature hypersurface the curvature will not only depend on the parameters $a$ and $b$ but vary with the spatial coordinate $x$ as well.

 Using  Fig.~\ref{fig:paraHypI}   to define $a$ and $b$ in terms  of the two quantities $\max$ and $T$ yields,
\begin{equation}
\frac{\max^2 - T^2}{\max^2\, T^2}\, x^2 - \frac{t^2}{T^2}= -1,
\end{equation}
for the  $H_+$ spacetime. For the $H_-$ spacetime the parameters $a$ and $b$ remain the same, but a shift in the origin is required: 
\begin{equation}
\frac{\max^2 - T^2}{\max^2\, T^2}\, x^2 - \frac{(t + \max - T)^2}{T^2}= -1.
 \end{equation}
The induced spatial distance  between two points $p=(t_p,\, x_p)$ and $q=(t_q,\, x_q)$ is therefore 
\begin{eqnarray}
\Delta s =&& \int_{x_q} ^{x_p}  dx \sqrt{\frac{b^2\, x^2}{a^2\, (a^2 + x^2)} - 1}\\
=&&  a\, \left(\, \mbox{EllipticE}\left(\, \arcsin{\left(\, i\, \frac{x_p}{a} \right)},1-\frac{b^2}{a^2}\right) - \mbox{EllipticE}\left(\, \arcsin{\left(\, i\, \frac{x_q}{a} \right)},1-\frac{b^2}{a^2}\right)\right).
\end{eqnarray}
The maximum curvature is found at $x=0$ and yields a curvature radius of $\ell_K = \frac{a^2}{b}$.

In the case of constant extrinsic curvature, (i.e., $a = b$) the parametrisation simplifies to
\begin{equation}
(x-x_0)^2 - (t-t_0)^2 = -c^2.
\end{equation}
Note that now we only have one free parameter, $c$ and  that $x_{max} \neq t_{max}$. 
The induced distance between any two points $p,\, q$ on such a hypersurface is 
\begin{equation}
\Delta s = c\left(\arcsin{\left(\frac{x_p}{c} \right)} -\arcsin{\left(\frac{x_q}{c} \right)}\right),
\end{equation}
and the radius of curvature is found to be $\ell_K= c = - T$, where $T < 0$, cf.~Fig.~\ref{fig:paraHypI}. As the maximum value for $T = -t_{max}$, we find, in the limit of an infinite spacetime $t_{max}\rightarrow\infty$, the radius of curvature  $\ell_K\rightarrow\infty$ or $K = 1/\ell_K {\rightarrow}  0$. However, as the size of our causal sets is limited, the maximum value for the radius of curvature $\ell_K = t_{max}$ is likewise limited by the size of the spacetime.

\subsubsection{$K=0$}\label{sec:2dflat}
\begin{figure}[!t]
\centering
\includegraphics[width=0.48\linewidth]{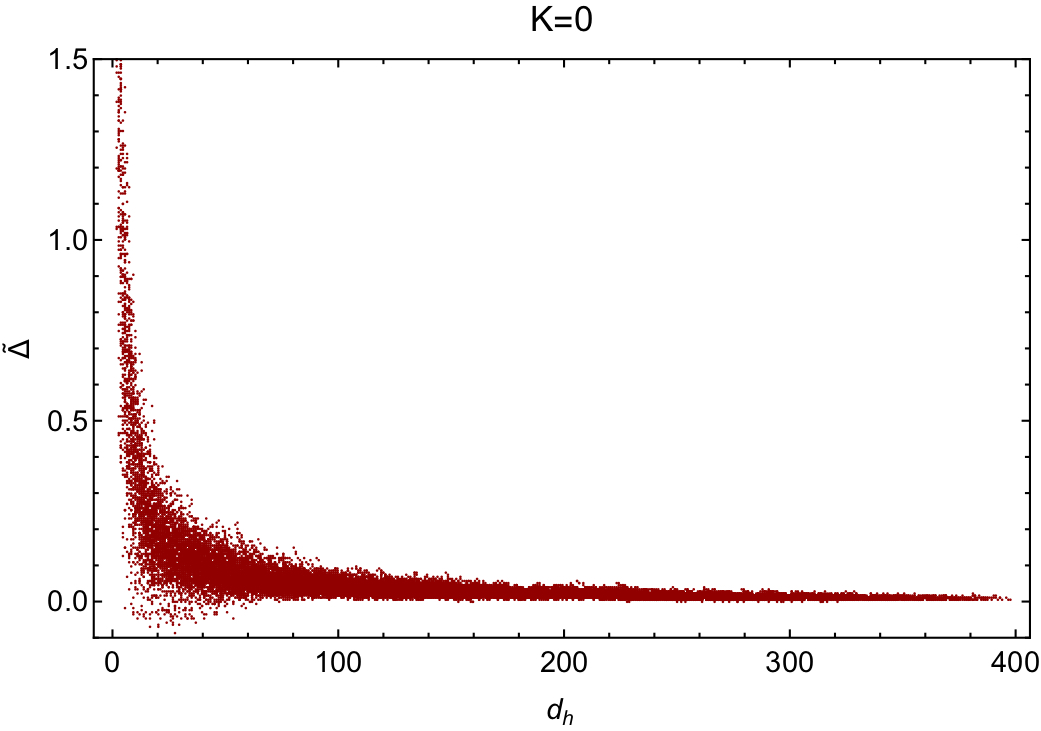} \quad
\includegraphics[width=0.48\linewidth]{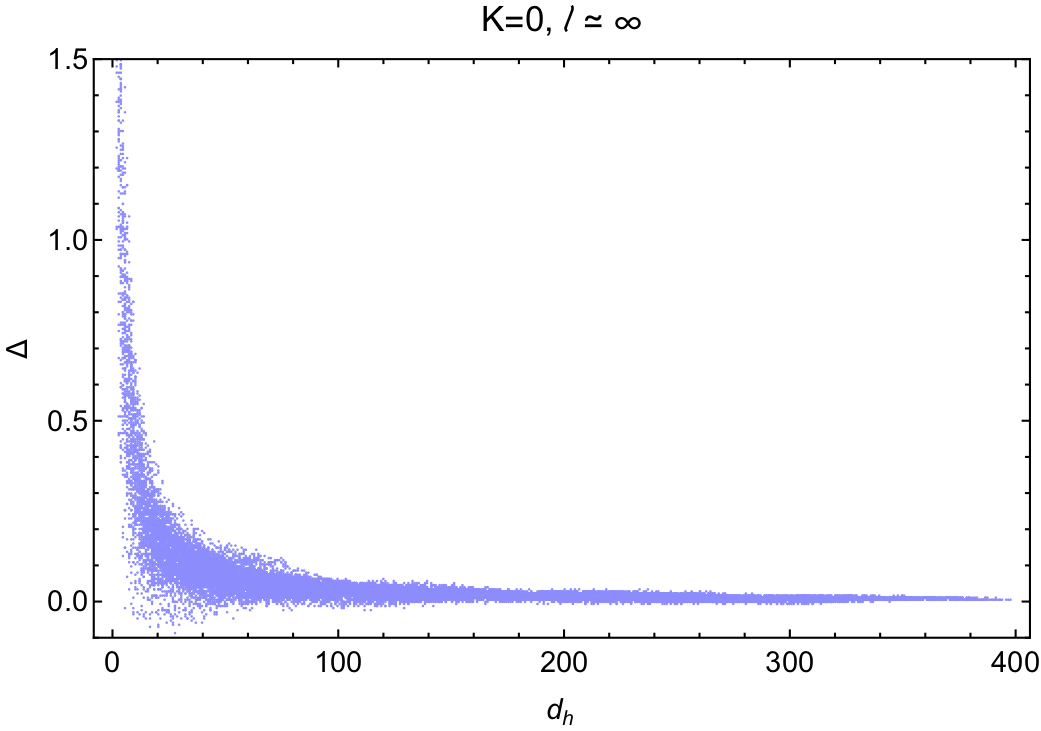}
\caption{\label{fig:flatpred} Left: Error in the  predistance as a function of the continuum distance for a rectangular region of $\bM^2$.  Right: Error in the  discrete distance as a function of the continuum distance for a rectangular region of $\bM^2$, using an ``infinite''  cut-off $\mco$.}
\end{figure}
\begin{figure}[!t]
\centering
\includegraphics[width=0.6\linewidth]{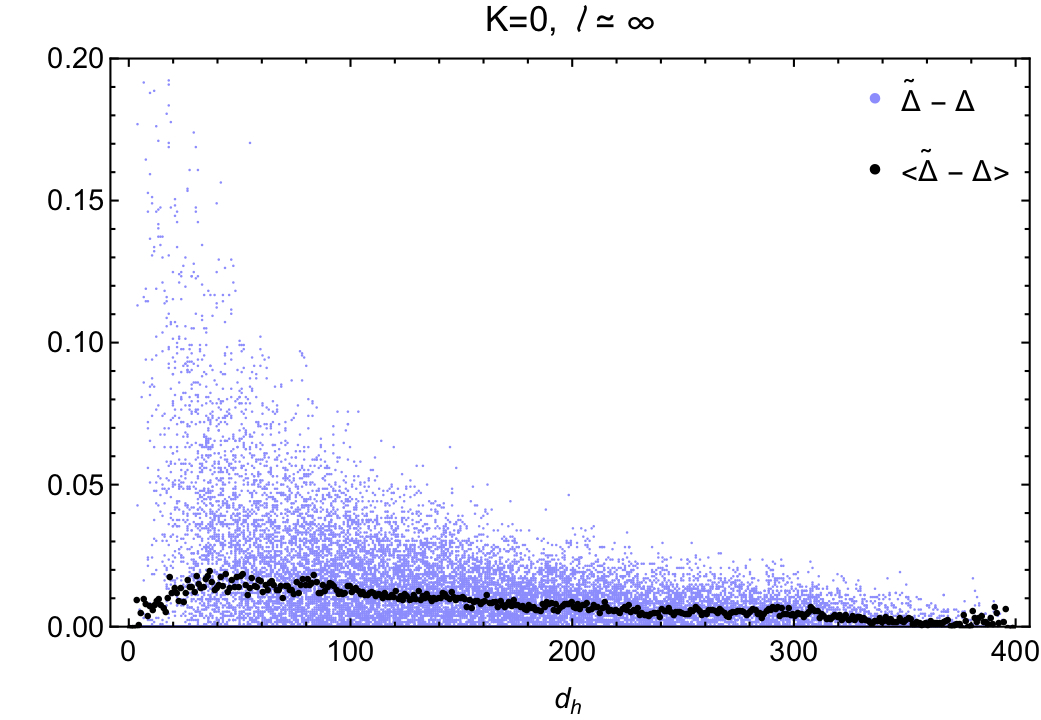}
\caption{\label{fig:flatdpreddis} Difference of $\tilde{\Delta}$ and $\Delta$ as a function of $\dcont$ (blue dots). }
\end{figure}

We sprinkle into the open set  $I \times I \subset \bM^2$, with the initial boundary at coordinate time $t=0$. 
The elements in the  minimal antichain  $\ca$  
hug the boundary at  $(0,x)$ so that $d_h$ can be assumed to be the coordinate distance $|x-x'|$ between any two elements in $\ca$ whose spatial coordinates are $x$ and $x'$.     

The error in the predistance function  $\wD$ is shown in  the left panel of Fig.~\ref{fig:flatpred}.   
At small $d_h$, the DAS effect  shown in \cite{Eichhorn:2017djq} is reproduced, whereas at larger $d_h$, $\dd$ converges rapidly to $d_h$. Thus for $K=0$, the predistance $\dd$ suffices as a distance function for distances larger than $\ell_{DAS}$, as expected.

However, in  preparation for the non-trivial cases, we proceed to find the distance function $\DD$ using the SWB/SIL algorithm. 
We show the error $\Delta$ in  the right panel of Fig.~\ref{fig:flatpred}, 
without imposing a mesoscale cut-off $\mco$. 

In Fig.~\ref{fig:flatdpreddis} we show  the difference $\tilde \Delta-\Delta$, which is always positive.  For clarity we include the average difference (black dots) $\langle \tilde \Delta - \Delta \rangle$. The larger $\dcont$ becomes, the smaller the relative deviation between the predistance and discrete distance, supporting the expectation that the predistance performs satisfactorily for $K=0$ and large $\dcont$.

For our simple case, since $K=0$, $\ell_K \rightarrow \infty$ and  hence the mesoscale $\mco$ can be made arbitrarily large. The right panel of Fig.~\ref{fig:flatpred} shows the error in $\DD$ when $\mco$ is allowed to be ``infinite'', which in practical terms simply means that  
$\mco$ is larger than the maximum $\dcont$ allowed by the spacetime.
For distances larger than $\ell_{DAS}$ the error becomes negligible. This is the first confirmation that our discrete distance function has the right continuum approximation.

\begin{figure}
\centering{
\includegraphics[width=0.45\linewidth]{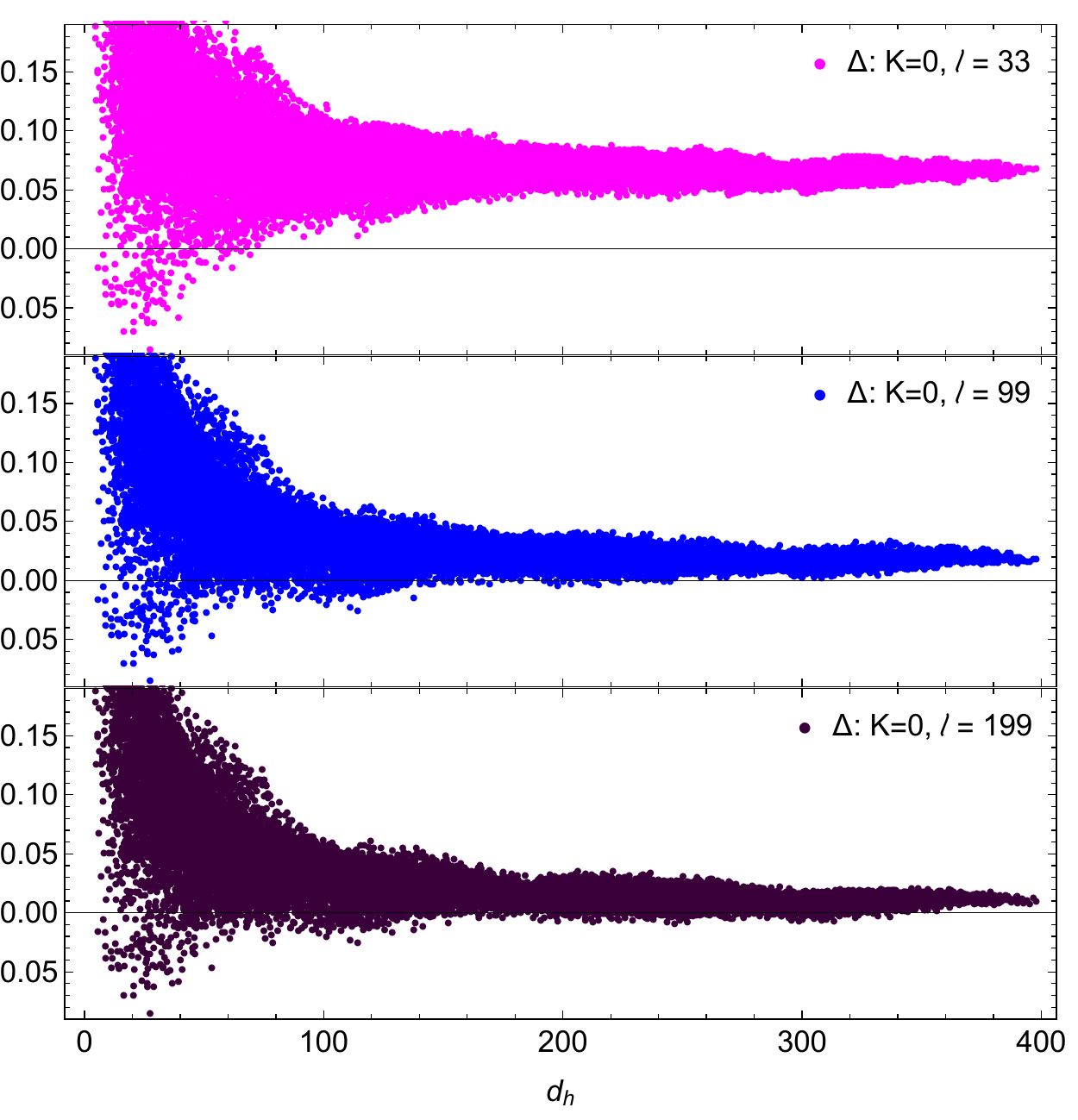}\quad \includegraphics[width=0.45\linewidth]{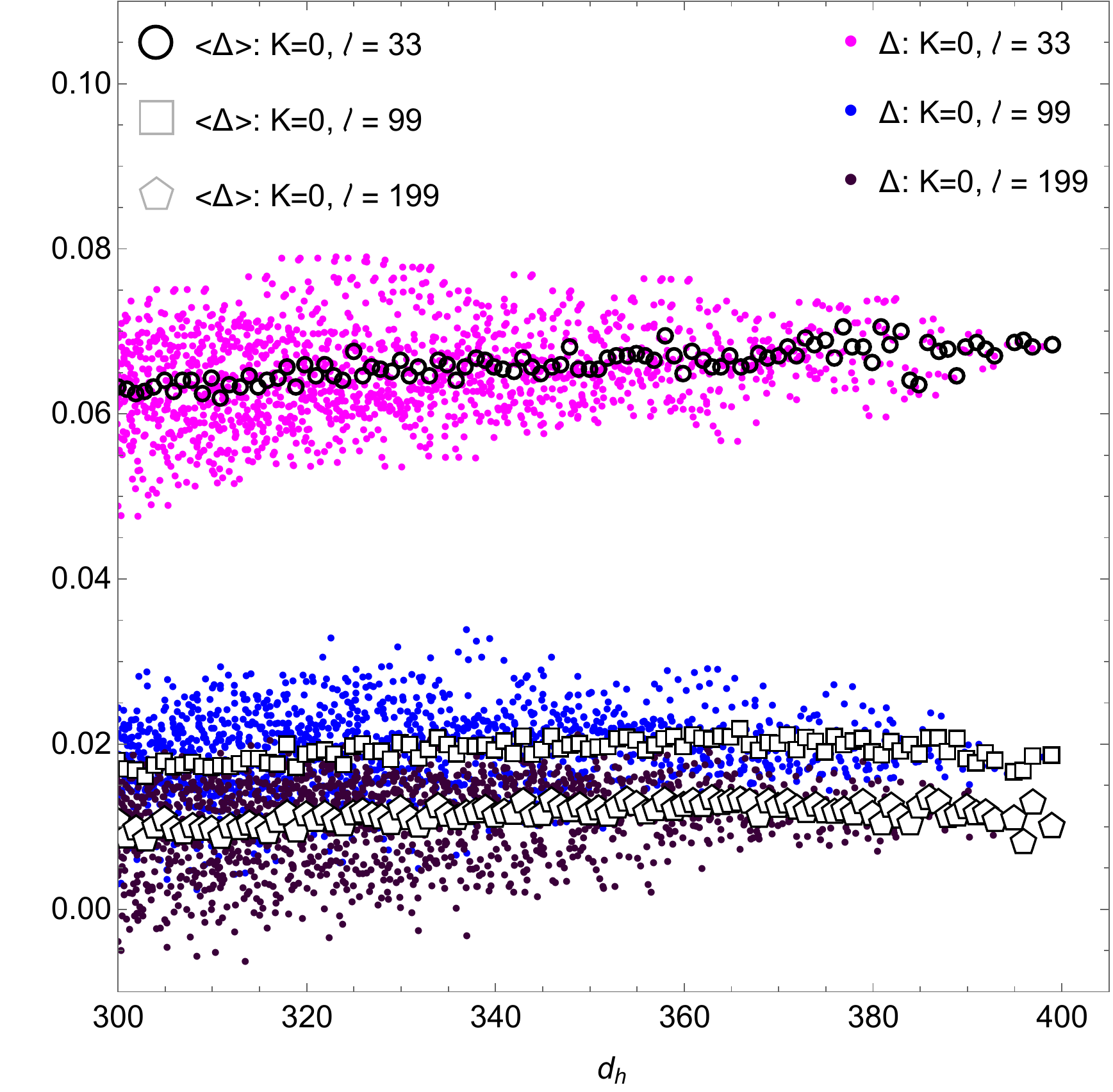}
\caption{\label{fig:flatcutoff}$\Delta$ as a function of $\dcont$ for an IR cutoff $\mco\approx 33$ (pink dots), $\mco\approx 99$ (blue dots) and $\mco\approx 199$ (purple dots). }}
\end{figure}

However, as we show in   Fig.~\ref{fig:flatcutoff} ,
$\mco$ cannot be made too small: for $\mco \sim \ell_{DAS}$, the deviation from the continuum becomes significant not only for small distances, but also for large distances. The latter is a result of an accumulation of small errors in the distance: for every pair $a,b$ for which $\mco \ll d_h(a,b) $, the shortest path from $a,b$ is made up of  several segments of length bounded from above by $\mco$. For each one of those pairs, $\DD(a,b)$ is significantly overestimated due to the effect of DAS.

\subsubsection{ $K\neq 0$ }

In the previous section we noted that as long as $\mco>\ell_{DAS}$ the distance function converges to the continuum distance. The challenge of course is when $\ell_K < \infty$  and one has to find the  optimal range of $\mco$ in which  $\ell_{DAS} \ll \mco \ll \ell_K$.

\begin{figure}[!t]
\includegraphics[width=0.5\linewidth]{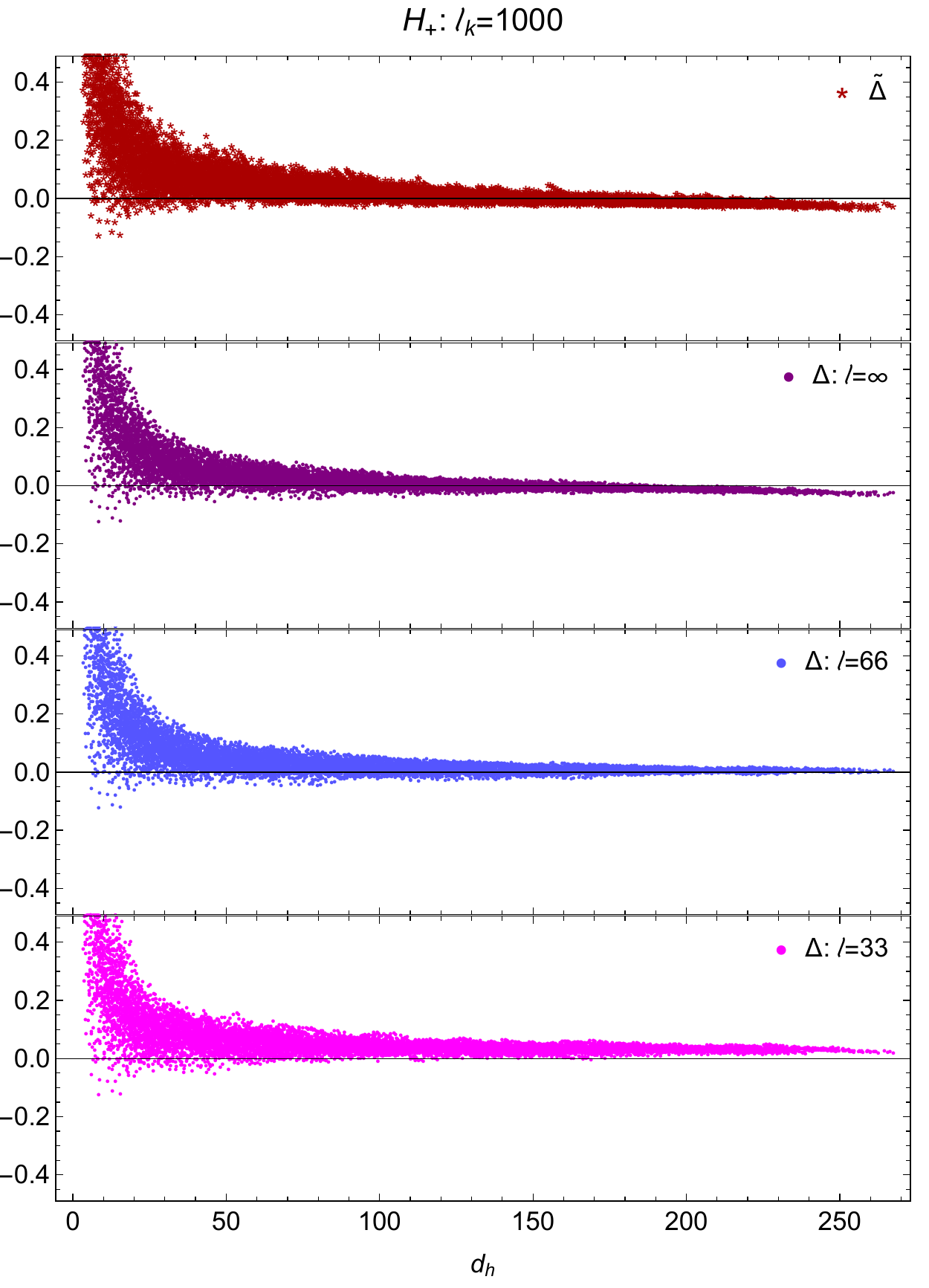}\quad\includegraphics[width=0.5\linewidth]{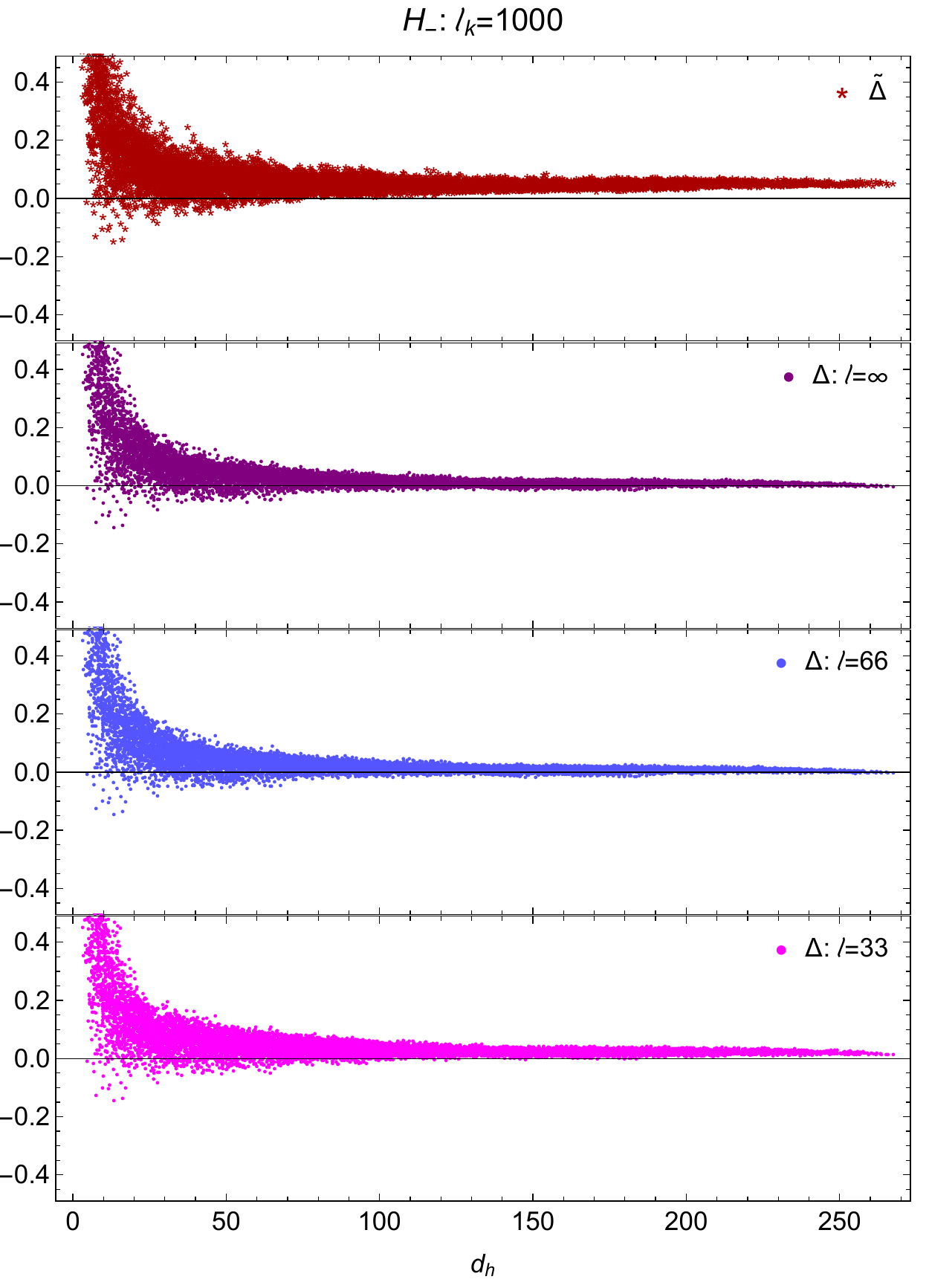}
\caption{\label{fig:CC} $\tilde{\Delta}$ (red stars) and $\Delta$ for different values of $\mco$ for $H_+$  (left panel) and $H_-$ (right panel). }
\end{figure}

We first consider the case of constant K hyperbolae, with positive or negative curvature with respect to the inward (or future time-like) directed normal, which we denote by $H_\pm$. Because of the obvious time asymmetry in our distance computation which uses the future of the antichain, one expects that the outcome for  $H_+$  and $H_-$  will differ. This is due to the consistent local underestimation of the predistance in the latter case as discussed in Section \ref{continuum}. Alternatively, the time-reflected version of our prescription could be used, employing the past instead of the future of the antichain.

In order for there to be  a  sufficient  separation of scales  between  $\mco$ and $\ell_K$,  we analyse cases with small  enough $K$ or conversely large enough $\ell_K$. To provide a contrast we also examine the case when there is no proper separation of scales i.e., any choice of $\mco$ is either too close to $\ell_{DAS}$ or $\ell_K$.  As expected,  $\DD(a,b)$ fails to reproduce $d_h$ correctly in these cases.   For $\ell_K$ sufficiently large, we find  a regime for $\mco$ where the error is small, and hence $\DD$ does a good job of reproducing  $d_h$. This observation goes hand in hand with the fact that the predistance is no longer a useful distance function. 

\begin{figure}[!b]
\includegraphics[width=0.5\linewidth]{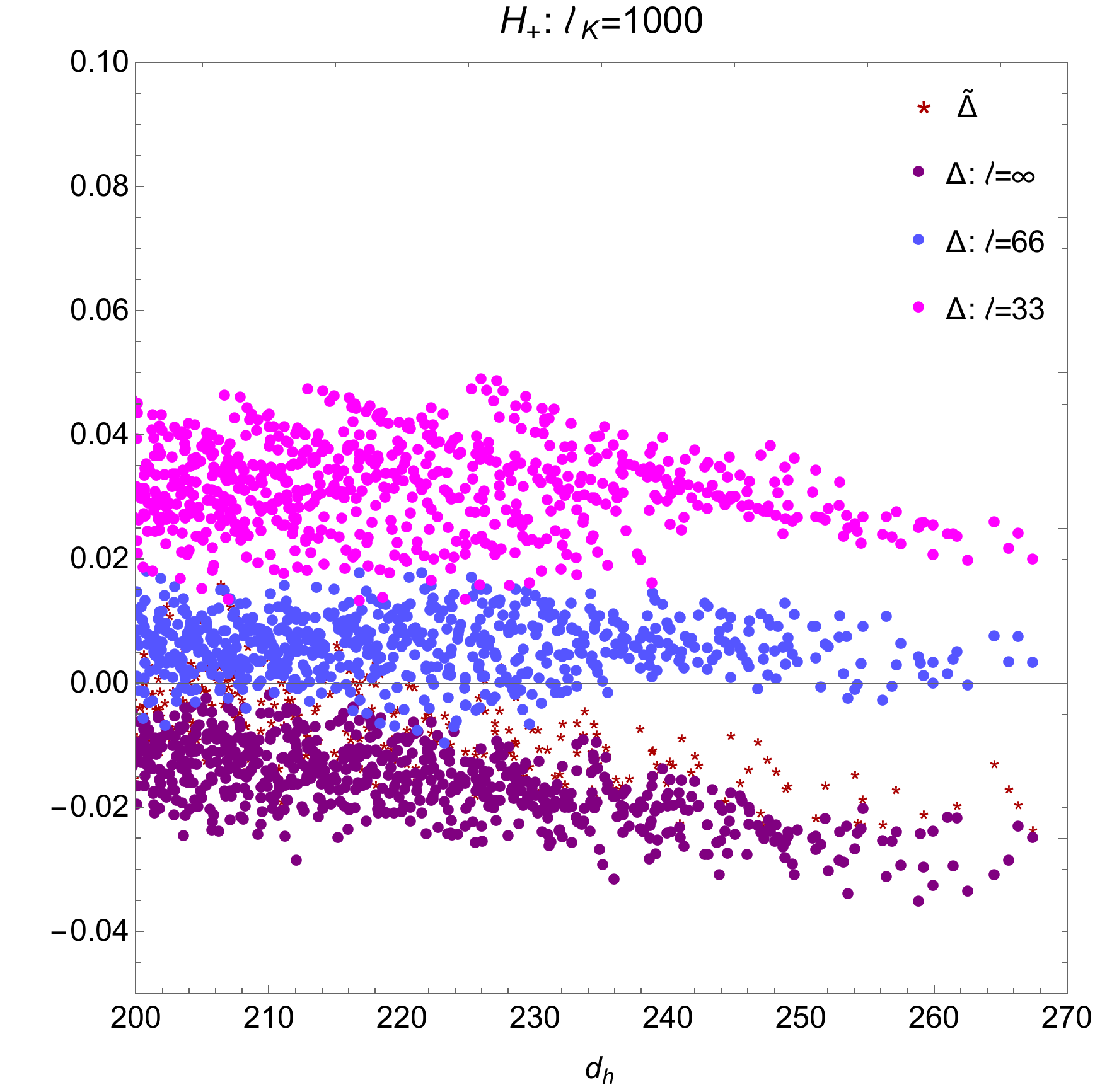}\quad\includegraphics[width=0.5\linewidth]{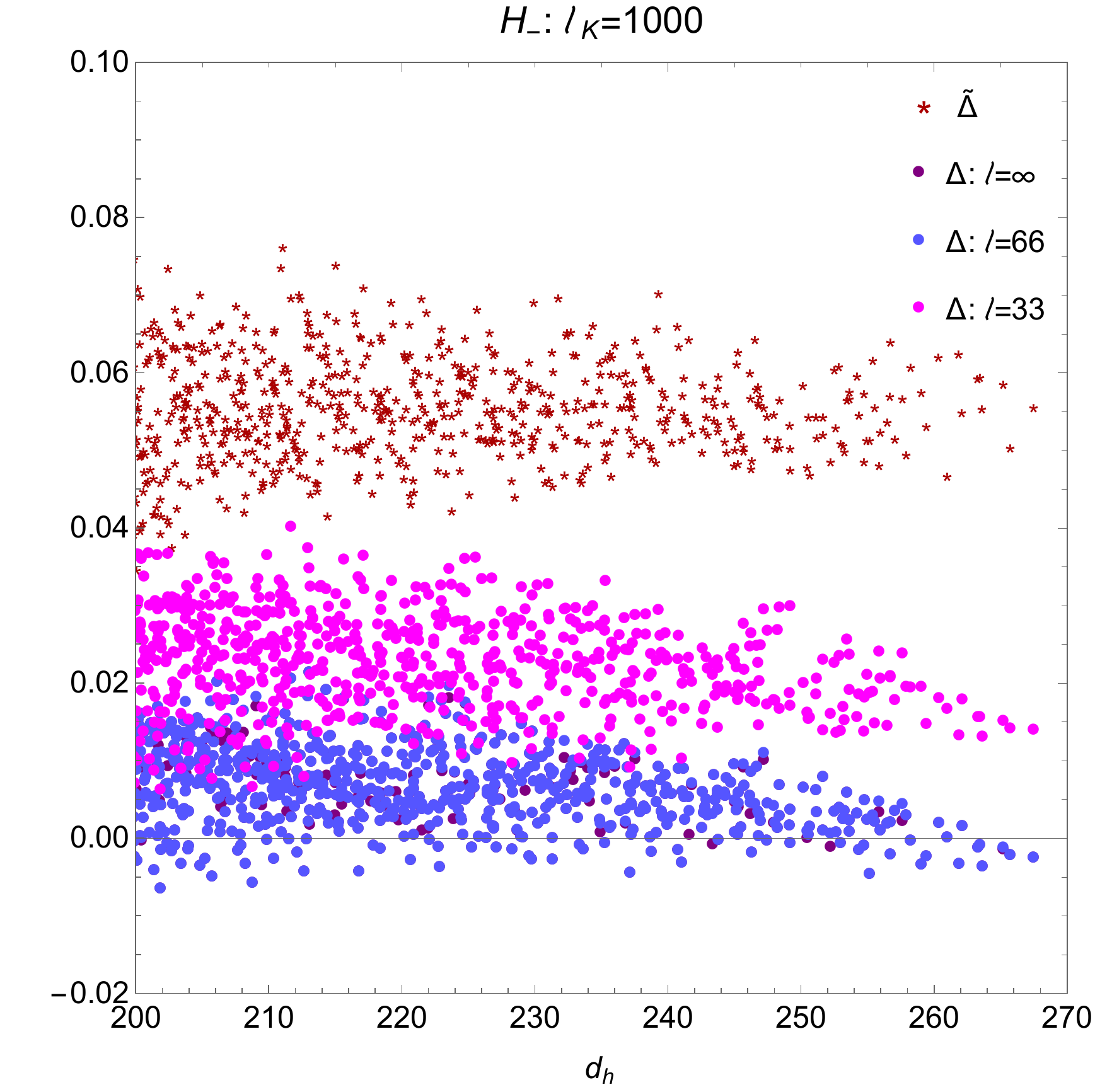}
\caption{\label{fig:CC_zoom} Zoom in of the distance function for different values of the cut-off and a large continuum distance, $d_h\gg d_{\rm DAS}$. For the $H_+$  case, a clear difference between the three different cut-offs is visible. The $H_-$  shows  a good convergence to $\dcont$ already for $\mco=\infty$, although lowering the cut-off to $\mco=66$ yields a slightly better convergence. In both cases a cut-off too deep within the DAS regime ($\mco = 33$), results in an overestimation of $\dcont$.}
\end{figure}

Here we present the outcome for simulations for $\ell_K=1000$ and $\ell_K=500$. 
We find that $\ell_K=1000$ admits a good separation of scales and hence that there is a range of mesoscales for which the error in the discrete distance is small, while for $\ell_K=500$, this is no longer the case. 
Fig.~\ref{fig:CC}  shows the errors for the predistance and distance functions for different choices of $\mco$ when  $\ell_K=1000$, both  for $H_+$ and $H_-$.  As expected, from the time asymmetry of our construction, the predistance shows an underestimation for $H_+$ and an overestimation for $H_-$. (If one had chosen to define the distance using the {\it past} of the antichain rather than the future, this would obviously be reversed).

For  $H_+$, and $\mco \rightarrow  \infty$, $\DD$ shows the same trend as $\dd $ and underestimates the continuum distance, $d_h$. By lowering the cut-off to $\mco=66$, $\DD$ is forced to larger values, leading to a better convergence. Lowering  the cut-off further   to $\mco = 33$ however takes us to the DAS regime, and  again  $\DD$ is pushed to towards larger values and less convergence. As the radius of curvature is $\ell_K=1000$ and $\ell_{DAS}\approx 33$ there is consistency when $\mco$ lies in the approximate range  $ [66, 100]$.  In all cases the increased errors in $\DD$ for large $d_h$ are a result of the cumulative accumulation of errors at the mesoscale. This is one reason why the prescription, even in the continuum, needs a compact Cauchy hypersurface, or equivalently a finite $\ca$.

\begin{figure}[!b]
\includegraphics[width=0.5\linewidth]{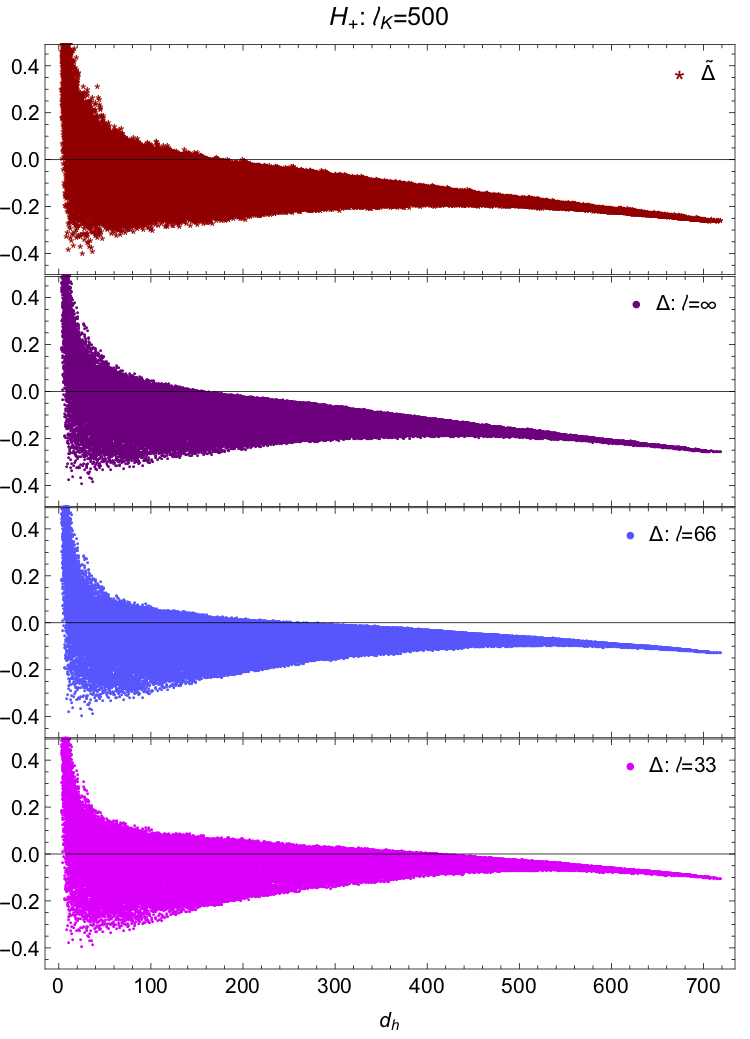}\quad\includegraphics[width=0.5\linewidth]{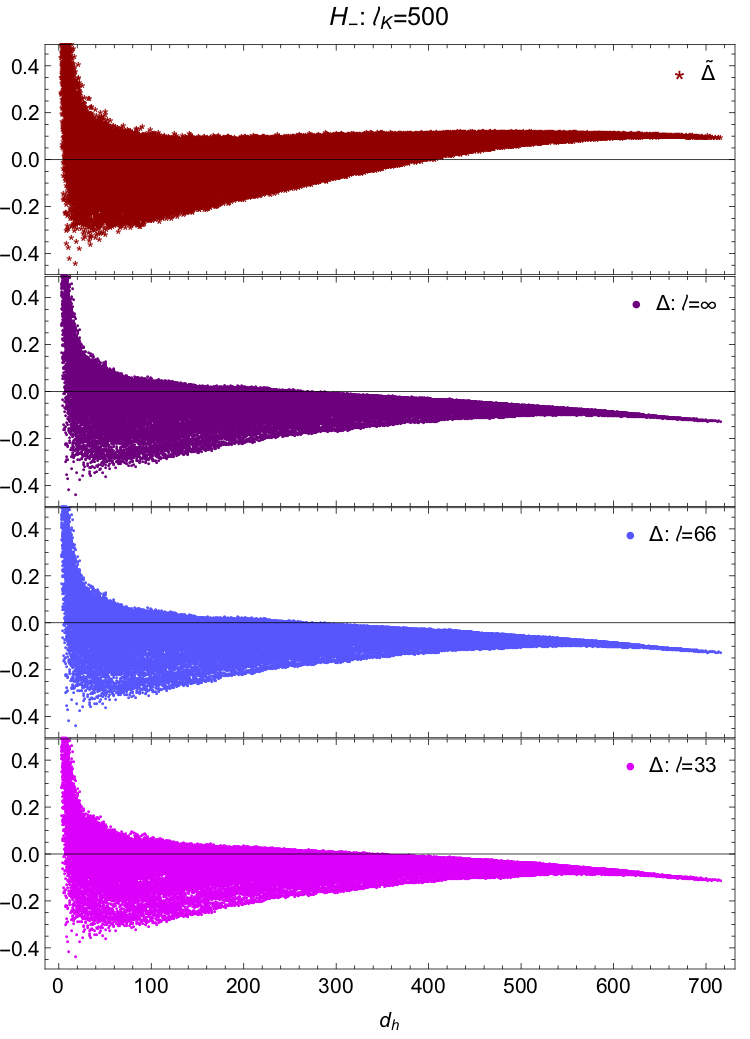}
\caption{\label{fig:CC_Klarge} $\tilde{\Delta}$ (red stars) and $\Delta$ (magenta, blue, purple) for different values of $\mco$, for the $H_+$  (left panel) and the $H_-$  (right panel) cases. Due to the smallness of the radius of curvature, $\mco_K$, it is not possible to find a value for $\mco$ that leads to a satisfactorily convergence, hence there is no adequate separation of scales. }
\end{figure}

For $H_-$,   $\dd$ overestimates $\dcont$ as does $\DD$ for $\mco\approx 33 \sim \ell_{DAS}$, as expected. Interestingly,  $\DD$ shows a good convergence to $d_h$ for all values of $\mco \gg \ell_{DAS}$,  including the infinite cut-off limit ($\mco=199$).  This can be understood by remembering that the distance function is constructed as a minimisation over all possible paths, for a given $\mco$ which limits the maximum step size. Consequently, once the smallest value for the cut-off is identified for which there is  agreement between $\DD$ and $d_h$,  increasing $\mco$  does not further minimise $\DD$.  This is because by construction the shortest path will be selected, which happens to already be present at small $\mco$ for the case of $H_-$.

If by increasing the cutoff it is possible to find a shorter path, then one starts to observe an underestimation in the infinite cut-off limit. This does not happen for $H_-$, but does occur for $H^+.$  A moderate manifestation of this effect can already be observed in the current case ($\ell_K=1000$), cf.~Fig.~\ref{fig:CC_zoom}, but becomes more apparent in the case of a smaller radius of curvature, cf.~Fig.~\ref{fig:CC_Klarge}. As mentioned earlier, this apparent asymmetry in the two cases is simply the result of our choosing the future of the antichain to determine $\dd$ and not its past.

\begin{figure}[!t]
\includegraphics[width=0.5\linewidth]{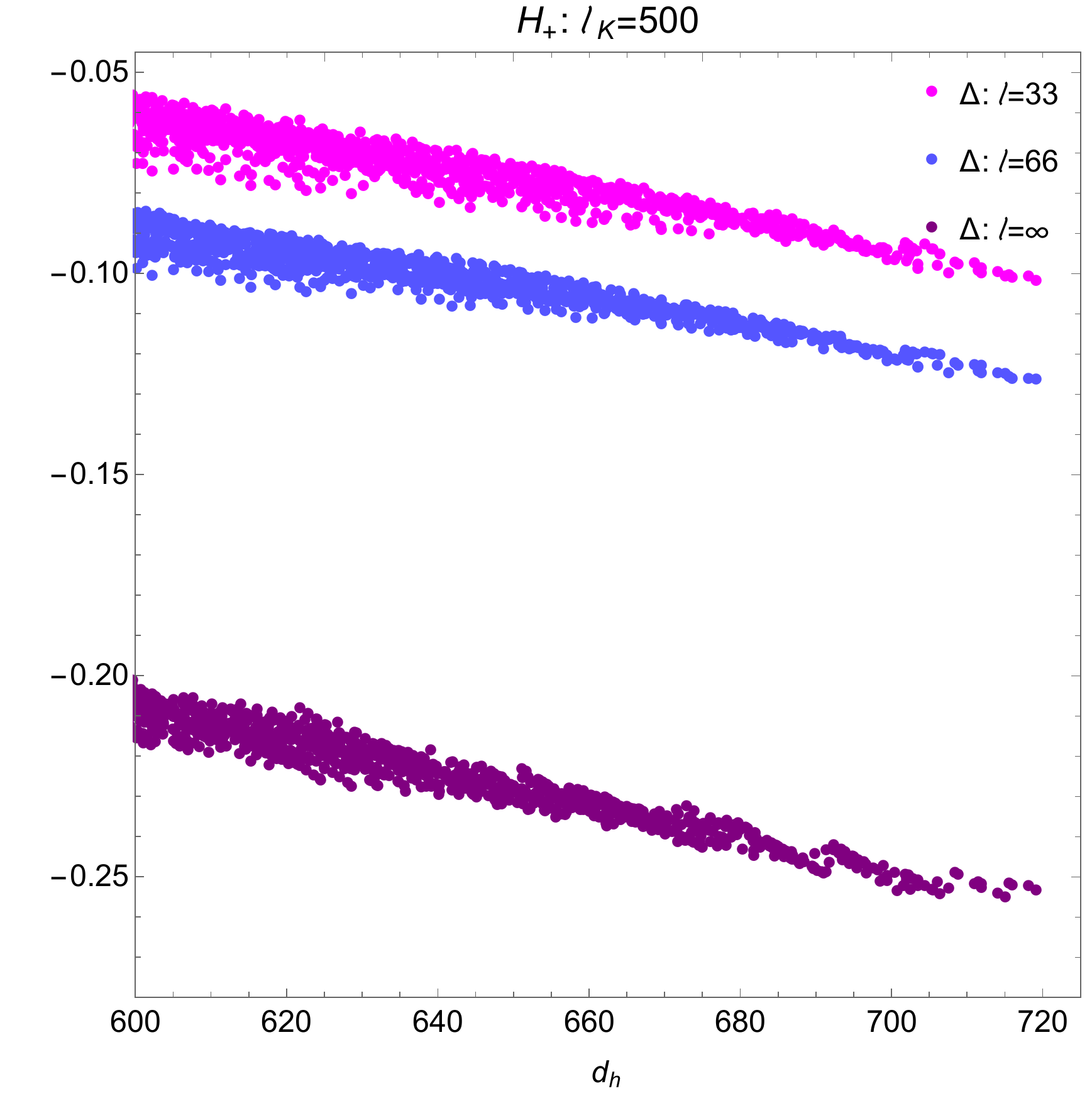}\quad\includegraphics[width=0.5\linewidth]{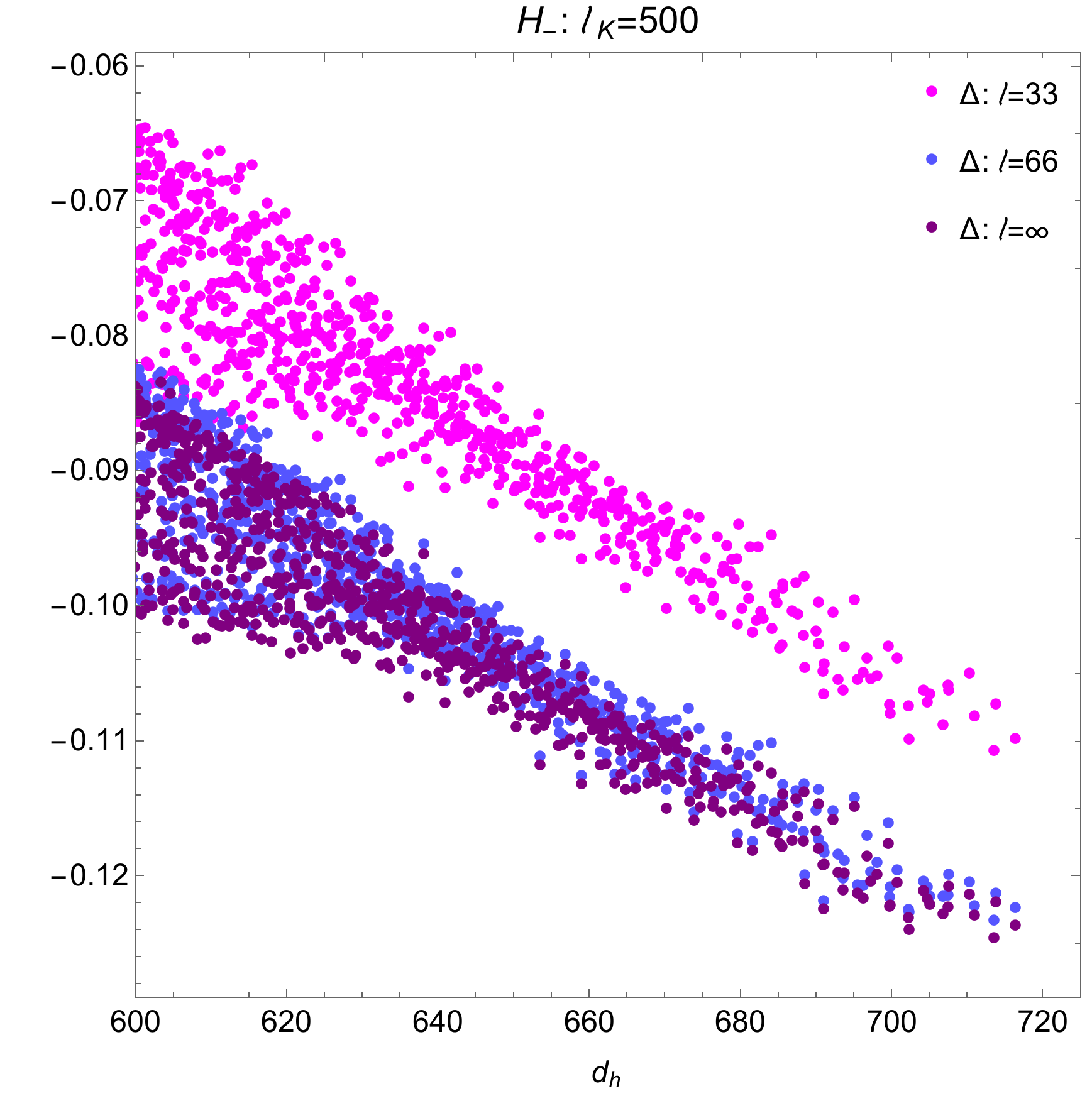}
\caption{\label{fig:CC_Klarge_zoom} $\Delta$ for different values of $\mco$, for both the $H_+$  (left panel) and the $H_-$  (right panel) cases, for large $d_h$.}
\end{figure}

 Next we consider the case when there is no separation of scales, i.e. when $\ell_K$ is too small. To that end we analyse a configuration with $\ell_K=500$.  Here we find that for $H_+$ and $H_-$, since there is no sufficient separation of scales,  no choice of $\mco$ gives a good convergence, cf.~Fig.~\ref{fig:CC_Klarge} and Fig.~\ref{fig:CC_Klarge_zoom}.  
This suggests that $\mco/\ell_K $  is required to be at most  $\sim 10^{-2}$  to obtain convergence. 
The larger the curvature, the more narrow the regime for admissible choices of $\mco$.  Thus there is a critical value of $\ell_K$ above which no value of $\mco$ can be found {for which}  $\Delta \rightarrow 0$ for large enough $d_h$.  Fig.~\ref{fig:CC_Klarge}  shows that when the extrinsic curvature is too large, $\dd$ either severely underestimates ($H_+$) or severely overestimates ($H_-$) the continuum distance, $\dcont$. 

In  App.~\ref{app:constK} we show more examples of our numerical simulations. In the constant $K$ case, we find that for $\ell_K =580$, there is a range of $\mco$ for which $\Delta \rightarrow 0$ for large enough $d$, suggesting that one has passed the critical value of $\ell_K$. Simulations confirm that the critical value of $\ell_K$ is in fact  $\mco/\ell_K \approx 0.12 $.  We also present several results for non-constant extrinsic curvature, and in each case find that there is convergence for a range of $\mco$ when there is a sufficient separation of scales.

\subsubsection{Three dimensions}

\begin{figure}[!t]
\includegraphics[width=0.45\linewidth]{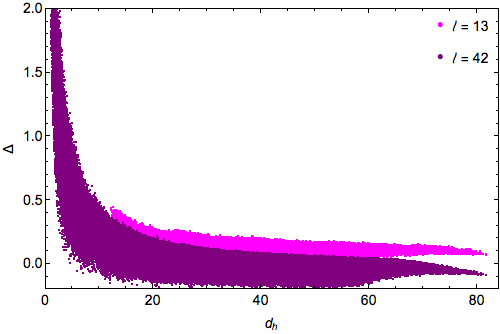}\quad\includegraphics[width=0.45\linewidth]{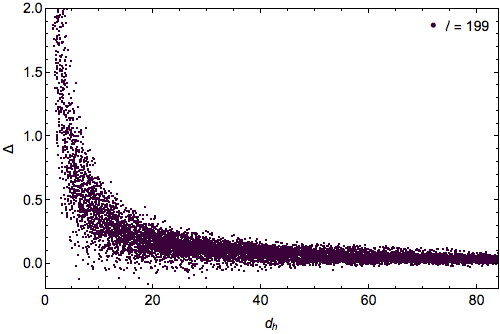}
\caption{\label{fig:3d_flat_ddisc_cont}Left: Plot of $\Delta$ for the flat, three-dimensional case. The effects of asymptotic silence become negligible at considerably smaller continuum distances than for the two-dimensional case, which is shown for comparison (right panel).}
\end{figure}

Here we  test the three-dimensional case for the simplest boundary $K=0$ in $\bM^3$,  cf.~Fig.~\ref{fig:3d_flat_ddisc_cont}.  Here, the effects of DAS  appear to be restricted to smaller continuum distances compared to that  in  two dimensions, as one can see by comparing the left and right panel in Fig.~\ref{fig:3d_flat_ddisc_cont} (see \cite{Eichhorn:2017djq}). 
Thus, $\mco$ can be chosen to be smaller than in the two-dimensional case, allowing us to probe curvature at smaller scales. 

If $\Vas$  represents a dimension independent  DAS volume,  then from Eq.~\eqref{predistance}
\be\label{eq:dAS}
\ell_{DAS}^{[ { n}]} = 2\left(\frac{\Vas}{\zeta_{ n}} \right)^{1/{ n}},
\ee
where $\ell_{DAS}^{[{ n}]}$ is the  associated length scale in $\dimm$ dimensions. Thus, 
\be
\frac{\ell_{DAS}^{[2]}}{\ell_{DAS}^{[3]}} = \frac{\zeta_3^{1/3}}{\zeta_2^{1/2}}\, \Vas^{1/6}.
\ee
As $\zeta_3 > \zeta_2>1$ and $\Vas>1$, the ratio is necessarily larger than one
\be
\frac{\ell_{DAS}^{[2]}}{\ell_{DAS}^{[3]}} > 1,
\ee
from which we conclude that the {\it linear} asymptotically silent regime becomes smaller as the dimension increases. 
Using the results of Sec.~\ref{sec:2dflat}, an estimate for  $\ell_{DAS}^{[3]}$ can be obtained by using $\Vas= \zeta_2\, \left( \ell_{DAS}^{[2]}\right)^2\approx 199^2\, \zeta_2$, yielding for Eq.~\ref{eq:dAS} in three dimension
\be
\ell_{DAS}^{[3]} \approx 42.
\ee

The analysis above is supported by  the numerical simulations:  $\mco\, =\, 42$ lies well outside of the DAS regime.
Analogous to the 2 dimensional case, choosing the mesoscale cutoff too low (cf.~magenta dots in left panel of Fig.~\ref{fig:3d_flat_ddisc_cont}) yields an overestimation of the $\DD$'s at large continuum distances, as most of the contributing piecewise distances are overestimated due to DAS. 
The behaviour of $\Delta$ at $d\approx  60$ in the left panel of Fig.~\ref{fig:3d_flat_ddisc_cont} could be due to boundary effects which, as mentioned earlier, do become more relevant as the dimension increases.

\subsubsection{A dimension estimation from an intrinsic comparison}
 
\begin{figure}[t!]
\centering
\includegraphics[width = 0.48\textwidth]{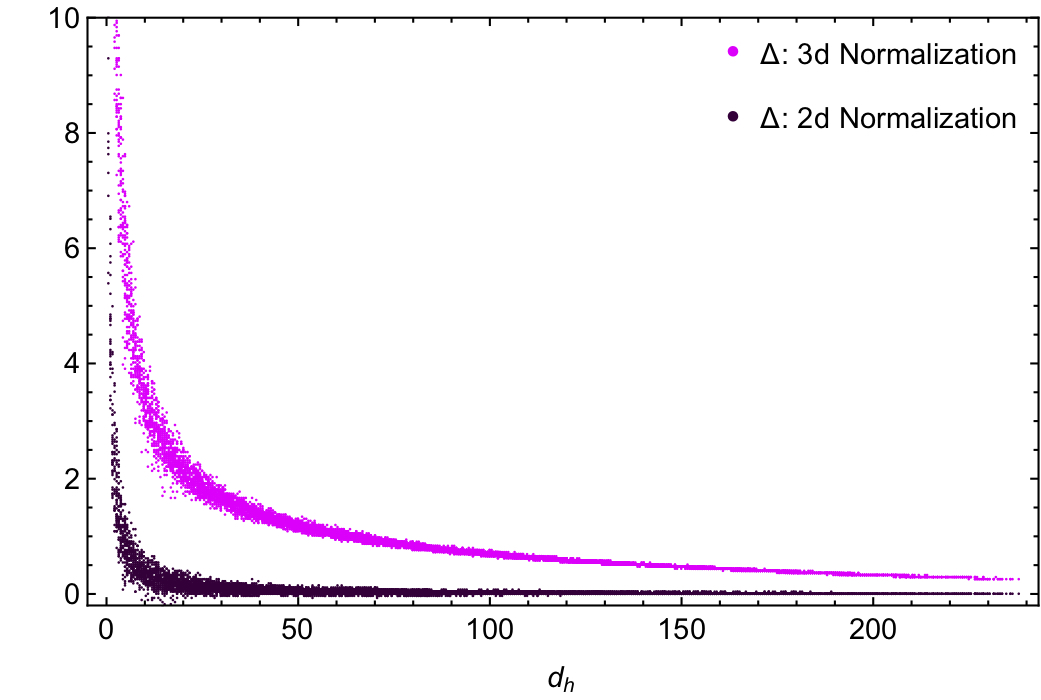}\quad \includegraphics[width=0.48\textwidth]{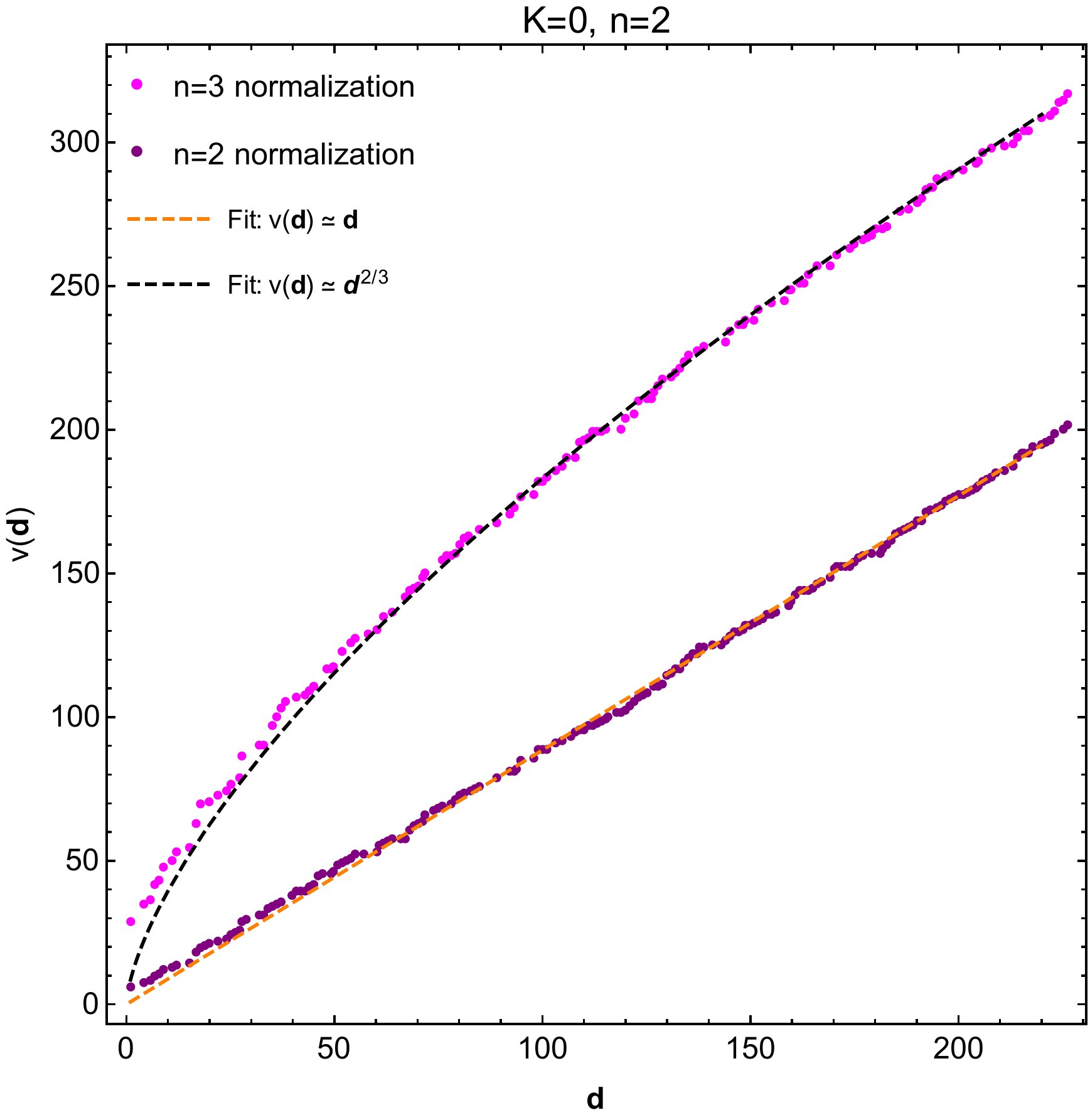}
\caption{ Left panel: For an antichain without extrinsic curvature for $\mathds{M}^2$, the error  between discrete and continuum distance is calculated using a two dimensional (magenta) and a three dimensional (purple) conversion between the discrete volume and the predistance. As expected, a three dimensional conversion factor yields an overestimation of $\DD$. Right panel: Spatial volume as a function of the distance in the ordered antichain for a 3d and a 2d conversion.}
\label{fig:dim_est}
\end{figure}

\begin{figure}[!t]
\includegraphics[width=0.45\linewidth]{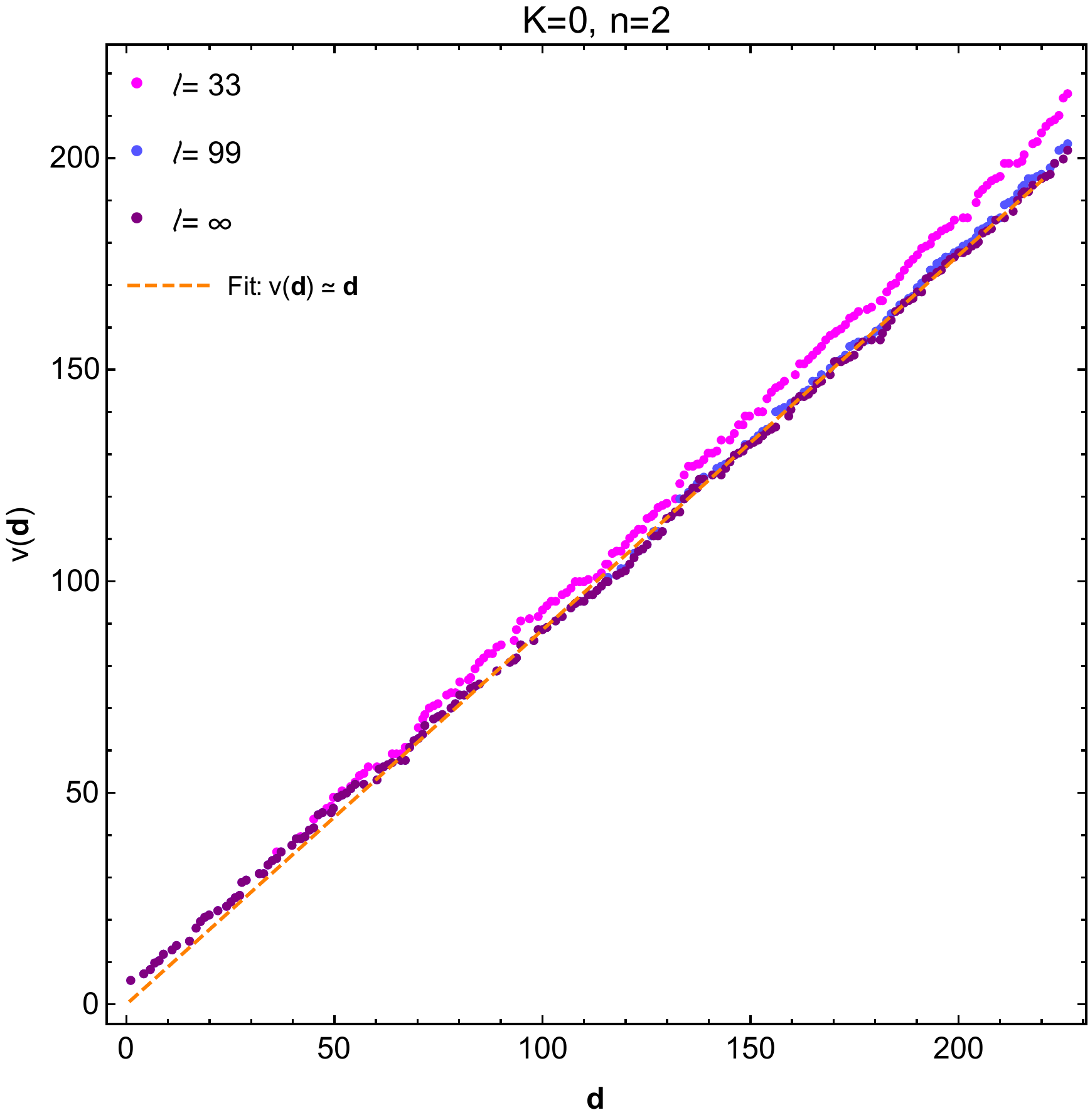} \quad \includegraphics[width=0.47\linewidth]{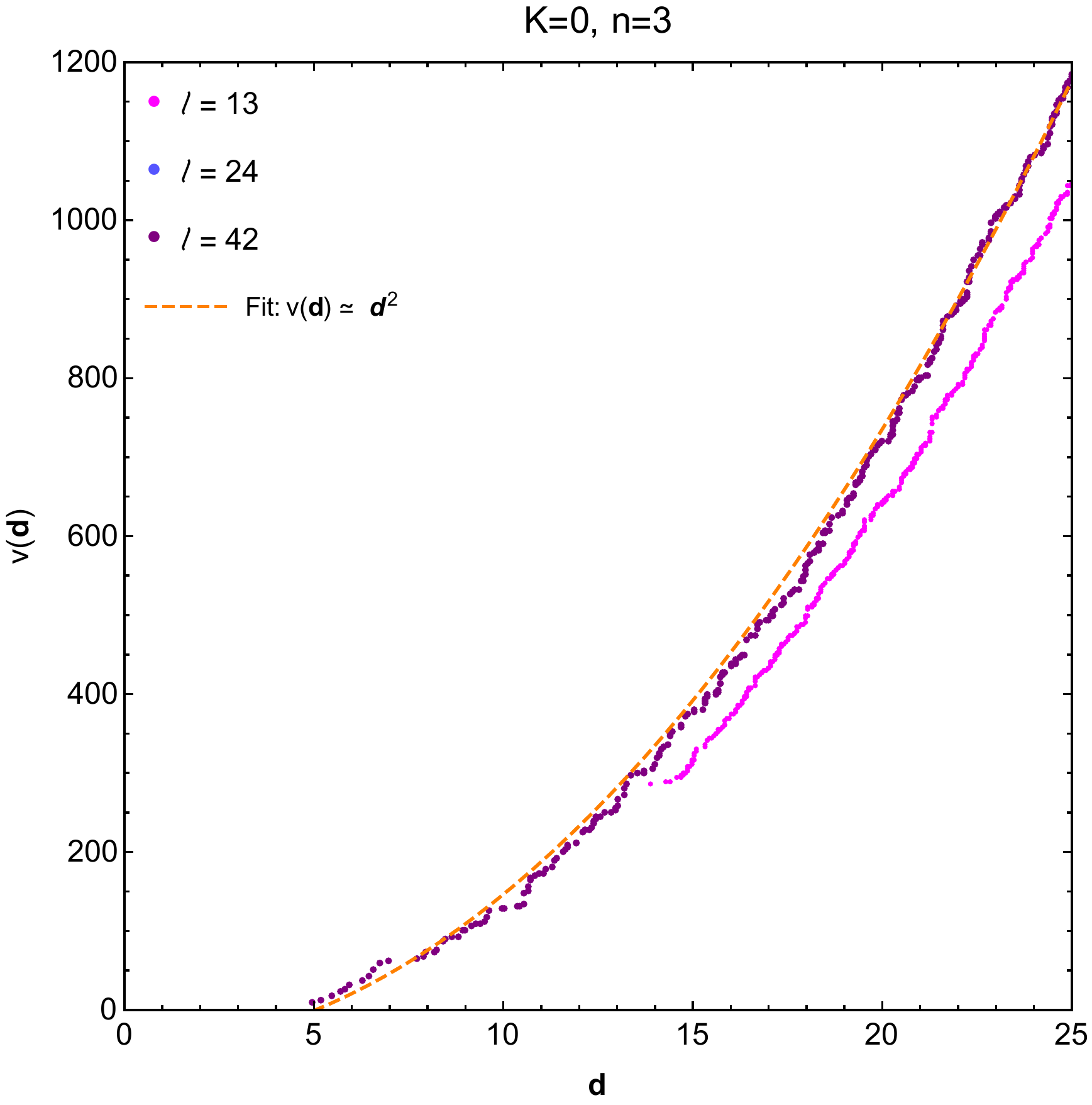}
\caption{\label{fig:Nd2mco} Scaling of the spatial volume with the distance for various choices of $\mco$. Left: in two dimensions the spatial volume scales linearly with the discrete distance. For a choice of $\mco = \mco_{DAS}$ (pink dots), there is a clear deviation from the expected  scaling (orange, dashed line). Right: In three dimensions the scaling is quadratic. Similarly to two dimension, when $\mco $ lies within the DAS regime, a jump in the spatial volume is observed. Note that the line $\mco = 42$ (purple dots) lies on top of $\mco = 24$ (blue dots), indicating that the latter already suffices as a cut-off in $3$ dimensions. }
\end{figure}

\begin{figure}[!t]
\includegraphics[width=0.45\linewidth]{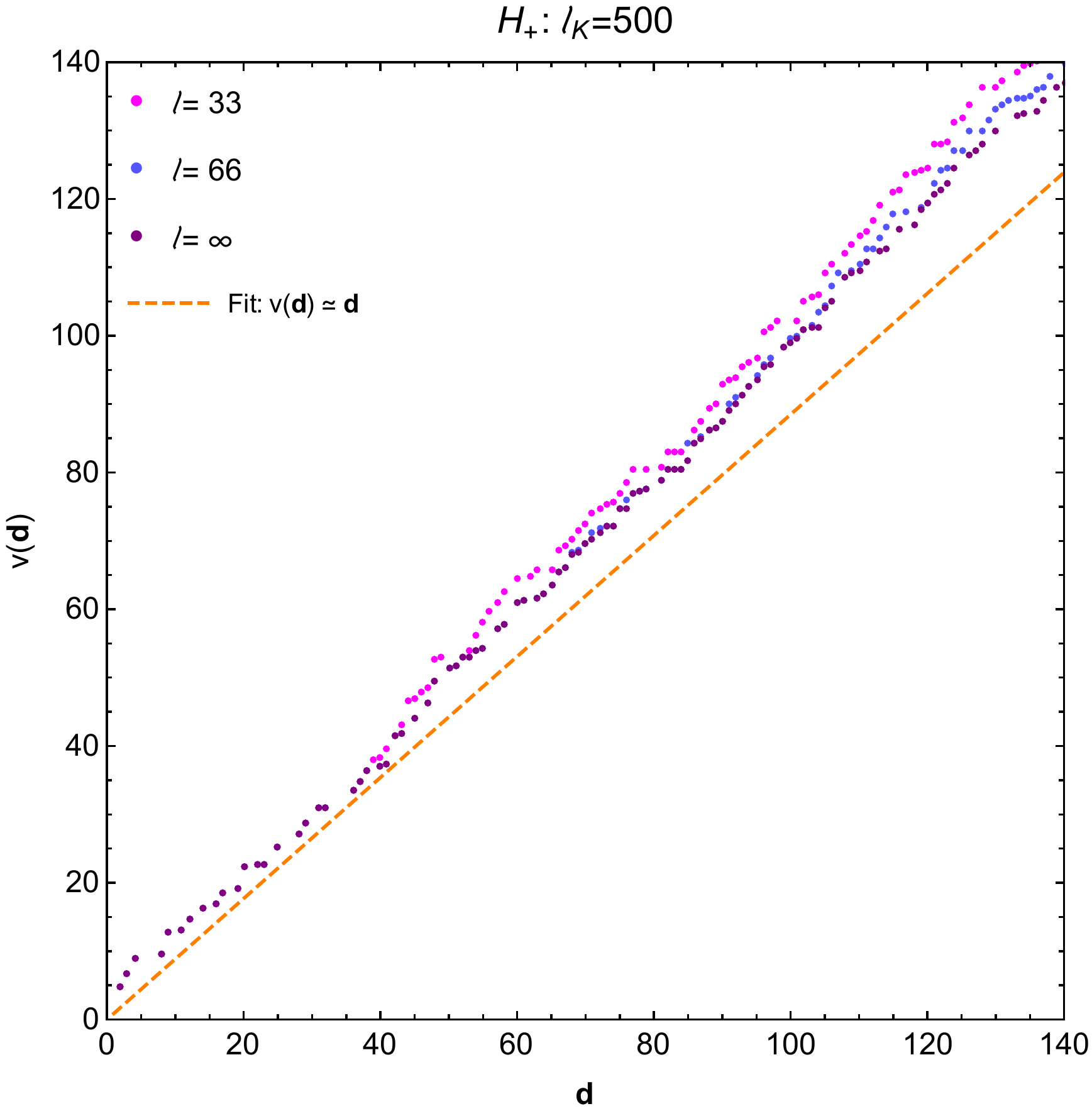}\quad
\includegraphics[width=0.45\linewidth]{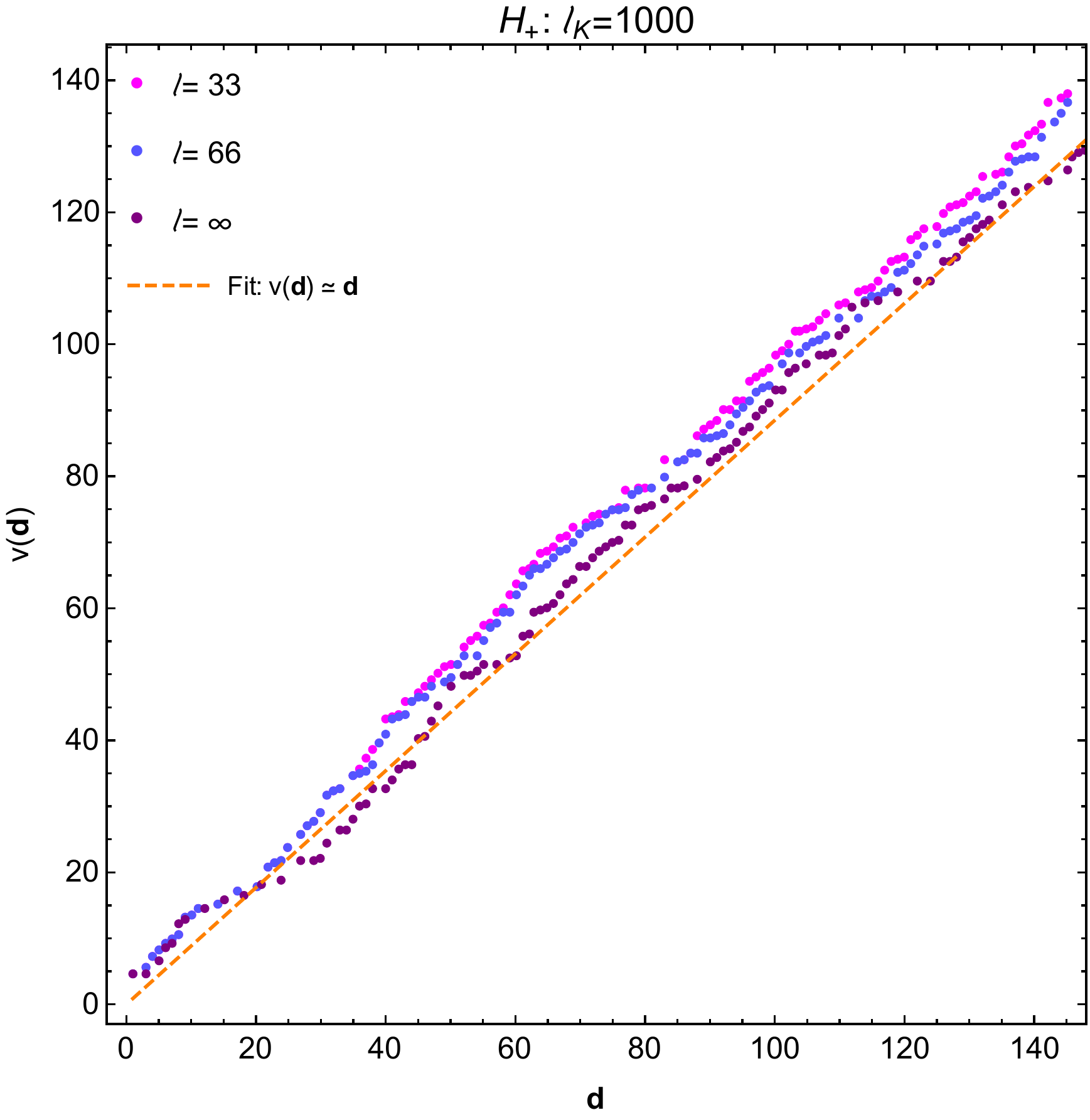}
\caption{\label{fig:Nd2mcolK}  { Two-dimensional case.} Left: At an insufficient separation of scales, the scaling of the spatial volume with the discrete distance is off compared to the flat case (orange, dashed line). Right:  At large enough separation of scales the expected scaling behaviour is present.}
\end{figure}

The distance function $\DD$ is of course intrinsic to the causal set, and as such we do not need to compare it with the continuum. We can instead {examine} an intrinsic measure of this distance function as follows. Let $a$ be a randomly chosen  element in $\cA$ and let $v(\DD_a)$ be the number of elements in $\cA$ which are within a distance $\DD$ from $a$. This ``spatial volume'' function  $v(\DD_a)$ should  monotonically increase  with $\DD_a$. In Fig.~\ref{fig:dim_est} and Fig.~\ref{fig:Nd2mco}  for $K=0$ boundaries and Fig.~\ref{fig:Nd2mcolK} for $K\neq 0$ in $\bM^2$ and $\bM^3$, we see the expected linear behaviour of $v(\DD)$ in $2$d and the quadratic behaviour in $3$d. 


In calculating $\DD$, however, we do use the continuum dimension.  In fact, using the wrong dimension in Eq.~\eqref{predistance} gives rise to a detectably different distance function.  In $2$ dimensions, for example, since $\left(\frac{V}{\zeta_3}\right)^{1/3}\, <\,  \left(\frac{V}{\zeta_2}\right)^{1/2}$, an underestimation/overestimation in both the predistance and distance is expected when a larger/smaller dimension than the actual dimension of spacetime is used, cf.~Fig.~\ref{fig:dim_est}.  The  use of the $d=3$ formula to calculate $\DD$ in $2$d  is also shown, cf.~Fig.~\ref{fig:dim_est}. The left panel shows that the  error $\Delta$ is significantly larger but of course this is based on the  continuum comparison. What is more striking is the intrinsic calculation of $v(\DD)$ which not only deviates from linearity, but fails to resemble the actual $3$d case, i.e. the right panel of Fig.~\ref{fig:Nd2mco}. In particular $v(\DD)$ becomes a concave function of $\DD$, as opposed to a convex one, expected for all higher dimensions. 

This suggests that one can use $v(\DD)$ as a new dimension estimator. 
An important  challenge will be to proceed {\it intrinsically}  at scales much smaller than the curvature scale $\ell_K$.  Indeed,  our results suggest a procedure of obtaining this scale. As in \cite{homologyone,homologytwo}, where there is a mesoscale regime where the homology is stable, one would also expect a stable regime for $\DD_\mco$, for which $v(\DD)$ does not change significantly  as function of $\mco$. However, as $\mco$ approaches $\ell_K$, one would expect a significant change in $v(\DD)$, thus marking, at least approximately,  the smallest curvature scale on $\cA$. We defer the fleshing out of some of these ideas in more detail to future work.

\subsubsection{A Non-manifoldlike causal set: The Lightcone Lattice}

\begin{figure}[!t]
\includegraphics[width = 0.45\textwidth]{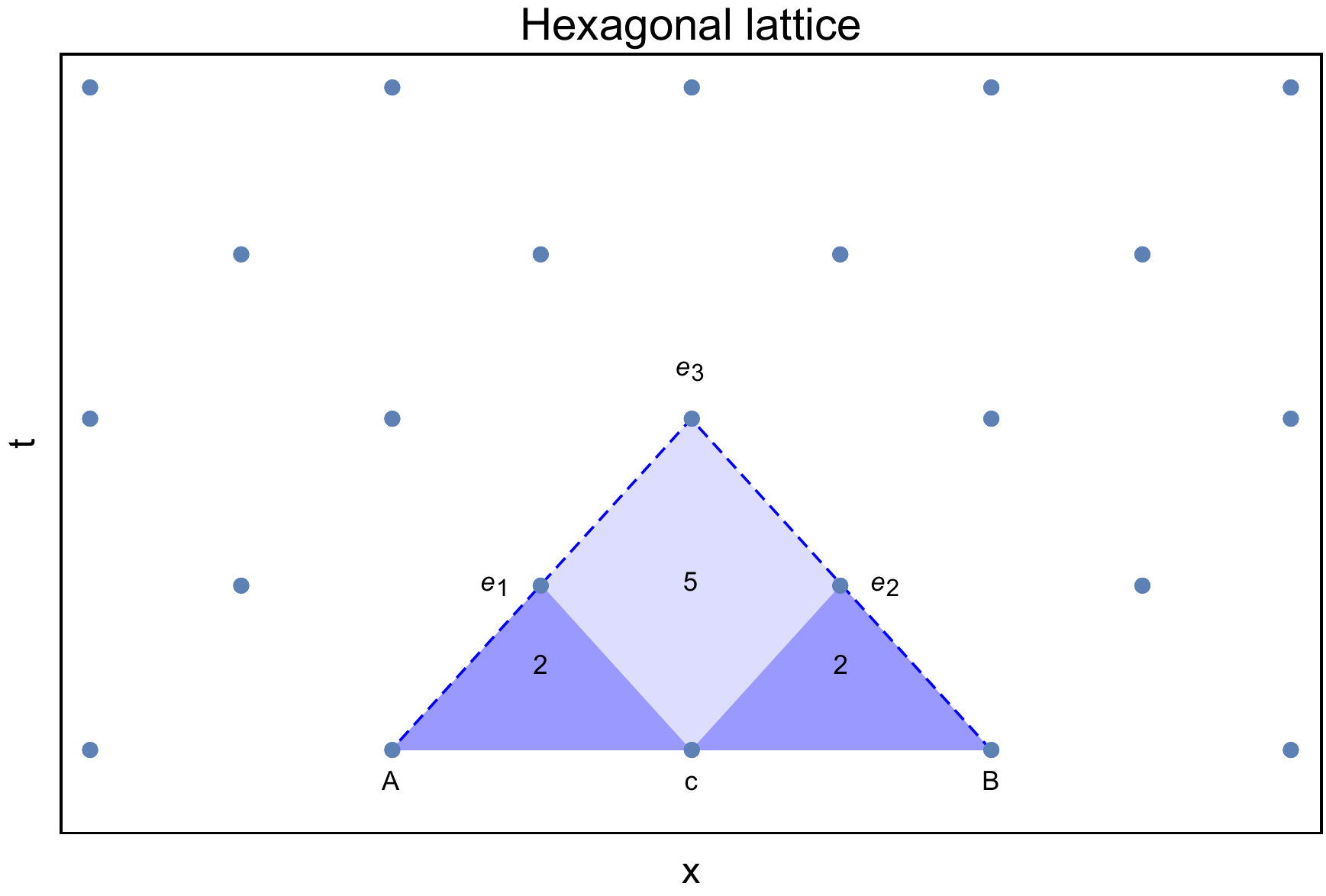}\quad \includegraphics[width = 0.45\textwidth]{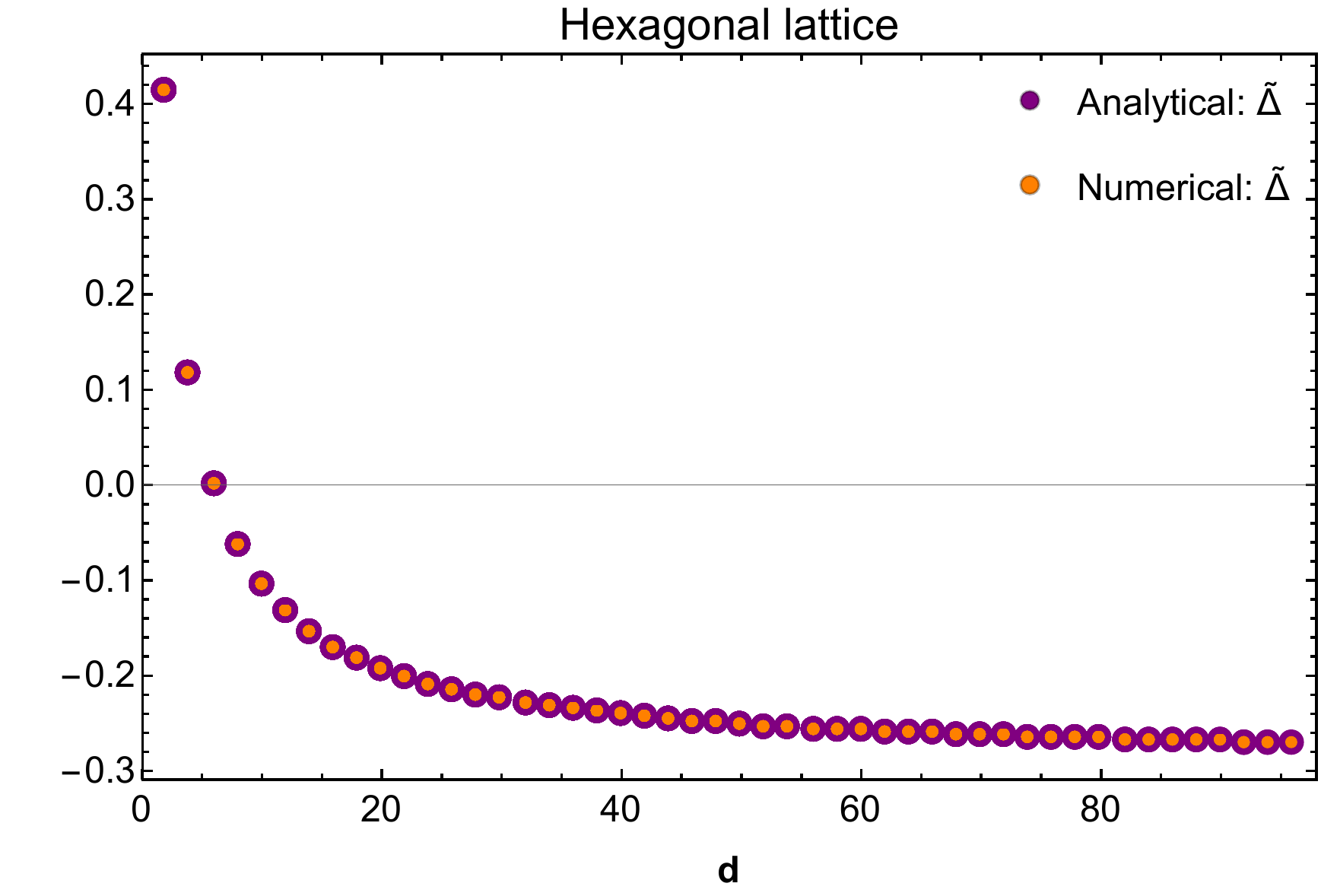}
\caption{  \label{lattice.fig} Left: For the hexagonal lattice the discrete volume given by the beam $e_3$ is equal to $5$ and corresponds to a direct jump from $A$ to $B$, whereas an intermediate jump to $c$ yields a discrete volume of $4$, given by the sum of the beams $e_1$ and $e_2$. Right: The numerical and analytical results for the predistance agree up to errors in the numerical precision, signalling the correct implementation of the predistance function.  As expected the lightcone lattice introduces an underestimation in the error of $\sim 0.3$.\label{fig:lattice}}
\end{figure}

A natural question to ask is what the  distance function gives us  for a causal set that is not manifold-like. As formulated in Eq.~\eqref{predistance}, there need be no reference to the continuum, except in the choice of a continuum dimension. As an example, let us consider the $2$-d lightcone lattice in Fig.~\ref{fig:lattice}. This is not a  manifold-like causal set since the apparent number to volume correspondence is in actual fact frame dependent. 

Such a lattice is characterised by its fundamental spatial length cut-off $\lco$, rather than by spacetime density $\rho$. Using our criterion for distance we see that the distance between two elements $p,q$ is simply given by $d(p,q)=2 \sqrt{V(r)}$ where again $r$ is the first element in the common future of $p,q$. Unlike in a  manifold-like causal set, $r$ is right where it should be in the continuum, i.e., it is $\rmin$,  and hence there can be no overestimation of the distance and hence no DAS. As shown in Fig \ref{fig:lattice}, if $p,q$ are at  $t=0$, then  $V(r)=t(r)^2$, where $t(r)$ is the coordinate time. Thus $d(p,q)=2\times t(r)$ and $\Delta$ is exactly zero, i.e. there is no error. Importantly, as a result, there is also no DAS, just as expected for the lightcone lattice.

Critically, $V(r)$ does not correspond to the number of elements since the correspondence $V \sim N$  depends on the inertial frame. We illustrate this by calculating the error $\Delta$ that arises if a frame-dependent number-to-volume correspondence is used.
Let us use a $V\sim N$ relation   {\it in this frame}  by counting the number of elements to the past of $r$ including $r$ itself.  For example, if $t(r)=\lco$, then $V(r)=3 \lco^2$. Using Eq.~\eqref{predist}, $\prd(p,q)=2 \sqrt{V(r)} \sim 2 \sqrt{3} \lco$, which is an overestimation of the continuum distance $d_h(p,q)=2 \lco$. This resembles DAS, though its origins lie in the inappropriate identification of $V(r)$ with the cardinality.  More generally, for $t(r)=n \lco$,  $V(r) $ is the sum $\sum_{i=0}^{n-1} i  \lco^2 = \frac{n(n-1)}{2} \lco^2$. This gives $\prd(p,q)= \sqrt{2 n(n-1)} \lco$,  which for large $n$ is $\sim \sqrt{2} n \lco$. On the other hand, the continuum distance $d_h(p,q)=2 (n-1) \lco \sim 2 n \lco $. Hence $\prd$ underestimates the distance for large $n$.  The error $\wD \sim 0.3$, which is not small, and most importantly remains finite at large $d_h$.  

\section{Summary and outlook}

\subsection{Synopsis of the key ideas}

In this work we have presented a key step in the derivation of spatial geometry from causal structure. Specifically, we have provided a definition of a  spatial distance on a spatial hypersurface purely from the causal structure and the local spacetime volume element, using a  piecewise linearisation of the Lorentzian structure.  The underlying idea has far-reaching conceptual consequences, as it suggests a radical order theoretic reinterpretation of spacetime.  Using the HKMM theorem, we can reimagine spacetime as a  partially ordered set with a local volume-element. Such a framework lends itself naturally to causal set theory, where the assumption of discreteness and the number to volume relation suffice to give us spacetime in the continuum approximation. The construction of the spatial distance from the continuum can thus be simply translated into the language of causal sets, with the spatial hypersurface replaced by its discrete  analogue,  namely an antichain.

Our construction in the continuum uses a piecewise linearisation of Lorentzian geometry.  At small enough scales, the extrinsic and intrinsic curvature effects become negligible and spacetime is approximately flat. In flat spacetime, the spatial distance function on an inertial  spatial hypersurface $\Sigma$ can readily be  obtained from the causal structure and volume as follows. For any  $p,q \in \Sigma$ there is a unique event $\rmin$ in their common lightlike (null) future which  minimises the proper time $T$ to $\Sigma$.  The past lightcone of $\rmin$ throws a ``beam" onto $\Sigma$ which is the intersection $J^-(\rmin) \cap \Sigma $, with $p,q$ on its boundary, separated by the spatial distance $2T$. The volume of the past of $\rmin$ upto $\Sigma$, $V(\rmin)=\vol(J^-(\rmin) \cap J^+(\Sigma))$
 is a simple function of $T$, Eq.~\eqref{zetadef}.  This means that $\rmin$ also minimises the past volume upto $\Sigma$, and hence $V(\rmin)$ can be used as a proxy for $T$, indicating that the  spatial distance between $p,q$ on $\Sigma$ is  a simple function of $V(\rmin)$,  Eq.~\eqref{predist}. This construction can be  generalised to  a spacetime  with non-vanishing intrinsic curvature containing  a hypersurface with non-vanishing extrinsic curvature using a piecewise linearisation of the Lorentzian structure. The flat spacetime construction only provides a predistance function and one has to minimise over all discretised paths between $p$ and $q$ in $\Sigma$ to get a  distance function which satisfies the triangle inequality.  Since curvature effects can lead to an over or underestimation of the actual induced distance we introduce an additional parameter, a  mesoscale cut-off $\mco$,   which limits the step-size of the paths, and hence the error in the piecewise linearisation.  For a compact Cauchy hypersurface we argue that the error can always be bounded by choosing $\mco$ small enough. 

In the causal set, there are important differences in the construction.  For $p,q$ in the antichain $\cA$,  $\rmin$ as defined above is almost surely not {an element of} the causal set, since in the continuum it is null related to $p,q$, and hence the intervals $J^+(p) \cap J^-(\rmin)$ and $J^+(q)\cap J^-(\rmin)$ have zero volume. From Eq.~\eqref{poisson}  this means that the probability of there being a non-zero number of elements in this region is zero. Instead, one considers an element $r$ in the common future of $p$ and $q$ which minimises the discrete past volume $V(r)$ to $\cA$,  and hence $V(r)$ is strictly greater than $V(\rmin)$. Here, we have associated a continuum volume with the discrete volume by using the statistical correspondence between the number of elements in a spacetime region and its volume at a given density $\rho$ or spacetime volume cut-off $\rho^{-1}$. In our simulations we have assumed $\rho=1$  in Planckian units.   For small distances, the difference in $V(\rmin)$ and $V(r)$ is more significant because of statistical fluctuations which generate larger local voids. Thus, since $V(r) > V(\rmin)$,  small continuum distances are overestimated in the causal set, or what we term ``discrete asymptotic silence (DAS)''  (see \cite{Eichhorn:2017djq}).  The associated scale $\ell_{DAS}$ at which this is significant  is larger than the discreteness scale. Thus $\mco$ is bounded from both above and below (hence the use of the term  ``mesoscale''), and one needs a sufficient separation of scales: $\ell_{DAS} \ll \mco \ll \ell_K$. Using detailed and extensive numerical simulations of two and three dimensional causal sets that are approximated by Minkowski spacetime, with inextendible antichains corresponding to hypersurfaces with and without extrinsic curvature, we have explored the interplay of these three scales. We have shown  that when there is a  sufficient separation of scales, $\mco>33$ in Planckian units, and $\mco/\ell_K \lesssim 10^{-2}$,  a stable range for the mesoscale cutoff can be found, such that the distance $\DD$ converges to the continuum induced  distance $d_h$. This is a clear demonstration of the practicality of our construction, suggesting that it can be used in diverse contexts in causal set theory.  

Finally, we have shown  evidence that the distance function can be used intrinsically as a dimension estimator by finding a spatial volume $v(\DD)$ on the antichain. Using the stability of $v(\DD_\mco)$ as a function of $\mco$ further allows us to obtain the intrinsic smallest curvature scale on $\cA$. A more detailed exploration of these new features will be studied in future work.

\subsection{Outlook}

Our work paves the way for a number of interesting questions.

Having a spatial distance function means that there is a  definition of  local neighbourhoods on the antichain as we have seen from our numerical simulations, and
thus a recovery of spatial locality of the type used in most constructions of physical theories.  This, we believe, is a  step towards understanding how causal sets relate to other approaches to quantum gravity. 

The distance function makes it possible, moreover,  to recover more aspects of the continuum spatial geometry from the causal set.  For example, the methods of  \cite{Klitgaard:2017ebu,Klitgaard:2018snm} can be used  to extract  a local {\sl spatial} Ricci scalar curvature on the antichain \cite{ESV}. This  would provide  a useful spatial  observable in the Hartle-Hawking calculation which could characterise, for example,  the spatial flatness of the final  antichain, of relevance to the early universe.  

The distance function also gives us a  one-parameter family of spatially connected graphs from the nodes of the antichain and associated simplicial complexes.  This is an alternative method of constructing topological invariants and it would be interesting to compare this  with the construction of \cite{homologyone, homologytwo}.  The  mesoscale corresponds to a ``maximum'' allowed thickening of the antichain, ala \cite{homologyone, homologytwo} and hence one would expect a similar ``stable'' regime of homology.

The construction of local neighbourhoods in the antichain provides us with the key element necessary to define a spatial diffusion process on the antichain. From this, the spectral dimension of the antichain can be extracted. As the size of the neighbourhood is reduced,  one expects a decreasing spectral dimension -- i.e.,  dimensional reduction at small scales.  In \cite{AstridSebastian} a diffusion 
process was set up on the entire causal set. However, due to the inherent non-locality, the spectral dimension was seen to diverge rather than decrease at smaller scales. The connected spatial graph on the antichain, on the hand {\it is} local. It would be useful therefore to see if causal sets also belong to the larger class of quantum gravity models that show dimensional reduction for spatial diffusion processes.

{Our construction of the distance suggests an intrinsic way of extracting a rough estimate of the extrinsic curvature. It is based on the observation that for choices of $\ell$ that satisfy $\ell_{DAS}<<\ell<<\ell_K$, the distance function $\DD(a,b)$ between the pair of points $a,b$ exhibits a stable regime. Specifically, an estimate of $\ell_K$ can thus be obtained by finding the maximum value of $\ell$, such that $\DD(a,b)$ is independent of the choice of $\ell$. As is suggested for example by Fig.~\ref{fig:CC}, once $\ell \lesssim \ell_K$, $\DD(a,b)$ will depend on $\ell$. Note that this construction is intrinsic, as it only requires $\DD_\ell(a,b)$  to be constructed for various $\ell$. 
Fig.~\ref{fig:CC} also highlights that our construction of the distance function using future lightcones is more sensitive to positive curvature than to negative curvature. Therefore, for a better detection of negative curvature, one should simply use a distance function defined in a time-reversed fashion, from the past lightcones. 
To test the quantitative precision of this construction requires future studies.}

Finally, our work provides an explicit example an order-theoretic recreation of spacetime geometry using its causal structure,  rather than  its metric. Given that causal structure is physical -- whereas due to diffeomorphism symmetry the metric is not -- such a viewpoint appears more physical. In particular, it could provide a promising starting point to understand the quantum structure of spacetime. Despite this  radical reinterpretation of Lorentzian geometry, we see that no relevant information encoded in the metric is lost. In particular, information on the spatial geometry of a spatial hypersurface (key, for example  in the extraction of gravitational waves from numerical simulations of dynamical spacetimes)  is actually encoded in the causal structure and the local volume element. Conversely,  since physical observations are {\it always} done using causal and hence spatio-temporal information, {\it practically speaking} spatial geometry is implicitly  reconstructed from this information. Our construction thus brings us closer to actual observation and experiment than the traditional metric based formulations.

{\bf Acknowledgments:} 
A.~Eichhorn and F.~Versteegen were supported by the DFG under the Emmy-Noether program, grant no.~EI-1037-1. 
This research was supported in part by Perimeter Institute for Theoretical Physics. Research at Perimeter Institute is supported by the Government of Canada through the Department of Innovation, Science and Economic Development and by the Province of Ontario through the Ministry of Research and Innovation. S.~Surya was supported by a FQXi grant FQXi-MGA-1510 and an Emmy Noether Fellowship, at the Perimeter Institute. A.~Eichhorn was supported by a visiting fellowship at the Perimeter Institute.

{
\appendix
\section{}
\label{app:num}
The sizes of the causal sets used in the simulations can be found in the tables below.  Here $L_t,\, L_x,\, L_y$ refer to the continuum dimension of the patch of spacetime used for the sprinkling. $N$ is the size of the causal set and $AC_{size}$ the size of the corresponding, inextendible and minimal antichain.

\subsection{$K = 0$}
\begin{center}
  \begin{tabular}{ | l || c | c | c | c | c |}
  \hline
  $\ell_K = \infty$ & $L_t$ & $L_x$ & $L_y$& $N$ & $AC_{size}$\\
    \hline
    n = 2 &     400 & 400 & -& 58674 & 227 \\ 
    n = 3 &     60 & 60 &60& 75510 & 1571 \\ 
        \hline
  \end{tabular}
\end{center} 

\subsection{$K = const.$}
\begin{center}
  \begin{tabular}{ | l || c | c | c | c | c |}
  \hline
  $ \ell_K= const.$ &$\ell_K$& $L_t$ & $L_x$ & $N$ & $AC_{size}$\\
    \hline
    $H_+$  &     1000&1200 & 300 & 129473 & 148 \\ 
    $H_-$  &     1000&1200 & 300 & 130150 & 169 \\ 
        \hline
  \end{tabular}
\end{center} 

\begin{center}
  \begin{tabular}{ | l || c | c | c | c | c |}
  \hline
  $ \ell_K= const.$ &$\ell_K$& $L_t$ & $L_x$ & $N$ & $AC_{size}$\\
    \hline
    $H_+$  &     500&2020 & 270 &  197443 & 146 \\ 
    $H_-$  &     500&2020 & 270 & 197729 & 141 \\ 
        \hline
  \end{tabular}
\end{center} 

\subsection{$K \neq const.$}
\begin{center}
  \begin{tabular}{ | l || c | c | c |c| c | c |}
  \hline
  $ \ell_K\neq const.$ & $a$ & $b$ & $L_t$ & $L_x$ & $N$ & $AC_{size}$\\
    \hline
    $H_+$  &     485 & 97 &200& 200 & 14550 & 113 \\ 
    $H_-$  &     485 & 97&200&200 &14522& 101 \\ 
        \hline
          \end{tabular}
\end{center} 
Here $a,\, b$ are the parameters defining the hyperbola in Eq. ~\ref{eq:hyp}. 

\begin{center}
  \begin{tabular}{ | l || c | c | c |c| c | c |}
  \hline
  $ \ell_K\neq const.$ & $R$ & $L_t$ & $L_x$ & $N$ & $AC_{size}$\\
    \hline
    $C_+$ &     585 &400 & 400 &55174 & 229 \\ 
    $C_-$ &    585 & 400 & 400 & 56910 & 224 \\ 
        \hline
  \end{tabular}
\end{center} 
In the above $R$ is the (Euclidean) radius of the circle. 
}

\section{} 
\label{app.difftop}
 
The definitions of $\cH(p,q)$ and $V(r)$ are given in Section \ref{continuum}. 

\noindent {\bf Claim:} $\cH(p,q)$ is spacelike. 
 
\noindent {\bf Proof:} 

Let  $r_1,r_2 \in \cH(p,q)$ with $r_1\prec r_2$. Let  $\pprec$ denote  the timelike relation and $\rightarrow$ the null relation \footnote{Null related points in $\dmp$ are order theoretically defined as zero volume totally ordered sets, i.e., $p\prec q$ such that $\vol(J(p,q))=0$. Timelike relations are causal relations which are not null.}.  If $r_1 \pprec r_2$ then since $p,q \rightarrow r_1$ this implies that  $p,q \pprec r_2$,  which is a contradiction. Next, if $r_1 \rightarrow r_2$  then there must be a single null generator $\gamma$ from both $p$ through $r_1$ and onto $r_2$ and similarly from $q$, otherwise $r_2$ would not be null related to $p$ or $q$ \cite{penrose}. This is not possible for both $p$ and $q$ simultaneously, since otherwise $p$ and $q$ would lie on the same null generator. Hence $\cH(p,q)$ is a spacelike hypersurface. 

\noindent {\bf Claim:} $V(r)$ takes on a minimum value in  $\cH(p,q)$.

  \noindent {\bf Proof:}

$V(r)$ is a continuous function since $(M,g)$ is globally hyperbolic.  Consider $V(r)$ as a continuous function on $\overline{\cH(p,q)} \subset N_p$. By the extreme value theorem $\exists \,\, \rmin \in \overline{\cH(p,q)}$ such that $V(\rmin)=V(p,q)$.  Let us assume that  any such $ \rmin \nin \cH(p,q)$, i.e., $\rmin \in \pd \overline{\cH(p,q)}$. Now, $V(r)$ is given by  Equation (\ref{GNCvol}), where the proper time  $T$ also corresponds to the GNC time coordinate $t$ which labels the foliations $\Sigma_t$ in $N_\Sigma$.  Equation (\ref{GNCvol}) implies that for $KT$ small enough, the leading order contribution to $V(r)$ is the same for any point in $U_t \equiv \overline {\cH(p,q)} \cap \Sigma_t$ and decreases with $t$. If $t_{\mathrm{min}}$ is the proper time of $\rmin$ this implies that $U_t$ is empty for all $t < t_{\mathrm{min}}$.  For any strictly smaller neighbourhood $M_p' \subset M_p$ of $p$ (i.e. such that $\overline{M_p'} \subset M_p $) which also contains $q$, $D(M_p') \subset D(M_p)$ and $\pd \overline{D(M_p)}\cap \overline{D(M_p')}=\emptyset$.   Again, defining $\cH(p,q)'\equiv \pd  J^+(p)\cap \pd  J^+(q) \cap D(M_p')$, we see that  it is simply the restriction of $\cH(p,q)$ to $D(M_p')$ and moreover cannot be empty since $p,q$ lie in an RNN.  Thus, $U_{t_{\mathrm{min}}}\cap D(M_p') \neq \emptyset$ and hence $\exists \, r'\in \cH(p,q)' \subset \cH(p,q)$ with $V(r') = V(p,q)$ which is a contradiction. 

\section{}
\label{app:constK}
\subsection{ \bf K$=$ const }

\begin{figure}[!t]
\includegraphics[width=0.5\linewidth]{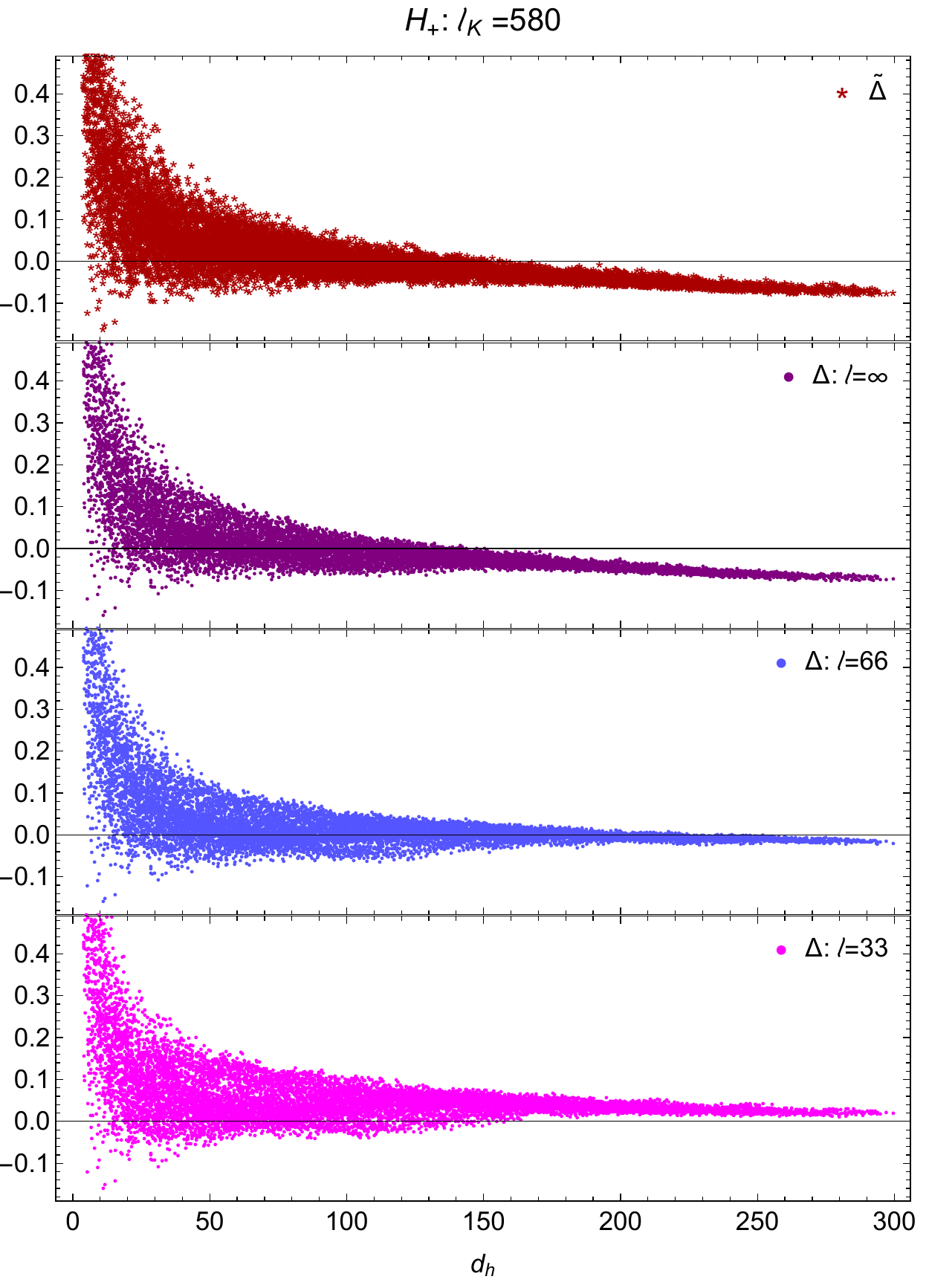}\quad\includegraphics[width=0.5\linewidth]{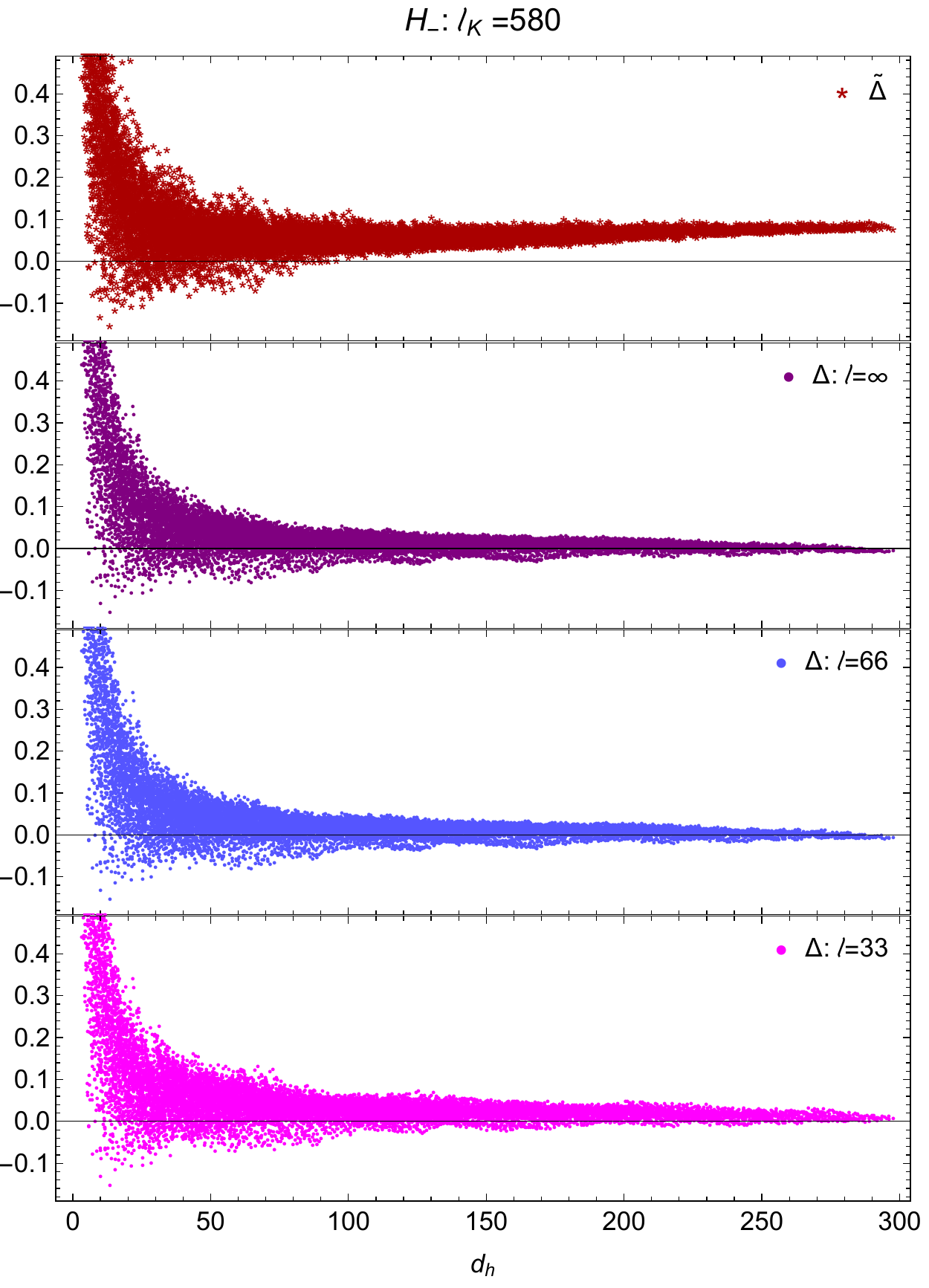}
\caption{\label{fig:CClK580}$\wD$ (red stars) and $\Delta$ (magenta, blue and purple dots)  for positive extrinsic curvature (left) and the negative extrinsic curvature (right) for the complete range of $d_h$ (left panel) at a radius of curvature $\mco_K = 580$.}
\end{figure}

 Here we show an additional K$=$const. case, for a radius of curvature that lies in between $\mco_K = 500$ and $\mco_K = 1000$, Fig.~\ref{fig:CClK580}. Already here the results converge towards $\dcont$ \sout{becomes apparent} for values of the cut-off larger than $\mco_{DAS}$.
 
\subsection{K $\neq$ const}
\noindent {\bf Hyperbolae}\\
We  now turn to boundaries with non-constant extrinsic curvature which are still hyperbolae. 
\begin{figure}[!t]
\includegraphics[width=0.5\linewidth]{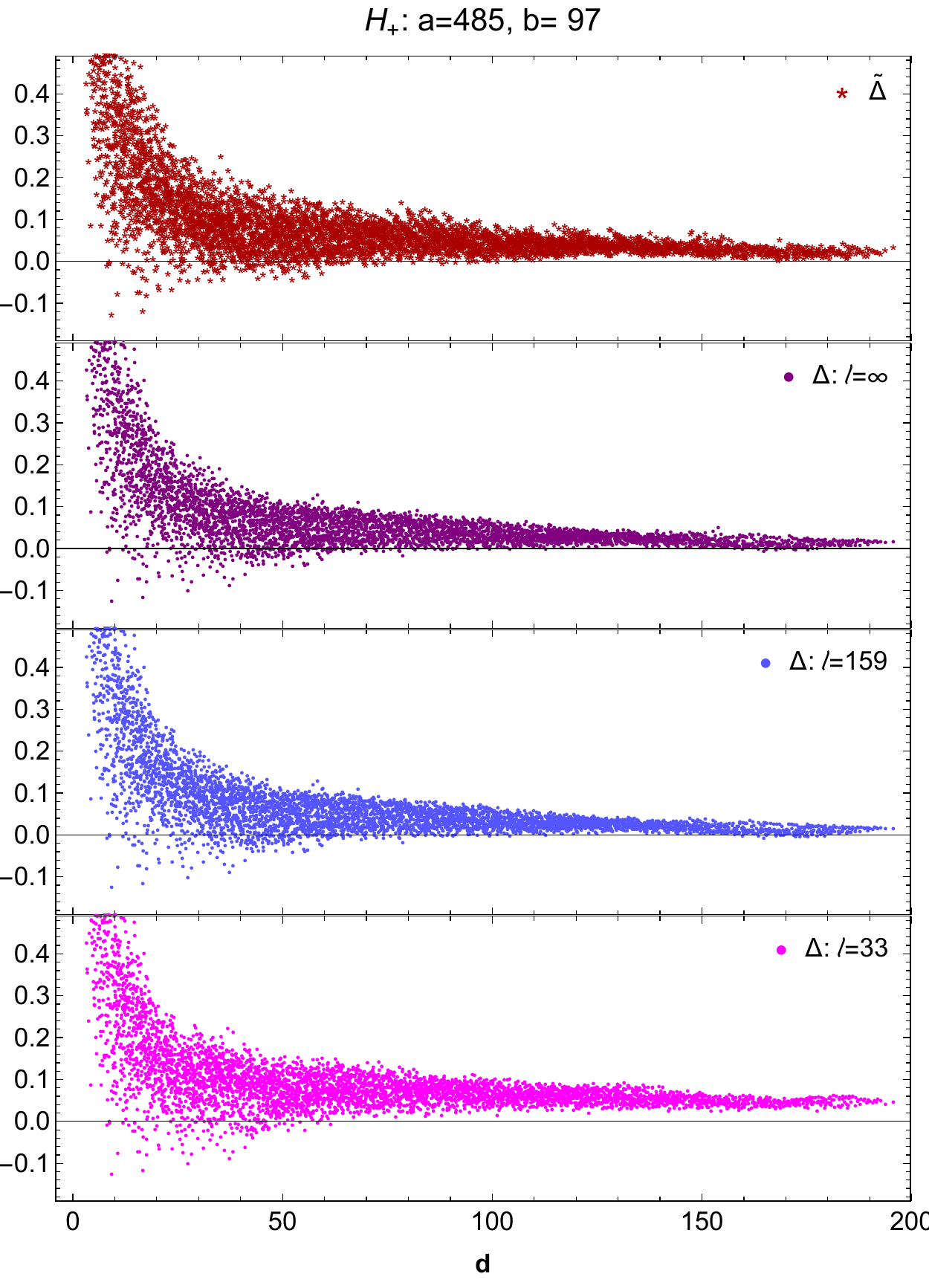}\quad\includegraphics[width=0.5\linewidth]{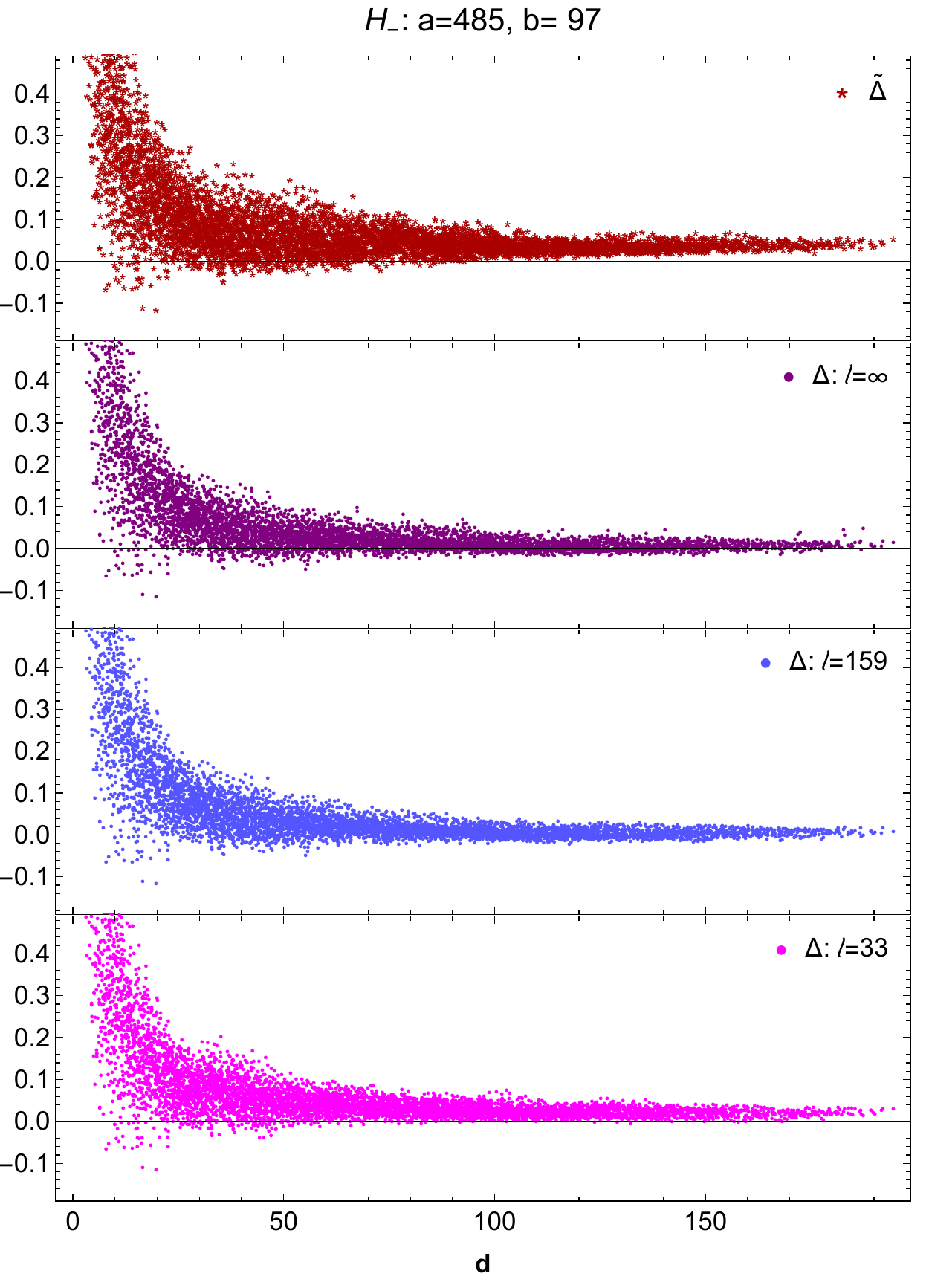}
\caption{\label{fig:Hyp}$\wD$ (red stars) and $\Delta$ (magenta, blue and purple dots)  for the hyperbolic case I (left) and II (right) for the complete range of $d_h$ (left panel).}
\end{figure}
As expected, there is an effect of a curvature scale $\ell_K$ even if it is not uniform:  increasing $\mco$ beyond a critical value results in a much larger value of $\Delta$, as shown in  Fig.~\ref{fig:Hyp}. For the cases we explore, where $a\sim 485,\, b\sim 97$, choosing $\mco \sim 164$  reduces the error $\Delta$,  whereas $\mco \approx 199$ results in a visible deviation,  cf.~Fig.~\ref{fig:Hyp}.

\noindent{\bf Circle}

In addition to the above cases of hyperbolic boundaries with non-constant curvature, we examine boundaries which are sections of circles. While the circle  has a constant extrinsic curvature in Euclidean space, it does not in a  Lorentzian spacetime.

\begin{figure}[!t]
\includegraphics[width =0.5\linewidth]{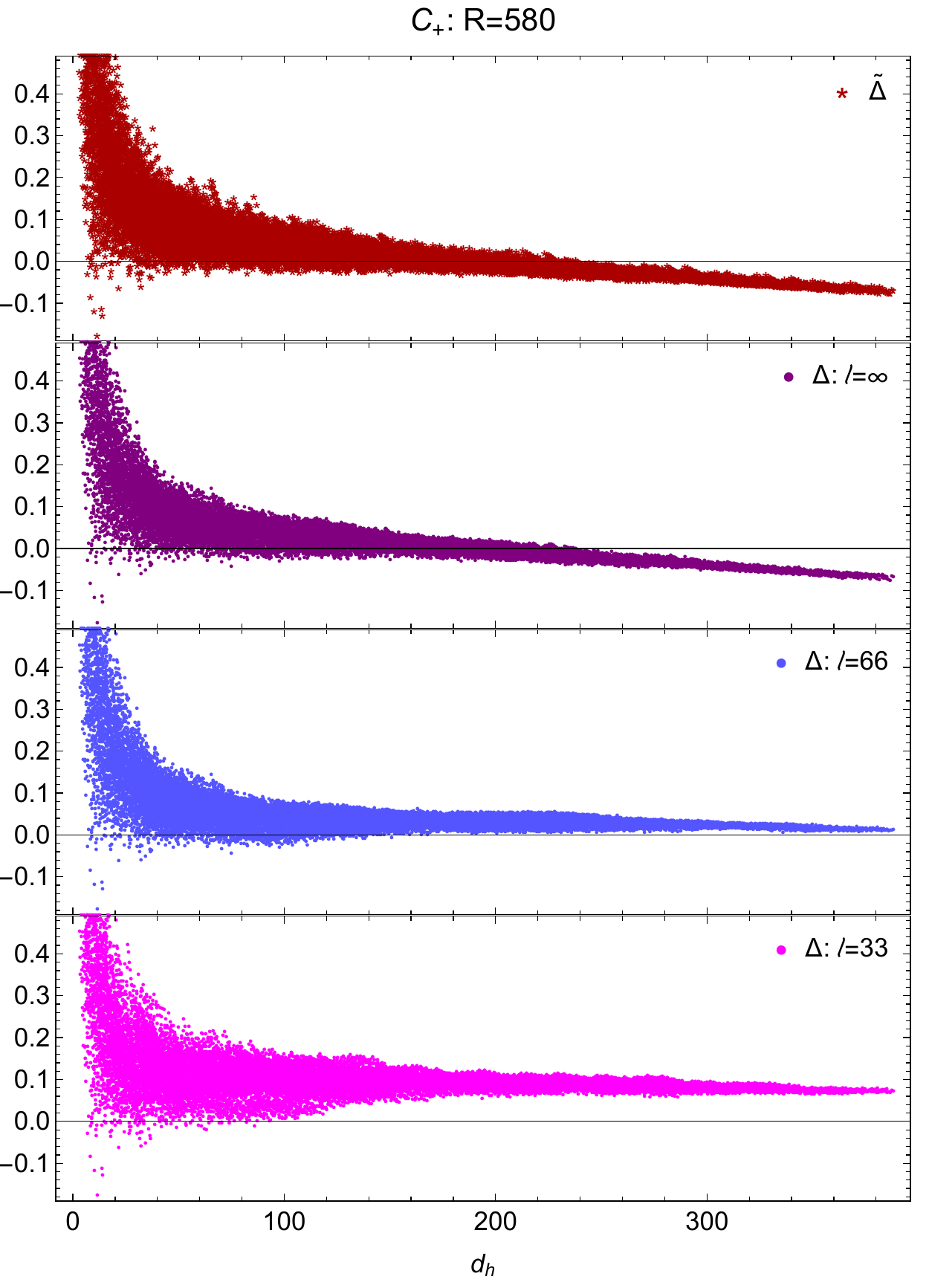}\quad\includegraphics[width=0.5\linewidth]{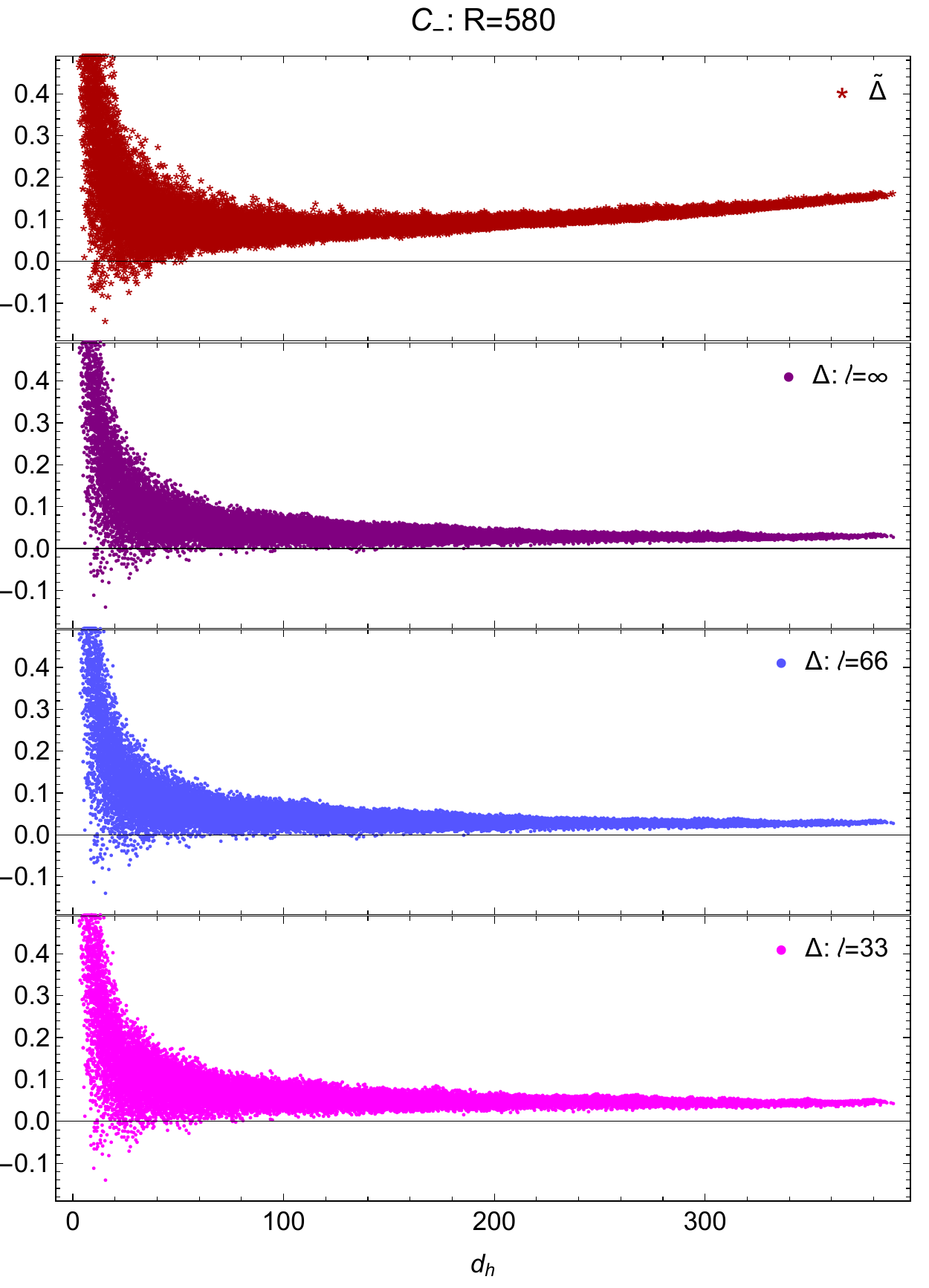}
\caption{\label{fig:Cir}The predistance (red asterisks) either underestimates or overestimates the distance at large values of $\dcont$ due to curvature effects, such that either $\Delta'<0$ or $\Delta'>0$. A large cutoff ($\mco=\infty$, purple dots) is insufficient for the $C_+$ case, such that $\Delta$ also drops below zero.}
\end{figure}

Our first result highlights the fact that the predistance is no longer a useful distance estimator  in this case, cf.~Fig.~\ref{fig:Cir}. As is obvious from the zoomed-in plot in the right panel of this figure, the error $\Delta'$ for the predistance goes to  zero around $d_h\approx 180$. Yet, one immediately sees that this is not due to a convergence of $\Delta' \rightarrow 0$, since $\Delta'$ continues to decrease past this point becoming increasingly negative as $d_h \gtrsim 200$.  In addition, the predistance does not fully  converge to $d_h$ even at large $d_h$ for $C_-$ (negative non-constant extrinsic curvature), cf.~Fig.~\ref{fig:Cir}, right panel, leading to an overestimation in this case. However, the size of the error appears to be similar than what was observed in the $K=0$ case, cf.~Fig.~\ref{fig:flatcutoff}.

These effects become more pronounced, the larger the curvature is chosen.
 As expected, if $\mco$ is chosen too large, the error in $\DD$ grows similar to that for the predistance. $C_+$ shows $\Delta<0$ at large continuum distances (cf.~Fig.~\ref{fig:Cir}), while $C_-$ is less sensitive to changes in the cut-off as long as one does not venture into the DAS regime. This is similar to the case of the hyperbolae.

\end{document}